\documentclass[11pt]{article}
\pdfoutput=1  

\usepackage[T1]{fontenc}
\usepackage[utf8]{inputenc}
\usepackage{newtxtext}                       
\usepackage[a4paper,top=1in,bottom=1in,left=1.25in,right=1.25in]{geometry} 
\usepackage{amsmath}
\usepackage{amssymb}
\usepackage{newtxmath}                        
\usepackage[scaled=0.92,varqu]{zi4}           
\usepackage{booktabs}
\usepackage{longtable}
\usepackage{array}
\usepackage{graphicx}
\usepackage{microtype}
\usepackage[table,svgnames]{xcolor}
\usepackage{caption}
\definecolor{rowgray}{gray}{0.94}
\definecolor{headcol}{HTML}{E7EEF3}   
\definecolor{accentcol}{HTML}{0B5563} 
\newcommand{\best}[1]{\textcolor{accentcol}{\textbf{#1}}} 
\newcommand{\hdr}[1]{\textbf{#1}}                         
\usepackage[round,sort&compress]{natbib}
\definecolor{linkblue}{HTML}{1A4FD6}     
\usepackage{hyperref}
\hypersetup{colorlinks=true,citecolor=linkblue,linkcolor=linkblue,urlcolor=black}
\usepackage{xurl}            
\usepackage{enumitem}
\usepackage{parskip}
\renewenvironment{abstract}{%
  \small
  \begin{center}{\large\bfseries\sffamily\abstractname}\end{center}%
  \begingroup\leftskip1.5em\rightskip1.5em\noindent\ignorespaces}{\par\endgroup}
\usepackage[htt]{hyphenat}  
\usepackage{titlesec}
\titleformat*{\section}{\Large\bfseries\sffamily}
\titleformat*{\subsection}{\large\bfseries\sffamily}
\titleformat*{\subsubsection}{\normalsize\bfseries\sffamily}
\titleformat*{\paragraph}{\normalsize\bfseries\sffamily}

\newcommand{\nDCG}{nDCG@10}
\newcommand{\model}[1]{\texttt{#1}}
\newcommand{\secref}[1]{\hyperref[#1]{\S\ref*{#1}}}
\newcommand{\appref}[1]{\hyperref[#1]{Appendix~\ref*{#1}}}
\graphicspath{{figures/}}

\title{\large\sffamily\bfseries HAKARI-Bench: A Lightweight Benchmark for Comparing\\ Retrieval Architectures and Efficiency Settings under Unified Conditions}
\author{Yuichi Tateno \textless hotchpotch@gmail.com\textgreater}
\date{}

\begin{document}
\maketitle

\begin{abstract}
With the rapid spread of retrieval-augmented generation (RAG) and semantic search, choosing the right text embedding and retrieval configuration has become both important and difficult. Large-scale retrieval benchmarks are comprehensive but too heavy to run repeatedly during development, and there is little infrastructure for comparing production-time settings---dimensionality reduction, quantization, and reranking---across many models under identical conditions. We present \textbf{HAKARI-Bench}, a lightweight evaluation infrastructure that reconstructs existing retrieval benchmarks into small evaluation datasets (Nano-sets) and handles $35$ benchmarks and $551$ retrieval tasks spanning $43$ languages in a unified format. Each task shares a common format of corpus, queries, relevance labels, and a fixed candidate set, enabling same-condition, model-agnostic evaluation of five retrieval families (BM25, dense, sparse, late interaction, and rerankers), together with efficiency variants: Matryoshka dimensionality reduction, int8/binary quantization, and float rescoring. Evaluating $55$ models (dense $33$, sparse $4$, late interaction $6$, reranker $11$, BM25 $1$), HAKARI-Bench acts as a high-fidelity ranking proxy: on the common models and intersecting tasks of each comparison, its overall ranking reproduces the official MTEB retrieval v2, MMTEB v2 retrieval, and English BEIR (full) at Spearman $>0.97$ ($0.983$, $0.975$, $0.973$, respectively). HAKARI-Bench is not a replacement for full evaluation; rather, it supports rapid model selection, regression detection, and reading the quality--efficiency Pareto frontier under the same conditions. The Nano-sets, evaluation pipeline, and a multi-axis leaderboard are released as open-source software under the MIT license.

\end{abstract}

\addtocontents{toc}{\protect\setcounter{tocdepth}{-1}}

\section{Introduction}
\label{sec:intro}

With the spread of retrieval-augmented generation and similarity search, the development of retrieval models, including text embedding models, is increasingly active. Text retrieval can be viewed as two broad stages. First is \emph{candidate generation}, which retrieves candidate relevant documents from the whole corpus; alongside lexical-matching methods such as BM25, this stage uses models that represent queries and documents as dense or sparse vectors, and late interaction models that use token-level representations \citep{khattab2020colbert}. Second is \emph{reranking}, which more precisely re-orders the top retrieved candidates; rerankers serve this stage \citep{nogueira2019passrerank}. Real retrieval systems are sometimes built from candidate generation alone, and sometimes from a two-stage configuration combining candidate generation and reranking.

To compare such retrieval models, large evaluation benchmarks such as MTEB \citep{muennighoff2023mteb} and MMTEB \citep{enevoldsen2025mmteb} have been developed, making it possible to measure performance across diverse tasks in a unified way. In information retrieval, BEIR extended to multi-domain zero-shot evaluation \citep{thakur2021beir} the test-collection format (corpus, queries, relevance labels) that TREC standardized and popularized at scale \citep{voorhees2005trec}. BEIR organized scattered retrieval datasets into a unified format and made it possible to consistently compare the five retrieval architectures---lexical, sparse, dense, late interaction, and re-ranking---under a single model-agnostic framework (i.e., a framework in which, given the same task format and metric, any retrieval model can be swapped in for comparison).

In multilingual retrieval, MIRACL \citep{zhang2023miracl} evaluates monolingual retrieval (queries and corpus in the same language) across $18$ languages, and MS MARCO \citep{bajaj2016msmarco}, derived from Bing's search logs, is widely used for passage retrieval. More recently, domain-specific benchmarks have proliferated rapidly: code retrieval CoIR \citep{li2024coir}, long-document retrieval LongEmbed \citep{zhu2024longembed}, and expert-domain instruction-following retrieval IFIR \citep{song2025ifir}. As described later, all of these are targets that our benchmark incorporates as Nano-sets (\secref{sec:tasks}).

On the other hand, evaluating retrieval models by measuring only overall performance on large benchmarks is not enough. In production, retrieval quality is balanced against compute, memory usage, and latency by means of dimensionality reduction or quantization of the candidate-generation embeddings, and reranking of the top candidate set. In particular, retrieval performance after output-dimension reduction based on Matryoshka representation learning \citep{kusupati2022matryoshka}, or after quantization from floating point to int8/binary \citep{shakir2024quantization}, is among the most-watched areas after the base model performance itself. Hence, in addition to base model performance, it is important to understand how performance changes when these efficiency settings are used, or when a two-stage configuration of candidate generation followed by reranking over its candidate set is adopted.

However, existing large benchmarks cover many tasks at a large data scale, so re-evaluating all other models under the same conditions after changing dimensionality reduction or quantization for one model is not easy. In particular, the scaled-up MMTEB \citep{enevoldsen2025mmteb} infrastructure can in principle handle dimensionality reduction based on Matryoshka representations \citep{kusupati2022matryoshka} and quantized embeddings \citep{shakir2024quantization}, but in practice these efficiency settings are rarely reported consistently per model, and there are almost no results comparing multiple models under the same conditions. Moreover, frameworks that systematically compare, under the same conditions and according to their respective roles, the architecture that generates candidates from the whole corpus and the architecture that re-orders top candidate sets, are limited. From the same motivation, in the English domain NanoBEIR lightweights each BEIR dataset and is widely used as a fixed-dataset ranking proxy \citep{camara2024nanobeir, aarsen2024nanobeir}.

In this paper we build \textbf{HAKARI-Bench}, a lightweight benchmark for evaluating multilingual, multi-domain retrieval models.\footnote{The evaluation and visualization implementation is open source under the MIT license: \url{https://github.com/hakari-bench/hakari-bench}. The leaderboard of evaluated models is public at \url{https://huggingface.co/spaces/hakari-bench/leaderboard}; the evaluation data (Nano-sets) is released on Hugging Face Datasets (\appref{app:G}).} The name ``HAKARI'' comes from the Japanese word for a weighing scale (\emph{hakari}, ``to measure''), reflecting the benchmark's aim of measuring and comparing retrieval models.

Specifically, we construct small evaluation datasets (hereafter \textbf{Nano-sets}) from existing retrieval benchmarks and develop an infrastructure that handles $35$ benchmarks and $551$ retrieval tasks uniformly. Each task is handled in a common format consisting of a corpus, queries, relevance labels, and a top candidate set, so that candidate-generation methods and reranking methods can be evaluated with the same metrics according to their respective roles. We further make it possible to evaluate, under the same conditions, candidate-generation methods such as BM25, dense, sparse, and late interaction, as well as reranker evaluation over the top candidate set, embedding dimensionality reduction, int8 quantization, and binary quantization.

In this paper, ``lightweight'' refers solely to \textbf{reduced evaluation cost} (the ease of repeated measurement enabled by Nano-set construction). Dimensionality reduction, quantization, and sparse pruning are evaluated as \textbf{reproducible proxies for storage and retrieval cost} (embedding dimension, quantization precision, number of non-zero dimensions); we \textbf{do not evaluate the inference speed itself} of each model, because fair measurement is difficult (\secref{sec:speed}).

The positioning of our benchmark is summarized in three points. First, it follows the consistent evaluation methodology established by BEIR \citep{thakur2021beir}, i.e., same-condition comparison of multiple retrieval architectures based on a unified format. Second, it extends this to many models and to many languages and domains. Third, by shrinking the data size of each task, it makes $551$ retrieval tasks repeatedly measurable at a realistic speed, and on top of that applies dimensionality reduction, int8 quantization, and binary quantization to all supporting models under the same conditions. This means providing, in a consistent manner across all target models, the comparison of efficiency settings that is possible on the evaluation infrastructure but has in fact been measured only sporadically per model.

We also verify the extent to which Nano-sets reproduce the model ranking of the original large-scale evaluation. Comparing HAKARI-Bench's Nano-set results against MTEB retrieval v2, MMTEB v2 retrieval, and English BEIR (full), the Spearman rank correlations were $0.983$, $0.975$, and $0.973$, and the Pearson correlations were $0.981$, $0.969$, and $0.974$, respectively. In addition to rank correlation itself, high correlation was obtained for the Borda score that aggregates per-task wins/losses, indicating that while HAKARI-Bench does not replace full evaluation, it reproduces model ranking with high fidelity and functions as a lightweight evaluation metric.

Given the established fact that neural retrieval models degrade substantially out of distribution \citep{thakur2021beir}, the true value of a multi-domain benchmark is not maximizing the overall score, but exposing ``which domains a model has not learned'' and providing material for use-appropriate model selection. Our benchmark is designed so that tasks can be sliced and compared along axes such as language, domain, and query length, supporting this perspective (\secref{sec:usecases}, \secref{sec:modelselect}).

Based on the above, the contributions of this paper are threefold.

\begin{enumerate}[leftmargin=*]
\item \textbf{A lightweight multilingual, multi-domain retrieval evaluation infrastructure.} We reconstruct existing retrieval benchmarks as Nano-sets and build an infrastructure that compares the five families of BM25, dense, sparse, late interaction, and reranker in a unified format under the same conditions over $35$ benchmarks and $551$ tasks (\secref{sec:design}, \secref{sec:method}).
\item \textbf{Empirical validation of ranking reproducibility of the lightweight evaluation.} We show, through three independent comparisons, that the overall ranking induced by Nano-sets reproduces the official MTEB retrieval v2 / MMTEB v2 retrieval and BEIR (full) at Spearman $>0.97$ in every case, on the common models and intersecting tasks of each comparison (\secref{sec:corr}, \secref{sec:validity}).
\item \textbf{Cross-model evaluation of efficiency settings and reranking.} We applied dimensionality reduction, int8/binary quantization, and rescoring to all supporting models under the same conditions, and evaluated reranking over a fixed candidate set on all tasks. These settings are in principle measurable on existing infrastructure, but in practice have been reported only sporadically per model. Our contribution is not opening a new measurability for the first time, but actually providing it, by applying it uniformly to all supporting models so that efficiency and reranking performance can be compared across models on the same basis. This makes concretely visible the differences that emerge only when settings are held fixed---for example, that robustness to binary quantization is determined by a model's training characteristics (not explained by size or dimension), and that whether a reranker beats dense changes with the task type and architecture (\secref{sec:dimquant}--\secref{sec:rerankanalysis}, \appref{app:F}).
\end{enumerate}

\section{Related Work}
\label{sec:related}

\subsection{Retrieval evaluation benchmarks and retrieval architectures}
The evaluation format for text embedding models was standardized when MTEB \citep{muennighoff2023mteb} unified eight tasks including retrieval, reranking, classification, and clustering. MMTEB \citep{enevoldsen2025mmteb} extended the scope to over $250$ languages and over $500$ tasks, and introduced quality review and correlation-based downsampling. Restricting attention to information retrieval, BEIR \citep{thakur2021beir} assembled a zero-shot IR suite of $18$ datasets and made it possible to consistently evaluate the five retrieval architectures---lexical, sparse, dense, late interaction, and re-ranking---under a single model-agnostic framework. This design of ``comparing different retrieval architectures under the same conditions'' is the direct origin of our evaluation methodology (\secref{sec:method}).

For multilingual monolingual retrieval, MIRACL \citep{zhang2023miracl} is widely used; for short-query passage retrieval, MS MARCO \citep{bajaj2016msmarco}; and for domain specialization, code retrieval CoIR \citep{li2024coir}, long-document retrieval LongEmbed \citep{zhu2024longembed}, instruction-following retrieval FollowIR \citep{weller2024followir}, and expert-domain IFIR \citep{song2025ifir} have been developed.

More recently, the official MTEB leaderboard introduced a retrieval-specific section centered on RTEB (Retrieval Embedding Benchmark; \citealp{liu2025rteb}). RTEB is a retrieval-focused benchmark that measures multilingual retrieval quality across production domains such as legal, finance, code, and medical; it combines public datasets with private (closed) datasets to be robust to training-data contamination and leaderboard overfitting. However, like MTEB, RTEB evaluates embedding models under a fixed protocol and does not primarily aim at cross-architecture comparison (lexical, dense, sparse, late interaction, re-ranking) or comparison of efficiency settings such as dimensionality reduction and quantization. Our HAKARI-Bench moves in step with this retrieval-focused trend, but is complementary in that it performs cross-architecture comparison and efficiency-setting evaluation on top of lightweight measurement via Nano-sets.

The retrieval architectures under evaluation divide broadly into two stages: candidate generation, which retrieves candidates from the whole corpus, and reranking, which re-orders top candidates. Candidate generation commonly uses lexical-matching BM25 \citep[a robust baseline even for zero-shot IR;][]{robertson2009bm25, thakur2021beir}, dense retrieval with bi-encoders \citep{reimers2019sbert, karpukhin2020dpr}, learned sparse SPLADE-family models \citep{formal2021splade}, and token-level late interaction \citep[ColBERT family;][]{khattab2020colbert, santhanam2021colbertv2}. Reranking, starting from two-stage retrieval that re-orders BM25 candidates with a BERT cross-encoder \citep{nogueira2019passrerank}, is a re-ordering of top retrieved candidates, not a retrieval over the whole corpus. Because the two have different roles, they should be evaluated according to their respective roles rather than compared in the same role.

These benchmarks have greatly contributed to the comprehensive comparison of retrieval models, but the more comprehensive and large-scale they become, the harder it is to repeat the full evaluation during development. We restrict our scope to retrieval and reranking precisely because a lightweight, same-condition infrastructure is needed to iteratively check retrieval-specific comparison axes such as candidate generation, reranking, and efficiency settings.

\subsection{Lightweight evaluation and Nano-set construction}
To lower the cost of repeated evaluation on large benchmarks, the evaluation data has been lightweighted. MMTEB \citep{enevoldsen2025mmteb} shrinks tasks through correlation-based downsampling, showing a policy for obtaining conclusions close to the full evaluation at low cost. As a smaller-scale lightweighting, NanoBEIR is a collection that shrinks each BEIR dataset to about $50$ queries $\times$ up to $10$K documents; it was introduced by Zeta Alpha for evaluation-cost reduction \citep{camara2024nanobeir} and unified into a single format by Sentence Transformers as the \texttt{NanoBEIREvaluator} \citep{aarsen2024nanobeir}. Negative documents are sampled with Pyserini's BM25 and a general-purpose dense model. For lightweight reranker evaluation, a derived collection with BM25 candidate scores attached to the Nano-sets has also been prepared \citep{sentencetransformers2024nanobeirbm25}. Multilingual extensions have been released as translated/improved versions by LightOn AI \citep{sourty2025nanobeir}, Liquid AI \citep{liquidai2025nanobeir}, and Sionic AI \citep{sionic2025nanobeir}.

Our Nano-set construction follows this idea of a ``ranking proxy on a small collection'' and the practice of fixing BM25 candidates, and is distinctive in extending the net to non-English languages, expert domains, and comparison of efficiency settings including dimensionality reduction and quantization.

\subsection{Evaluating embedding efficiency settings}
\label{sec:related-eff}
In production, embedding dimensionality reduction and quantization are widely used to balance retrieval quality against compute, memory, and latency. Matryoshka representation learning \citep{kusupati2022matryoshka} is a method that trains embeddings so that the dimensions can be truncated while preserving the leading dimensions, giving an axis for comparing retrieval performance after dimensionality reduction. Embedding Quantization \citep{shakir2024quantization} combined int8/binary quantization with float rescoring to reduce storage and retrieval cost. In particular, the two-stage configuration of ``generating candidates efficiently with binary codes and re-ranking (rescoring) accurately with continuous vectors'' traces back to the Binary Passage Retriever \citep{yamada2021bpr}. Production approximate nearest neighbor (ANN) search uses more advanced quantization, such as Product Quantization \citep{jegou2011pq}, Optimized PQ \citep{ge2013opq}, RaBitQ with a theoretical error bound \citep{gao2024rabitq}, Better Binary Quantization with correction terms \citep{trent2024bbq}, Optimized Scalar Quantization \citep{veasey2026bbqturbo}, and TurboQuant, which combines Hadamard rotation with re-normalization and calibration \citep{pijpelink2026turboquant}.

However, evaluation infrastructure that can compare the impact of these efficiency settings on retrieval quality across many models under the same conditions is limited. Large-scale infrastructure such as MTEB / MMTEB \citep{muennighoff2023mteb, enevoldsen2025mmteb} can in principle handle dimensionality reduction, but performance changes after quantization or dimensionality reduction are often confined to per-model initialization settings, and cross-model same-condition comparisons are rarely reported. We treat these efficiency settings as first-class records of the evaluation results and compare quality and efficiency side by side in the same table (\secref{sec:dimquant-method}, \secref{sec:dimquant}).

\subsection{Positioning relative to existing benchmarks}
\label{sec:positioning}
We summarize the relationship between the existing benchmarks discussed above and HAKARI-Bench in Table~\ref{tab:compare}. In the table, $\bigcirc$ = consistently provided as a first-class feature across retrieval tasks, $\triangle$ = limited (possible on the infrastructure but not consistently reported across models, or restricted to dedicated tasks / separately distributed data), $\times$ = out of scope.

\begin{table}[t]
\centering
\caption{Comparison of existing retrieval evaluation benchmarks and HAKARI-Bench.}
\label{tab:compare}
\scriptsize
\setlength{\tabcolsep}{3pt}
\renewcommand{\arraystretch}{1.2}
\begin{tabular}{@{}>{\raggedright\arraybackslash}p{2.6cm}>{\raggedright\arraybackslash}p{2.0cm}>{\raggedright\arraybackslash}p{2.0cm}>{\raggedright\arraybackslash}p{2.0cm}>{\raggedright\arraybackslash}p{2.3cm}>{\raggedright\arraybackslash}p{2.4cm}@{}}
\toprule
\rowcolor{headcol} \hdr{Aspect} & \hdr{BEIR} & \hdr{MTEB} & \hdr{MMTEB} & \hdr{NanoBEIR} & \hdr{HAKARI-Bench} \\
\midrule
Retrieval task scale & 18 datasets & retr.\ $\approx$15 & 500+ tasks & 13 datasets & 551 tasks / 35 bench. \\
Languages & English-centric & English-centric & 250+ langs & English (multiling.\ derivs) & multilingual (43) \\
Consistent cross-architecture & $\bigcirc$ 5 fam. & $\triangle$ dense & $\triangle$ dense & $\triangle$ dense+BM25 & $\bigcirc$ 5 fam. \\
Reranker eval.\ (on task cand.) & $\triangle$ ad-hoc BM25 & $\triangle$ dedicated only & $\triangle$ dedicated only & $\triangle$ separate BM25 set & $\bigcirc$ fixed set, all tasks \\
Dim.\ reduction (Matryoshka) & $\times$ & $\triangle$ & $\triangle$ & $\times$ & $\bigcirc$ \\
Quantization (int8/binary) & $\times$ & $\triangle$ & $\triangle$ & $\times$ & $\bigcirc$ \\
Leaderboard & $\bigcirc$ & $\bigcirc$ & $\bigcirc$ & $\triangle$ (dataset set) & $\bigcirc$ (multi-axis) \\
Repeated-measurement cost & heavy & medium--heavy & medium (downsamp.) & light & light \\
\bottomrule
\end{tabular}
\end{table}

Regarding reranker evaluation: BEIR's re-ranking is described as re-ranking the top $100$ first-stage BM25 hits \citep{thakur2021beir}, and is not, as in this paper, a design that distributes a single fixed candidate set to all models for rescoring. MTEB / MMTEB reranking consists of a few dedicated tasks and is not a re-ordering over the candidate set of the retrieval task itself \citep{muennighoff2023mteb, enevoldsen2025mmteb}, and NanoBEIR requires a separately distributed BM25-candidate-augmented derived collection \citep{sentencetransformers2024nanobeirbm25}. By contrast, HAKARI-Bench ships a fixed hybrid candidate set for all $551$ retrieval tasks and evaluates rerankers consistently on the same candidate set as candidate generation (\secref{sec:cands}, \secref{sec:reranker-method}). On repeated-measurement cost, MTEB retrieval originally uses BEIR's full corpora (hundreds of thousands to millions of documents) and is heavy; MTEB v2 retrieval partly adopts MMTEB-derived hard-negative downsampling to shrink the corpus (the v2 version is what we compare against in \secref{sec:corr}), and is moderately lightweighted within that scope.

Overall, BEIR established the evaluation design of consistent cross-architecture comparison, but is English and full-scale, with cross-model evaluation of efficiency settings out of scope. MTEB / MMTEB extended to multilingual, multi-task settings and reduced measurement cost via downsampling \citep{enevoldsen2025mmteb}, but dimensionality reduction and quantization, though handleable on the infrastructure, are not reported consistently per model, and reranking is limited to dedicated tasks. NanoBEIR achieved lightweighting in the English domain \citep{camara2024nanobeir, aarsen2024nanobeir}, but efficiency settings are out of scope. HAKARI-Bench inherits these strengths---BEIR's consistent methodology, MMTEB's multilinguality, NanoBEIR's lightness---while integrating reranker evaluation on the retrieval task candidate set and cross-model evaluation of efficiency settings into a single evaluation infrastructure.

\section{Design of HAKARI-Bench}
\label{sec:design}

HAKARI-Bench is not merely a dataset collection but an evaluation infrastructure that handles the task set, candidate-generation evaluation, reranking evaluation, and efficiency settings as a whole. It is a five-stage pipeline.

\begin{enumerate}[leftmargin=*]
\item \textbf{Task specification.} The dataset location and version (commit SHA), language, and domain category are written as a declarative configuration file (\secref{sec:format}).
\item \textbf{Common task format.} Each task is aligned to a corpus, queries, relevance labels (qrels), and a fixed top candidate set (by default the hybrid top $100$ obtained by fusing BM25 and dense with RRF) (\secref{sec:tasks}, \secref{sec:cands}).
\item \textbf{Evaluation.} The five families of BM25, dense, sparse, late interaction, and reranker, together with efficiency variants (dimensionality reduction, int8, binary, rescore, sparse pruning), are run on the same tasks (\secref{sec:method}).
\item \textbf{Result records.} All runs are stored in a single schema (per-query top ranking, various @k scores, variants, resolved versions, diagnostic records).
\item \textbf{Aggregation and display.} Results are aggregated into a DuckDB warehouse and displayed as a leaderboard with macro/micro averages and multi-axis filters (\secref{sec:metrics}).
\end{enumerate}
This section describes that design.

\subsection{Task set and Nano-sets}
\label{sec:tasks}
The basic unit of the benchmark is a \textbf{retrieval task}. Each task adopts the test-collection format that TREC standardized and popularized at scale (corpus, queries, relevance labels \texttt{qrels}; \citealp{voorhees2005trec}), augmented with a fixed top candidate set (by default the hybrid top $100$ fusing BM25 and dense, \secref{sec:cands}). Just as BEIR unified scattered IR datasets into this format \citep{thakur2021beir}, our benchmark adopts the same format as a common interface and consistently advances multilingual, expert-domain, and Nano-set development.

Each task is constructed as a small evaluation dataset (Nano-set) shrunk from the original benchmark to about $50$--$200$ queries and about $1$K--$10$K documents. This shrinking, inspired by MMTEB's downsampling \citep{enevoldsen2025mmteb} and NanoBEIR's Nano-sets \citep{camara2024nanobeir, aarsen2024nanobeir}, aims to lower the cost of repeated evaluation.

Nano-set construction is twofold by provenance. First, already-published Nano collections such as the NanoBEIR family are referenced by name and version on the Hugging Face Hub without re-implementing the individual shrinking logic. Second, families that we reconstruct from the official MTEB / MMTEB full evaluation (NanoMTEB-v2, NanoMMTEB-v2, etc.) are made into Nano-sets by a common shrinking procedure: (i) select up to $200$ deduplicated queries that have at least one positive qrel, and (ii) for the corpus, after including all positive documents of the selected queries, cap it at about $10$K documents, preferentially adding any hard negatives present in the original data (documents explicitly labeled non-relevant, i.e., qrels with \texttt{score $\le$ 0}) in a query-crossing round-robin, and filling the remainder with documents in the original corpus order.

For tasks whose original data has no hard negatives, the filler documents make up most of the candidate space, so the retrieval space becomes relatively easy (irrelevant documents are unlikely to be incidental hard negatives), making it easier to distinguish positives from queries. This construction can be applied uniformly to many tasks at low cost, but there is room to raise the discriminative power of Nano-sets, e.g., by adding hard negatives (\secref{sec:conclusion}). Even so, the NanoMTEB-v2 / NanoMMTEB-v2 reconstructed this way retain a sufficient rank correlation with the official evaluation, as confirmed in \secref{sec:corr}. Storing a fixed BM25 top-$100$ candidate set on the dataset side follows the same idea as Sentence Transformers' BM25-candidate-augmented derived collection \citep{sentencetransformers2024nanobeirbm25}, decoupling reranker and learned-sparse evaluation from per-run BM25 computation differences.

The benchmark contains $35$ benchmarks and $551$ retrieval tasks. Task selection follows the four criteria BEIR identified (task diversity, domain diversity, task difficulty, coexistence of annotation strategies; \citealp{thakur2021beir}), extended to multilingual and expert domains. The task set comprises the following five families. This taxonomy is a convenient organization based on benchmark provenance and target, not a distinction in the evaluation implementation. We give the main source benchmark for each Nano-set here; details of version, provenance, and number of languages are organized in \appref{app:A1} (Table~\ref{tab:a1}).

\begin{itemize}[leftmargin=*]
\item \textbf{BEIR family:} MNanoBEIR, a multilingual collection integrating the English version that Sentence Transformers reformatted \citep{aarsen2024nanobeir} from Zeta Alpha's original NanoBEIR collection \citep{camara2024nanobeir} with the translated/extended multilingual derivatives by LightOn AI \citep{sourty2025nanobeir}, Liquid AI \citep{liquidai2025nanobeir}, and Sionic AI \citep{sionic2025nanobeir} (original datasets are BEIR; \citealp{thakur2021beir}), comprising $13$ BEIR datasets $\times$ $14$ language editions. At aggregation time it is grouped hierarchically by language and dataset and treated as one benchmark like the others (\secref{sec:metrics}).
\item \textbf{Official MTEB family:} aligned with the official MTEB / MMTEB v2 \citep{muennighoff2023mteb, enevoldsen2025mmteb} and separated per official family: NanoMTEB-v2, NanoMMTEB-v2, NanoCMTEB, NanoJMTEB-v2, NanoFaMTEB-v2, NanoRuMTEB, NanoVNMTEB, NanoMTEB-Misc, and per-language NanoMTEB-\{Dutch, French, German, Korean, Polish, Scandinavian, Spanish, Thai\}. The per-language source benchmarks each family references (C-MTEB, MTEB-NL, MTEB-French, SEB, ruMTEB, VN-MTEB, etc.) are given in Table~\ref{tab:a1}.
\item \textbf{Multilingual general:} NanoMIRACL \citep{zhang2023miracl}, NanoMLDR \citep{chen2024bgem3}, NanoIndicQA \citep{doddapaneni2023indicxtreme}, NanoMuPLeR (built by MTEB from the EU DGT multilingual parallel corpus; Table~\ref{tab:a1}). Each spans multiple languages.
\item \textbf{Long-document, instruction-following, expert-domain, reasoning:} NanoLongEmbed \citep{zhu2024longembed}, NanoIFIR \citep{song2025ifir}, NanoChemTEB \citep{shiraee2024chemteb}, NanoR2MED \citep[][R2MED]{zhangx2025r2med}, NanoBIRCO \citep{wang2024birco}, NanoBRIGHT \citep{su2024bright}, NanoRARb \citep{xiao2024rarb}, NanoRTEB \citep[RTEB;][here English production-domain retrieval (legal, finance, code, etc.), not multilingual]{liu2025rteb}, NanoBuiltBench (BuiltBench; Table~\ref{tab:a1}), NanoDAPFAM (DAPFAM; Table~\ref{tab:a1}), and the composite tasks NanoLaw (legal IR composite; AILA, LegalBench, etc., Table~\ref{tab:a1}) and NanoMedical (medical IR composite; CURE, etc., Table~\ref{tab:a1}).
\item \textbf{Code:} NanoCoIR \citep{li2024coir}, NanoCodeRAG \citep{wang2025coderag}.
\end{itemize}

The task set contains duplicate tasks that derive from the same original dataset across families (e.g., \texttt{scidocs}, \texttt{trec\_covid}). Because the Nano-set sampling differs by family, the same original task can become a different evaluation surface, so we keep duplicates as independent tasks rather than merging or removing them. Duplicate tasks may be double-counted in the equal-weight micro average over all tasks, but this effect is mitigated in the per-benchmark macro average that our analysis uses as the primary basis (cross-benchmark micro/macro and the default display are discussed in \secref{sec:metrics}). The list and details of duplicates are in \appref{app:A3}.

The overall picture of benchmarks/tasks and the distribution of languages and document counts are shown in \secref{sec:dist}; the provenance and version of each Nano-set are organized in \appref{app:A}.

\subsection{Common evaluation format}
\label{sec:format}
All tasks are aligned to a common format of corpus, queries, relevance labels, and top candidate set. This unification makes it possible to compare candidate-generation methods (BM25, dense, sparse, late interaction; retrieving the top $k$ from the whole corpus) and reranking methods (re-ordering the candidate set) with the same metrics according to their respective roles (\secref{sec:method}). Evaluation results are stored in a single schema so that swapping models, evaluation methods, prompts, and efficiency variants can be handled by a common pipeline; task specifications are managed as declarative configuration files whose required metadata are the dataset location and version, language, domain category, and citation information (a collection of task specifications becomes one benchmark on the leaderboard).

\subsection{Top candidate set}
\label{sec:cands}
Reranking is not whole-corpus retrieval but a re-ordering over a top candidate set. To make this premise explicit, we fix and share the candidate set per task. The candidate set defaults to a hybrid candidate set (top $100$) fusing the BM25 top and the dense-retrieval top via RRF (Reciprocal Rank Fusion), and the reranker and the candidate-generation baselines share the same candidate set. The construction is as follows. For each query we retrieve the top $500$ from the whole corpus with BM25 and the top $500$ with a fixed dense model (\model{microsoft/harrier-oss-v1-270m}, using the dedicated prompt \texttt{web\_search\_query}), then fuse the two rankings with RRF (each document scored by $\sum 1/(\mathit{rrf\_k} + \mathrm{rank})$, $\mathit{rrf\_k}=100$) and take the top $100$. We use a fixed dense model for dense retrieval to decouple candidate-set construction from the models under evaluation and to fix a reproducible candidate pool consistently across all tasks. The dataset side also stores a BM25-only candidate set (top $100$) as a lexical baseline, switchable as needed. Fixing the candidate set on the dataset side makes a reranker's improvement less dependent on candidate-generation bias. In particular, because the hybrid candidate set contains not only BM25 but also the dense top, it is a shared re-ordering target that is not skewed to a single candidate-generation method (BM25-only or dense-only); it nonetheless depends on the specific hybrid construction (the BM25/dense tops, RRF, and the positive-append safeguard), as discussed in \secref{sec:cands-limit}.

This fixed candidate set has a \textbf{safeguard rule}: only when the top $100$ contains no positive at all do we append one positive at the tail (rank $101$), ensuring that every query has at least one relevant document in the candidate set (query coverage $100\%$). Since passing a candidate set with no positives to a reranker yields no meaningful evaluation signal, this is a design decision that prioritizes isolating reranker evaluation to ``ranking accuracy over the candidates.'' Inclusion of all relevant documents (relevant-document coverage) is not guaranteed (about $87\%$ on dense average; \secref{sec:rerankanalysis}), and candidate-generation failures are observed as an axis independent of reranking evaluation (\secref{sec:rerankanalysis}, \appref{app:E4}). This is also due to the exceptional situation that, in addition to candidate generation missing some positives, for tasks with many positive documents per query it is in principle impossible to include all of them in a capped $100$-document candidate set (the safeguard adds only one). The implications of this design, including its difference from real-world two-stage retrieval, are discussed in \secref{sec:cands-limit}. Also, to treat BM25 fairly across languages in multilingual monolingual retrieval, the BM25 computation for the candidate set uses per-language tokenizers (morphological analyzers for CJK, Thai, and Vietnamese; Unicode regular expressions plus stemming for some languages otherwise). Details of the candidate-set construction, including the without-safeguard metric and the tokenizer breakdown, are in \appref{app:E3}.

\subsection{Evaluation targets}
\label{sec:targets}
The benchmark evaluates candidate generation, reranking, dimensionality reduction, quantization, and sparse-representation pruning on the same task set. Candidate-generation methods (BM25, dense, sparse, late interaction) are evaluated as methods that retrieve candidates from the whole corpus. Rerankers are evaluated as re-ordering over the top candidate set. For dense embeddings, we derive, as variants, leading-dimension-preserving dimensionality reduction, int8 quantization, binary quantization, and their combinations, comparing quality and efficiency side by side. For sparse representations, we evaluate how far the query-side and document-side representations can each be pruned. In evaluation, the dataset version (commit SHA) can be specified explicitly, and the resolved SHA is recorded in the results, so correspondence with past numbers is preserved even when a dataset is updated (\appref{app:A2}).

\section{Evaluation Methodology}
\label{sec:method}
This section describes which models are evaluated as candidate generation, which as reranking, and with what metrics. The evaluation modes the benchmark provides align with the five retrieval architectures BEIR identified (lexical / sparse / dense / late interaction / re-ranking; \citealp{thakur2021beir}); every mode takes the same task specification as input and outputs the same result schema.

\subsection{Evaluating retrieval models}
\label{sec:retr-method}
Retrieval models are evaluated as methods that retrieve candidate relevant documents from the whole corpus. BM25 is a lexical-matching method; by default the stored BM25 top $100$ is evaluated (local computation is also switchable). Dense retrieval encodes queries and documents with an embedding model and retrieves the top $k$ from exact similarity over the whole corpus. For each model--task pair we compute both cosine and inner-product similarity and report whichever yields the higher task \nDCG{}; this is a per-task best-of-similarity upper bound over the two functions (an oracle over the similarity choice), applied uniformly to all dense models (\appref{app:C3}) (dimensionality-reduction and quantization variants are in \secref{sec:dimquant-method}). Sparse retrieval scores by the inner product of learned sparse representations (pruning settings in \secref{sec:sparse-method}); late interaction scores by the token-to-token MaxSim of ColBERT-family token-level embeddings. In every method, the top $100$ retrieval results per query can be stored, so downstream reranking and error analysis can be run without recomputing embeddings.

\subsection{Evaluating rerankers}
\label{sec:reranker-method}
A reranker is evaluated not by comparison in the same role as candidate generation, but as a re-ordering over the top candidate set. In this paper, a \textbf{reranker} is a model that takes a query--document pair as input, directly scores their relevance, and re-orders the candidate set. The representative example is a BERT/XLM-R cross-encoder \citep{nogueira2019passrerank}, but we also include LLM-style rerankers based on a large language model (decoder) that use the predicted logit of the ``yes / no'' token as the relevance score. We collectively call these rerankers, whether cross-encoder or LLM-style. A reranker re-orders the fixed candidate set (by default the hybrid top $100$) and we compute post-reranking metrics. Because the candidate set contains at least one relevant document for every query under the safeguard rule (\secref{sec:cands}; query coverage $100\%$), reranker evaluation focuses on ranking accuracy over the candidates.

Not only dedicated rerankers but also \textbf{retrieval models} (dense, sparse, late interaction) can be scored as rerankers by rescoring the same fixed candidate set. Hence reranking performance can be measured under the same conditions for both dedicated rerankers and retrieval models. In particular, the improvement when a retrieval model re-evaluates its own candidate set and the performance when a reranker re-orders the candidate set can be read separately on the same candidate set (\secref{sec:rerankanalysis}).

\subsection{Dimensionality reduction and quantization of dense embeddings}
\label{sec:dimquant-method}
Efficiency settings for dense embeddings are generated as derived variants by post-encoding transformations after computing the base embedding once. This derives multiple efficiency settings from a single inference under the same conditions and compares quality and efficiency. The variants are:
\begin{enumerate}[leftmargin=*]
\item \textbf{Dimensionality reduction (truncation):} a leading-dimension-preserving dimension slice (assuming the Matryoshka family; \citealp{kusupati2022matryoshka}). We compare performance when truncated to, e.g., $256$ dimensions.
\item \textbf{Quantization:} int8 and binary. int8 is not a type cast to \texttt{float16} but a scalar quantization that linearly quantizes each dimension to an 8-bit integer ($256$ levels). Concretely, we take per-dimension min/max from the corpus-side embeddings and map each value into one of $256$ buckets spanning that range (per-dimension affine quantization). Calibration is done only on the distribution-stable corpus side; queries are not used for calibration (to avoid fitting buckets to evaluation queries), and out-of-range values are clipped. No separate calibration sampling or training is performed (same family as the quantization of \citealp{shakir2024quantization}). Binary keeps only the sign of each dimension (1-bit).
\item \textbf{rescore:} the simplest two-stage retrieval, which rescores the top $100$ retrieved by quantized search using the original floating-point embeddings.
\item \textbf{Combinations:} the cross product of dimensionality reduction $\times$ quantization $\times$ rescore.
\end{enumerate}

Each variant is stored side by side as a separate record for the same task, and the leaderboard's ``delta vs.\ base'' column directly shows quality degradation. This lets the leaderboard be read not as a single score column but as a Pareto frontier of quality and efficiency. Here a Pareto frontier is the set of settings on the two axes of quality and efficiency (embedding dimension, quantization precision, etc.) that cannot be beaten without worsening one of the two; i.e., the locus of best quality reachable for a given efficiency, and conversely the locus of minimum cost reachable while preserving a given quality.

Note that our quantization is a simple post-hoc quantization for measuring a model's own quantization robustness; the gap to advanced production ANN methods (\secref{sec:related-eff}) is discussed in \secref{sec:prod}. Technical details of the variants are organized in \appref{app:E1}.

\subsection{Sparse-representation pruning settings}
\label{sec:sparse-method}
Because learned sparse representations are inherently sparse, it is common to keep only the top-absolute-value dimensions per row (a \texttt{max active dims} limit). We measure, on the same evaluation surface, single variants that independently specify the query-side and document-side \texttt{max active dims}, plus their combination variants. The query-side value determines the number of non-zero dimensions at search time and is directly tied to search latency. The document-side value, in addition to latency, is directly tied to the size of the inverted index and embedding matrix, i.e., the production-time memory/disk footprint. Listing the two independently lets us read the relationship between pruning settings and retrieval quality for a given operating environment (latency budget, memory/storage budget) (\secref{sec:sparse}, \appref{app:E2}).

\subsection{Metrics and aggregation}
\label{sec:metrics}
The benchmark's main metric is \nDCG{}, following the primary metric BEIR adopted \citep{thakur2021beir}. A key design point is that during each task's evaluation, the per-query top $100$ ranking is stored as an artifact. With the top $100$ ranking stored, various retrieval metrics (nDCG, recall, accuracy, MRR, MAP, etc.) can be recomputed at any time from the stored rankings when building the leaderboard (DuckDB warehouse). The viewer/leaderboard default display is the main metric \nDCG{}, recorded in $[0,1]$; co-reporting recall@$100$ as a secondary metric follows BEIR's convention. Metric definitions are detailed in \appref{app:B1}.

To robustly aggregate benchmark groups with skewed task counts and scales, we follow these rules. Per benchmark, we display the simple average over tasks ($\times 100$). For cross-benchmark aggregation, we co-report the equal-weight \textbf{micro} average over all tasks and the \textbf{macro} average that equally weights each benchmark. The leaderboard/viewer default display is the micro average, with macro equally switchable. We default to micro because, combined with language/category filters or Nano-set narrowing, ``equal-weight average over all tasks in the displayed range'' is a simple, easy-to-understand interpretation. Which aggregation is appropriate depends on ``what data, at what granularity, one wants to see,'' so the two are placed side by side and switchable. In our analysis, however, to avoid the overall score being dominated by benchmark groups with skewed task counts and scales (especially the $182$-task MNanoBEIR and cross-family duplicate tasks), we report the per-benchmark macro average as the primary aggregation basis. In macro aggregation, the BEIR-family MNanoBEIR ($13$ BEIR datasets $\times$ $14$ languages) is first averaged over the $14$ languages within each BEIR dataset, and the $13$ dataset averages are then averaged into a single benchmark score (hierarchical aggregation by language and dataset). This prevents the high-row-count MNanoBEIR from dominating aggregation by task-count weight. Ranking targets only models that have the entire expected task set within the selected display range. Aggregation details and handling of missing tasks are in \appref{app:B2}, \ref{app:B3}.

Results can be displayed through multi-axis filters based on task and model metadata. Representative axes are (i) language tags (e.g., comparing a Japanese-specialized model side by side with multilingual general models), (ii) domain category (code/natural language, expert domain), (iii) per-task average query length and average document length (e.g., excluding long-document tasks when a model not trained on long context produces an extremely low score that distorts the overall ranking), and (iv) model embedding dimension and parameter count. These are means of reconstructing the leaderboard along use-appropriate cuts, supporting the separation of a model's strong and weak domains that is hard to see in a single score over the whole task set (\secref{sec:modelselect}).

\section{Results}
\label{sec:results}
The evaluation values reported below are based on a fixed HAKARI-Bench snapshot as of \textbf{2026-06-09} (DuckDB warehouse \texttt{hakari-bench/leaderboard\_database} commit \texttt{1f0d59d}, build 2026-06-09, schema v8); the official \texttt{mteb/results} used for the rank-correlation comparison (\secref{sec:corr}, \appref{app:D}) is fixed at commit \texttt{1e8ab5d}, as of 2026-06-08. These two snapshots are the fixed reference basis of the paper. Hereafter ``the present snapshot'' refers to this data snapshot (build 2026-06-09) and is used without further qualification.

The most important result of this section, stated up front: the overall ranking induced by Nano-sets reproduces the official full evaluations (MTEB retrieval v2 / MMTEB v2 retrieval and BEIR) at Spearman $>0.97$ (\secref{sec:corr}), confirming that the lightweighting does not damage the ranking proxy. The overall values in this section use the per-benchmark macro average as the primary basis to suppress task-count skew (the leaderboard/viewer default display is the micro average; when the difference between the two affects interpretation we note it explicitly; \secref{sec:metrics}). Below we first show the evaluation targets and task distribution (\secref{sec:dist}), then model performance and efficiency-setting results (\secref{sec:overview}--\secref{sec:rerankanalysis}), the rank correlation that grounds their validity (\secref{sec:corr}), and real-data use cases (\secref{sec:usecases}).

The evaluation targets include base rows of \textbf{$55$ models\footnote{The fixed DuckDB snapshot itself contains $57$ models ($35$ dense); we exclude two unreleased dense models from all pools, aggregations, and figures/tables, leaving the $55$ ($33$ dense) analyzed here (\appref{app:C1}).} $\times$ $35$ benchmarks $\times$ $551$ tasks}. The models comprise, as candidate-generation methods, $33$ dense embeddings, $4$ learned sparse, $6$ late interaction (ColBERT family), and $1$ lexical-baseline BM25, plus $11$ rerankers ($10$ cross-encoders, $1$ LLM-style) that re-order the top candidate set. Except for BM25, all are small-to-medium models of about $1$B parameters or fewer; the benchmark mainly targets the band that distributes at about $1$B or fewer on the MMTEB leaderboard. This lets all five families BEIR defined (lexical / sparse / dense / late interaction / re-ranking) be compared on the same task set. The model composition and references are in \appref{app:C1}.

\subsection{Task-set distribution}
\label{sec:dist}
The benchmark is not a lightweight version of a single domain but a multilingual, multi-domain evaluation surface. The category distribution is $526$ natural-language tasks ($68{,}920$ queries, $3{,}069{,}418$ documents) and $25$ code tasks ($4{,}408$ queries, $157{,}468$ documents). Five benchmarks contain code tasks, of which NanoCoIR and NanoCodeRAG are code-only and NanoBRIGHT, NanoRTEB, and NanoRARb are mixed with natural-language tasks (per-benchmark task counts are in Table~\ref{tab:a1}). The languages tagged in the task metadata span $43$ languages in total. The top $10$ languages by per-language task count (counting a task under each of its languages when it spans multiple) are English $201$, Vietnamese $40$, German $30$, French $29$, Dutch $28$, Japanese $27$, Spanish $26$, Thai $24$, Korean $20$, and Arabic/Persian $19$ each; the cumulative number of tasks for non-English languages exceeds $450$. Note these are task counts, not language counts; the number of distinct target languages is $43$. All $551$ tasks have complete metadata for query count, document count, and average character length.

\subsection{Overview of model performance}
\label{sec:overview}
The per-benchmark task average ($\times 100$, on a $33$-dense-model basis) varies greatly across benchmarks. High benchmarks include NanoCodeRAG $77.94$, NanoRuMTEB $74.04$, NanoChemTEB $72.61$, NanoCoIR $72.31$, and NanoMIRACL $68.55$; low benchmarks include NanoRARb $22.00$, NanoR2MED $23.98$, NanoDAPFAM $26.67$, NanoBIRCO $26.73$, and NanoBRIGHT $30.74$. Even on the Nano-set task collection, differences of over $50$ points are observed across benchmarks. This reflects, rather than wins/losses of individual models, the exposure of existing embedding models' weaknesses on expert-domain, instruction-following, and complex-reasoning tasks, and on natural-language tasks a model does not support (e.g., English-only models degrade greatly on multilingual tasks): performance differences vary greatly by task, domain, and supported language (\secref{sec:modelselect}).

Looking at the overall ranking of dense models by per-benchmark macro average ($\times 100$), the top are \model{jinaai/jina-embeddings-v5-text-small} \citep[jina-embeddings-v5;][]{akram2026jinav5} $64.93$, \model{jinaai/jina-embeddings-v5-text-nano} $63.80$, \model{microsoft/harrier-oss-v1-0.6b} $63.68$, \model{perplexity-ai/pplx-embed-v1-0.6b} $63.64$, and \model{google/embeddinggemma-300m} $62.58$ (the equal-weight micro average gives $62.18$, $61.18$, $60.42$, $61.14$, $59.48$, respectively, with a stable top composition). Because the macro average is less pulled by large benchmarks, we use it as the primary overall basis (the leaderboard default display is micro, with macro switchable; \secref{sec:metrics}). The lexical-baseline BM25 scores macro $50.24$ (micro $47.64$) under full-corpus retrieval; though below the top dense group, it is co-reported on all tasks as the baseline for same-condition comparison across architectures. Fine distinctions between nearby models cannot be settled by a single Nano-set with limited queries, and should be read as a ranking proxy (\secref{sec:noise}).

\subsection{Performance change from dimensionality reduction and quantization}
\label{sec:dimquant}
int8, binary, and their rescore variants are complete over $33$ dense models $\times$ $551$ tasks. Matching these variants against the base rows on the same tasks and taking the all-model mean of the delta vs.\ base ($\nDCG{} \times 100$, i.e., points), binary is $-6.50$ points, int8 $-1.95$, binary\_rescore $-0.93$, and int8\_rescore $-0.09$. That is, binary quantization alone has the largest quality drop, int8 is mild, and adding rescore restores int8 to almost lossless ($-0.09$) and binary to $-0.93$. Here, rescore means rescoring the top $100$ retrieved by the quantized vectors using the original floating-point (e.g., fp16) embeddings retained before quantization, then re-ordering (\secref{sec:dimquant-method}). Production search engines often retain the original vector values; then the rescoring targets only a few top candidates, so the additional compute is small (one may even recompute with the original model's non-reduced, non-quantized vectors). The recovery by rescore shows that quality is almost entirely regained at this small cost. This is a trend that can only be confirmed cross-model by applying the efficiency settings to all supporting models under the same conditions.

The number of models supporting dimensionality reduction (leading-dimension preserving) differs by dimension: $768$ ($3$ models), $512$ ($8$), $384$ ($1$), $256$ ($11$), $128$ ($9$), $64$ ($7$), $32$ ($5$), each with coverage over all $551$ tasks. Combination variants of quantization and dimensionality reduction also align the same coverage over all $551$ tasks for supporting models. Because these variant rows are stored side by side in the result table, the ``delta vs.\ base'' column directly shows quality degradation, and one can compare under the same conditions which models are strong at $256$ dimensions and how much int8/binary quantization degrades performance. A figure comparing quality degradation from quantization (per-model macro delta of int8/binary) and the retention rate of dimensionality reduction in native-dimension ratio, across all models, is in \appref{app:F4} (Figure~\ref{fig:dimquant}). Variant naming/correspondence and rescore details are in \appref{app:E1}.

\subsection{Performance change from sparse-representation pruning}
\label{sec:sparse}
The evaluation targets include $4$ learned sparse models (\model{naver/splade-v3} \citep[SPLADE-v3;][]{lassance2024spladev3}, \model{prithivida/Splade\_PP\_en\_v2}, \model{ibm-granite/granite-embedding-30m-sparse}, \model{opensearch-project/opensearch-neural-sparse-encoding-multilingual-v1}). For the SPLADE family \citep{formal2021splade}, we pruned with combinations of query-side and document-side \texttt{max active dims}, $q \in \{8,16,24,32\} \times d \in \{64,128,256,512\}$, and measured the performance change. For \model{naver/splade-v3}, the average score decreases monotonically from $34.16$ at $q=32, d=512$ to $29.31$ at $q=8, d=64$. The document side shows almost no improvement beyond $256$ dimensions ($+0.01$--$0.04$ for $d{=}256{\to}512$), whereas query-side reduction is more sensitive ($+2.5$--$3.6$ for $q{=}8{\to}32$ at the same $d$). This shows the practical compression headroom: aggressive document-side pruning (reducing memory and inverted-index size) barely harms quality, while cutting the query side below $16$ has a large quality cost. The full pruning grid in base ratio and the operating envelope that keeps quality $\ge 99\%$ ($q \ge 24$ and $d \ge 128$) are in \appref{app:F6} (Table~\ref{tab:f2}). Note that the SPLADE family is English-centric (from MS MARCO), so its average over all $551$ tasks including multilingual tasks comes out low, and absolute comparisons should be read with language held fixed. Pruning details are in \appref{app:E2}.

\subsection{Analysis of reranking and the candidate set}
\label{sec:rerankanalysis}
Each task's evaluation carries diagnostic records for analyzing reranker and candidate-set behavior. The default candidate set is the hybrid candidate set (top $100$, \secref{sec:cands}) fusing the BM25 and dense-retrieval tops via RRF, with the safeguard that appends one positive at the tail for any query that contains none. The records include the base and reranker scores and the improvement, the candidate-set origin, query coverage (fraction of queries with at least one relevant document), relevant-document coverage (fraction of relevant documents in the top candidates), and the runtime breakdown. On the $33$-dense-model average, query coverage was $100.0\%$ and relevant-document coverage $86.6\%$. While the safeguard ensures every query has at least one relevant document, about $14\%$ of all relevant documents do not reach the top $100$ candidates.

\paragraph{Retrieval models re-evaluating their own candidate set.}
When a dense model re-evaluates its own hybrid candidate set, the improvement is small: $+1.9$ points on dense average ($+1.5$ without the safeguard metric). It is small because the hybrid candidate set already contains the dense-retrieval top, so the model rescores almost exactly the documents it ranked at the top under full-corpus retrieval. The remaining small improvement comes from the BM25-derived candidates (lexical-match documents the dense model missed under full-corpus retrieval) entering the search target, and from the safeguard always including a positive. An advantage of sharing the hybrid candidate set is that the large apparent improvement arising from the compatibility between a BM25-only candidate set and dense, which occurs when the candidate set is BM25-only, is unlikely to be mixed in.

\paragraph{Reranker evaluation.}
The $11$ rerankers ($10$ cross-encoders, $1$ LLM-style; \secref{sec:reranker-method}) all align re-ordering results over the fixed candidate set on all $551$ tasks. Most of these rerankers---especially the encoder cross-encoders---are trained mainly for the general retrieval task of finding semantically close documents for short queries, and are most effective on tasks matching that assumption (LLM-style rerankers such as Qwen3-Reranker are an exception, as shown below). Indeed, restricting to tasks where both query and document are short (query $<70$ chars and document $<1000$ chars), the multilingual cross-encoder \model{BAAI/bge-reranker-v2-m3} \citep[BGE-M3 base;][]{chen2024bgem3} reaches macro $67.4$ on short multilingual tasks, above the best dense scored directly as a reranker on the candidate set ($65.9$), and the English-only cross-encoder \model{cross-encoder/ettin-reranker-400m-v1} \citep[Ettin;][]{weller2025ettin} reaches macro $70.2$ on short English tasks, above the best dense ($68.6$). That is, in the ``short query/document retrieval'' use case rerankers assume, using a reranker suited to multilingual or English respectively improves quality.

On the other hand, over all $551$ tasks including code, reasoning, instruction-following, long documents, and $40+$ languages, the only reranker that exceeds the best dense (\model{jinaai/jina-embeddings-v5-text-small}, $65.51$) in reranking macro over the candidate set is the LLM-style \model{Qwen/Qwen3-Reranker-0.6B} \citep[][Zhang Y.\ et al.]{zhangy2025qwen3} at $68.03$; classical multilingual cross-encoders (\model{BAAI/bge-reranker-v2-m3} $63.07$, \model{Alibaba-NLP/gte-multilingual-reranker-base} \citep[mGTE;][]{zhang2024mgte} $62.97$, etc.) all fall slightly below the best dense. This is because many of these rerankers are trained (often exclusively) for the short semantic-search queries above and do not generalize as broadly as the best dense to the full diversity of this benchmark.

The top reranker ($68.03$) also exceeds the top full-corpus dense model ($64.93$), showing that reranking over the hybrid candidate set functions as a configuration that surpasses full-corpus dense retrieval under the fixed-candidate reranking protocol (with the safeguard; this is not an end-to-end production retrieval comparison, \secref{sec:prod}). A detailed decomposition of rerankers by type, scope, and query type ($z$-score comparison; Table~\ref{tab:e1}, Figures~\ref{fig:zscore}, \ref{fig:shortlong}) is in \appref{app:E4}. Note these reranker scores are computed under the premise that the safeguard includes a relevant document in the candidate set for every query, so the ``degradation when the candidate set contains no positive'' that real two-stage retrieval faces is isolated.

We summarize the per-scope best models in Table~\ref{tab:scope}. What exceeds the best dense (\model{jinaai/jina-embeddings-v5-text-small} scored as a reranker on the same candidate set) is: over all $551$ tasks, only the LLM-style \model{Qwen/Qwen3-Reranker-0.6B} ($68.03$); on the short-multilingual scope (query $<70$ chars and document $<1000$ chars, non-English), the multilingual cross-encoder \model{BAAI/bge-reranker-v2-m3} ($67.41$; Qwen3-Reranker also exceeds it slightly at $66.48$); and on the short-English scope (same condition, English), the English-only cross-encoder \model{cross-encoder/ettin-reranker-400m-v1} ($70.23$). Thus whether a reranker beats dense depends not on ``rerankers in general'' but on the scope and reranker type.

\begin{table}[t]
\centering
\caption{Per-scope reranking macro for four representative models (over the candidate set, $\nDCG{}\times 100$; \best{teal bold} = value exceeding the best dense in each scope). ``Short'' = query $<70$ chars and document $<1000$ chars; ``short multilingual'' = non-English, ``short English'' = English; both are per-benchmark macro. The dense row (shaded) is the best dense scored as a reranker on the same candidate set, and is the best dense in all three scopes. Model IDs are abbreviated (Hugging Face org prefix omitted; full IDs in the text). A finer decomposition by query/document length is in \appref{app:E4}.}
\label{tab:scope}
\footnotesize
\setlength{\tabcolsep}{6pt}
\renewcommand{\arraystretch}{1.15}
\begin{tabular}{@{}llrrr@{}}
\toprule
\rowcolor{headcol} \hdr{Model} & \hdr{Type} & \hdr{All 551} & \hdr{Short ML} & \hdr{Short EN} \\
\midrule
\model{Qwen3-Reranker-0.6B} & LLM reranker & \best{68.03} & \best{66.48} & 66.59 \\
\model{bge-reranker-v2-m3} & multilingual CE & 63.07 & \best{67.41} & 63.29 \\
\model{ettin-reranker-400m} & English-only CE & 60.82 & 58.91 & \best{70.23} \\
\rowcolor{rowgray} \model{jina-v5-small} & dense (reference) & 65.51 & 65.91 & 68.59 \\
\bottomrule
\end{tabular}
\end{table}

\paragraph{Per-benchmark advantage/disadvantage.}
The ``reranker top $-$ dense top'' gap is large on multilingual, expert-domain, and reasoning benchmarks (e.g., NanoMLDR $+13.33$), and even on benchmarks where the reranker top falls below, the downside is small. The per-benchmark breakdown and the decomposition of rerankers by type (cross-encoder / LLM-style), scope, and query type are organized in \appref{app:E4}, where we show that multilingual cross-encoders are strong on short factual queries and collapse on long queries, while LLM-style rerankers are robust to length.

\subsection{Rank correlation with MTEB / MMTEB retrieval}
\label{sec:corr}
\begin{figure}[t]
\centering
\includegraphics[width=\linewidth]{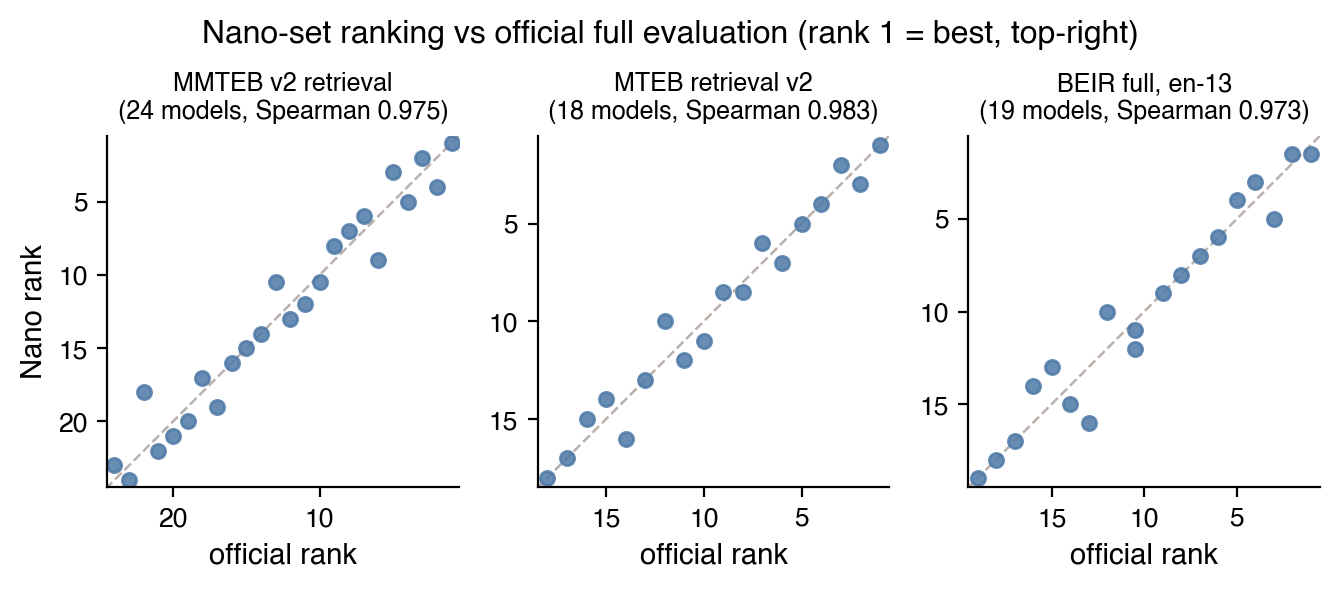}
\caption{Correspondence between the official evaluation and the Nano-set overall rankings (MMTEB / MTEB-v2 / BEIR-en).}
\label{fig:scatter}
\end{figure}

To empirically show how well Nano-sets reproduce the ranking of the original benchmarks, we independently compared \textbf{NanoMMTEB-v2}, \textbf{NanoMTEB-v2}, and \textbf{NanoBEIR-en} against, respectively, \textbf{MMTEB v2 retrieval}, \textbf{MTEB retrieval v2}, and English \textbf{BEIR (full)} from the official \texttt{mteb/results} (commit \texttt{1e8ab5d}, reflected up to 2026-06-08). The analysis uses only the same base rows of the \secref{sec:results} results on the Nano side, and excludes from the common model set any model for which the official side does not have all tasks as a single-revision single measurement, isolating pure ranking reproducibility. The aggregation assigns a rank by descending score within a task (ties get the average rank) and computes the overall ranking by averaging each model's Borda score $100 \times (N - \mathrm{rank})/(N-1)$ ($N$ = number of models) over all tasks. The results are in Table~\ref{tab:corr}, and the scatter of official vs.\ Nano ranks is in Figure~\ref{fig:scatter}.

\begin{table}[t]
\centering
\caption{Rank correlation between Nano-sets and the official evaluation. Columns: NanoMMTEB-v2 vs.\ MMTEB v2; NanoMTEB-v2 vs.\ MTEB v2; NanoBEIR-en vs.\ BEIR (full).}
\label{tab:corr}
\footnotesize
\renewcommand{\arraystretch}{1.15}
\begin{tabular}{@{}lrrr@{}}
\toprule
\rowcolor{headcol} \hdr{Metric} & \hdr{MMTEB} & \hdr{MTEB-v2} & \hdr{BEIR-en} \\
\midrule
Common models & 24 & 18 & 19 \\
Tasks & 18 & 10 & 13 \\
Spearman rank correlation & 0.975 & 0.983 & 0.973 \\
Spearman 95\% CI (model bootstrap) & [0.915, 0.995] & [0.912, 0.998] & [0.882, 0.997] \\
Pearson correlation (Borda score) & 0.969 & 0.981 & 0.974 \\
Mean absolute rank difference & 1.208 & 0.722 & 0.895 \\
Median rank difference & 1.000 & 1.000 & 0.500 \\
Max rank difference & 4.000 & 2.000 & 3.000 \\
\bottomrule
\end{tabular}
\end{table}

For all three pairs, the Spearman rank correlation exceeds $0.97$, with rank differences of about $1$ on average and at most $4$ (MMTEB), $3$ (BEIR-en), and $2$ (MTEB-v2). Because the common model counts ($24$/$18$/$19$) are limited, we obtained $95\%$ confidence intervals for Spearman by bootstrap ($10{,}000$ resamples with replacement) over the common model set (Table~\ref{tab:corr}). Even at the interval lower bounds, the correlation stays at $\ge 0.91$ for MMTEB and MTEB-v2 and $\ge 0.88$ for BEIR-en, all high. The Borda-score Pearson correlation is also $\ge 0.96$, so even from the perspective of aggregating per-task wins/losses, every Nano-set faithfully reproduces the official overall ranking. In particular, for both MMTEB v2 retrieval and MTEB retrieval v2, the top model is rank $1$ on both the official and Nano sides (rank difference $0$), a representative agreement on top-rank reproduction. Rank swaps exist, but no large movement crossing the boundaries of the top/middle/bottom groups is observed, and they are not of a scale that changes model-selection judgments.

The per-model ranking tables, per-task mean/variance differences, and the discussion of the factors behind the differences are gathered in \appref{app:D}.

From these results, Nano-sets are not a final evaluation replacing the official full retrieval, nor do they guarantee absolute-score agreement. However, for iterative ranking judgments such as model selection, separating the top from the middle group, and pre-release regression detection, they function as a proxy that provides conclusions close to the official full evaluation at low cost, as confirmed through three independent comparisons. Fine distinctions between nearby models and conclusions that depend on a particular task still require referring to the official full tasks.

\subsection{Real-data use cases}
\label{sec:usecases}
The overall ranking answers only ``which model is best on average,'' but practical model adoption is made under conditions such as target language, document length, latency budget, and index size. Because the benchmark measures many models $\times$ tasks $\times$ architectures $\times$ efficiency settings under the same conditions, it directly answers such conditional adoption decisions. For example, on a pool of $38$ first-stage retrieval systems (dense $33$, learned sparse $4$, BM25 $1$), contrasting each scope's top-$1$ system with its overall macro rank: for code RAG, instruction-following, and medical reasoning, the overall-rank-$1$ (\model{jinaai/jina-embeddings-v5-text-small}) is also the scope top-$1$; whereas for multilingual semantic search (NanoMIRACL) the overall-rank-$10$ \model{BAAI/bge-m3}, for the two long-document series (NanoMLDR, NanoLongEmbed) the overall-rank-$24$ BM25, and for Japanese (NanoJMTEB-v2) the overall-rank-$28$ Japanese-specialized model \model{cl-nagoya/ruri-v3-310m} \citep[Ruri;][]{tsukagoshi2024ruri} are each top-$1$. Scopes where the overall score is a good guide coexist with scopes where it is a wrong guide, and which is which can only be determined by per-scope measurement. The full picture of per-scope ranks is in \appref{app:F1} (Figure~\ref{fig:retr-scope}).

The observation that ``the overall best model is not necessarily indicated, and the best model/architecture changes with the target scope'' generalizes to three questions, each detailed as a real-data use case in \appref{app:F}.

First, \textbf{which model/architecture to choose depends on the target scope} (\appref{app:F1}, \ref{app:F2}). Changing scope swaps the best model not only among dense models but also across architectures (dense / sparse / late interaction, etc.). For example, restricting to English BEIR, late interaction---not top overall---takes first place, and learned sparse enters the top quartile.

Second, \textbf{different architectures can be compared on the same footing} (\appref{app:F3}). Scoring all models as rerankers over the same fixed candidate set lets embedding models and rerankers be placed side by side. On the overall macro, only one modern general reranker exceeds the dense top, and the advantage of multilingual cross-encoders concentrates in the multilingual semantic-search scope (\secref{sec:rerankanalysis}, \appref{app:E4}).

Third, \textbf{the cost of efficiency settings can be read separately from quality} (\appref{app:F4}--\ref{app:F6}). Dimensionality reduction and int8 quantization are predictable small costs, robustness to binary quantization depends on a model's training characteristics, float rescoring nearly preserves cross-model comparison, and sparse pruning has a cheap document-side knob and an expensive query-side knob---each setting's cost can be evaluated separately.

All three are material for adoption decisions that cannot be read from a single overall score, and can be extracted only when a single harness, a single task format, same-condition measurement, and a consistent aggregation basis (macro as primary in this paper; \secref{sec:design}, \secref{sec:metrics}) are aligned.

\section{Discussion}
\label{sec:discussion}

\subsection{Validity as a lightweight evaluation}
\label{sec:validity}
How well Nano-set construction preserves the original benchmark ranking is the most important empirical question for this benchmark. In \secref{sec:corr}, for the two independent comparisons MMTEB v2 retrieval and MTEB retrieval v2, we obtained Spearman $0.975$ / $0.983$, Borda Pearson $0.969$ / $0.981$, and max rank difference $4$ / $2$. That the same level of rank preservation reproduced on two independent benchmarks strongly supports the validity of Nano-sets as a ranking proxy, on par with the post-downsampling correlation analysis MMTEB showed \citep{enevoldsen2025mmteb}. Furthermore, comparing NanoBEIR-en with the original BEIR (full) on the $13$ tasks and $19$ common models they share gave Spearman $0.973$ (model bootstrap $95\%$ CI $[0.882, 0.997]$) and Borda Pearson $0.974$, confirming the same level of rank preservation on a third independent comparison. However, Nano-sets are not a substitute for absolute scores. As shown in \secref{sec:corr} and \appref{app:D}, due to the shrunken retrieval space and the shortage of hard negatives, per-task mean scores and variance do not match the original benchmark. Indeed, in \appref{app:D1}, \ref{app:D2}, the Nano-side mean nDCG is about $-7$ points relative to MMTEB but about $+7$ points relative to MTEB-v2: even the sign (direction) of the discrepancy reverses depending on the reference benchmark. This plainly shows that reading absolute scores against a reference benchmark is a mistake, and that this metric is strictly a ranking proxy. Nano-sets should be read only as an evaluation metric for lightly viewing model ranking and setting differences.

\subsection{Use-appropriate model selection}
\label{sec:modelselect}
That neural retrieval models degrade greatly out of distribution has been shown repeatedly since BEIR \citep{thakur2021beir}. Given this, the value of reading a multi-domain benchmark as a single overall score is limited. Its proper role is to expose ``which domains a model has not learned'' and to provide material for use-appropriate model selection. The $50$-plus-point differences across benchmarks in the simple average (\secref{sec:overview}) reflect, rather than difficulty differences, directly the domains each model has not learned.

From this perspective, the implication is that ``a model high on average across all tasks'' is not necessarily best for production. In real model selection, retrieval performance on the actual query and document forms and domains is what contributes to solving the problem. Empirically too, on NanoMIRACL the overall-rank-$10$ \model{BAAI/bge-m3} is rank $1$ among $38$ first-stage systems, and on Japanese NanoJMTEB-v2 the overall-rank-$28$ \model{cl-nagoya/ruri-v3-310m} is rank $1$; restricting the scope, the overall-top general model is not required (\secref{sec:usecases}, \appref{app:F1}). Our benchmark enables filtering by language tag or category (\secref{sec:metrics}) precisely to iteratively support the selection of ``a model strong on tasks close to the retrieval situation one wants to use.''

\subsection{The quality--efficiency trade-off}
\label{sec:tradeoff}
Efficiency via candidate-generation dimensionality/quantization and quality improvement via a reranker are originally different axes. By storing the quantization/dimensionality-reduction/rescore variants and the candidate-set reranker evaluation side by side in the same result table, the benchmark lets them be read together. This lets the leaderboard be read not as a single score column but as a Pareto frontier of quality $\times$ dimension $\times$ quantization precision (as defined in \secref{sec:dimquant-method}, the set of settings that cannot be beaten on the two axes without worsening one; i.e., the locus of best quality for a given efficiency), supporting variant selection for a given operating environment (device, memory, latency requirements). For example, a configuration combining compact candidate generation (reduced dimension and quantized) with a lightweight reranker over the candidate set can be evaluated including operating cost, not only retrieval quality. As an empirical example, a side-by-side comparison of $11$ MRL-capable models at the fixed operating point $256$ dimensions + binary + rescore ($32$ bytes/vector, $1/128$ the size of float $1024$ dimensions) is in \appref{app:F5}. Multi-axis filters (\secref{sec:metrics}) combine with these to extract use-appropriate models and settings.

\subsection{Caveats in moving from benchmark results to production}
\label{sec:prod}
HAKARI-Bench results should be read not as production performance directly, but in light of task closeness, candidate set, model size, latency, and memory constraints. In particular, for quantization variants, the benchmark is limited to simple post-hoc scalar/binary quantization (\secref{sec:dimquant-method}). Advanced methods used in production large-scale ANN search (\secref{sec:related-eff}) each use only some of the elements such as space partitioning, learned codebooks, correction terms, asymmetric quantization, and residual re-ranking (not all methods combine all of these at once), and none are handled by the benchmark. Hence the final quantization robustness in production may differ from the benchmark values; advanced production quantization methods can improve it, but this direction is not measured here. Precise performance evaluation combined with production methods is out of scope and belongs to dedicated benchmarks such as ANN-Benchmarks.

Likewise, reranker evaluation values are candidate-set ranking accuracy under the premise that the safeguard gives every query at least one relevant document (query coverage $100\%$; \secref{sec:cands}), and do not include the degradation when production candidate generation misses positives. Candidate-generation failures should be read separately as query coverage and relevant-document coverage (\secref{sec:rerankanalysis}).

\section{Limitations}
\label{sec:limitations}

\subsection{Difference between Nano-sets and the original evaluation}
\label{sec:nano-diff}
Because Nano-sets shrink not only the query count but also the corpus to about $10$K documents, they cannot necessarily reproduce the difficulty of the original full benchmark pulling relevant documents from hundreds of thousands to millions. Shrinking the retrieval space makes incidental matches easier, and on hard-negative-dependent tasks the Nano-ized evaluation tilts to the ``easy'' side (e.g., the score range of \texttt{fever\_hard\_negatives} is official $27.5$--$92.9$ $\to$ Nano $74.1$--$99.1$; details in \appref{app:D3}). Overall-rank preservation is empirically shown in \secref{sec:corr}, but using Nano-sets as a substitute for absolute scores is inappropriate. Likewise, since rank swaps from quantization noise are less likely the smaller the corpus, the int8/binary degradation in \secref{sec:dimquant} and \appref{app:F4} should be read as an optimistic lower bound relative to full-corpus operation; its verification is future work (\secref{sec:conclusion}).

\subsection{Evaluation noise and comparison of nearby models}
\label{sec:noise}
At a scale of $50$--$200$ queries, the standard error of the evaluation values is larger than the original benchmark, and fine distinctions between nearby models cannot be guaranteed by a single Nano-set. To quantify this noise, recomputing the macro average of the $33$ dense models by bootstrap ($2{,}000$ resamples) over tasks within each benchmark, the half-width of the macro $95\%$ CI averaged $\pm 2.1$ points (max $\pm 2.3$) (\appref{app:D4}, Figure~\ref{fig:macroci}). Rank stability depends on the gap to neighboring models: pairs differing by about $1$ point or more virtually never swap (the rank-$1$ Jina-v5-small never falls to rank $2$ over $2{,}000$ resamples), while ranks $2$--$4$ differing by around $0.1$ point swap with probability $31$--$45\%$. That is, a macro-average difference of under $1$ point should not be read as a rank. This bootstrap quantifies task-sampling noise within a fixed Nano-set and does not capture the variance of the query/document sampling at Nano-set construction time. Verification of construction-seed variance and reliability metrics such as nAUC are not treated here; we take the above task bootstrap as our reliability description.

\subsection{Candidate-set-dependent evaluation}
\label{sec:cands-limit}
As in \secref{sec:cands}, the candidate set has a safeguard rule, so every query has at least one relevant document (query coverage $100\%$). On the other hand, not all relevant documents are in the candidate set; relevant-document coverage averages about $87\%$ on dense (\secref{sec:rerankanalysis}). This is a simplification to preserve the meaning of reranker evaluation and does not match production two-stage retrieval. Because how the candidate set is built affects reranking evaluation, candidate-generation failures and reranker ranking accuracy should be read separately. Also, the fixed candidate set is a BM25--dense hybrid; combining dense mitigates but does not fully remove the lexical bias from BM25 alone. As BEIR revealed through manual annotation analysis ($980$ query--document pairs on TREC-COVID), because the candidate pool depends on first-stage retrieval, methods that return lexically non-matching relevant documents can produce unjudged positives \citep{thakur2021beir}. That both dense and rerankers can be underestimated when lexically non-matching relevant documents are missed from the candidates should be read as the same premise as BEIR.

\subsection{Scope of evaluated models}
\label{sec:modelscope}
The benchmark currently evaluates mainly open-published models of about $1$B parameters or fewer, and does not include paid commercial models (e.g., OpenAI, Cohere, Voyage, Google embedding APIs). Hence direct comparison with very large public models or commercial models is out of scope. The current targets include $33$ dense, $4$ learned sparse \citep[SPLADE family, etc.;][]{formal2021splade}, $6$ late interaction \citep[ColBERT family;][]{khattab2020colbert, santhanam2021colbertv2}, $11$ rerankers ($10$ cross-encoders, $1$ LLM-style), and $1$ lexical-baseline BM25, letting all five BEIR families be placed in a single ranking table. BM25 is evaluated as a full-corpus lexical baseline and is also used as the lexical component of the hybrid candidate set.

In addition, there is potential for data contamination (overlap or proximity between evaluation data and training data). For example, Multilingual E5 \citep{wang2024me5} is reported to include MS MARCO and MIRACL in its supervised fine-tuning data mixture, so the NanoMIRACL and NanoBEIR-en numbers may be affected. Likewise BGE-M3 \citep{chen2024bgem3} uses MS MARCO and MIRACL training data and itself built the long-document retrieval dataset MLDR for training, so NanoBEIR-en (msmarco), NanoMIRACL, and NanoMLDR may be similarly affected. Note that what these models use for training is in most cases the train split of these datasets, not the relevance labels (positives) of the test/dev split used for evaluation. That is, it is not leakage in the strict sense of directly memorizing the evaluation positives. However, since they learn the same dataset's, same domain's query/document distribution through the train split, they may be indirectly advantaged and score higher on that benchmark. Our numbers should be read on the premise that this indirect domain-adaptation effect cannot be excluded. We do not investigate or verify the training-data disclosure status of the evaluated models. Since similar mixtures are possible for other public models, our numbers should be read on the premise that data contamination cannot be excluded.

Furthermore, the E5/BGE-M3 examples above are cases where the training data is explicitly disclosed; a harder-to-detect contamination source is unintended inclusion at the pre-training stage. Recent embedding models, like large language models, are often pre-trained on large public web corpora that effectively reach near-web scale, so a benchmark's test data (queries/documents) can be taken into training regardless of train-split use. Such benchmark data contamination has been pointed out as becoming almost unavoidable as corpora grow \citep{xu2024contamination}, and cannot in principle be fully excluded regardless of whether providers disclose training data. As a countermeasure, designs such as RTEB \citep{liu2025rteb} combine private (closed) datasets: RTEB scores on both public and undisclosed private datasets and infers contamination/leaderboard-overfitting from their gap. By contrast, our NanoRTEB, prioritizing reproducibility and ease of reference, by construction only Nano-izes RTEB's public datasets (English $14$ tasks). The private datasets cannot be obtained or incorporated, so NanoRTEB numbers cannot determine how contaminated a model is in that domain. The contamination separation RTEB achieves with private sets is not inherited by our benchmark, which uses only public Nano-sets.

Finally, prompt settings also have fairness limitations in how models are treated. We currently perform no fine instruction control over the query/document transformation; following each model's official documentation or the SentenceTransformers standard prompt format, we apply at most one prompt specification each for query and document. Retrieval models do not necessarily use natural-language instructions correctly, and performance can drop especially as instructions get longer \citep{weller2024followir}. The list of applied prompt settings and the fairness limitations for models that assume task-specific prompts are stated in \appref{app:C2}.

\subsection{Inference-speed comparison}
\label{sec:speed}
While the benchmark evaluates reproducible efficiency proxies (embedding dimension, quantization precision, number of non-zero sparse dimensions), it does not include the inference speed itself of each model (encoding throughput or retrieval latency). Speed strongly depends on hardware, batch size, sequence length, and parallelism, and the availability of recommended implementations and optimized attention (FlashAttention-2; \citealp{dao2023flashattention2}) differs per model, so drawing out every model's ``best speed'' fairly under the same conditions is difficult, and inappropriate measurement would mislead model selection. As a rough proxy for inference cost, the benchmark records both active and total parameters for each model. Active parameters are the number of parameters doing per-token computation (self-attention, feed-forward, etc.) and are a first-order proxy for the inference compute/speed of a Transformer architecture. Total parameters include the static word-embedding table and indicate model size and memory usage, but since looked-up word embeddings do not contribute to per-token matrix operations, active parameters are more appropriate as a speed proxy. Both are useful as a first-order approximation of relative compute scale across families/models, but actual inference speed strongly depends, as noted, on implementation, hardware, sequence length, batch size, and optimization, and is not uniquely determined by parameter count alone. Because speed is, alongside quality, important for production model selection, this is an important limitation of the benchmark; speed comparison belongs to dedicated benchmarks with fixed implementation/hardware/optimization (e.g., ANN-Benchmarks; \secref{sec:prod}) and is placed out of scope here.

\section{Conclusion}
\label{sec:conclusion}
We built HAKARI-Bench, a lightweight benchmark for evaluating multilingual, multi-domain retrieval models. It inherits the context of MTEB / MMTEB / BEIR / MIRACL / NanoBEIR while integrating into one infrastructure an evaluation that handles candidate-generation and reranking methods according to their respective roles under the same conditions, and that compares efficiency settings such as dimensionality reduction, quantization, and sparse pruning side by side in the same table. We showed integrated handling of the result set of $55$ models (dense $33$, learned sparse $4$, late interaction $6$, reranker $11$ ($10$ cross-encoders, $1$ LLM-style), lexical-baseline BM25 $1$) $\times$ $35$ benchmarks $\times$ $551$ tasks (\secref{sec:results}; results as of 2026-06-09). Furthermore, against the three official evaluations MMTEB v2 retrieval, MTEB retrieval v2, and English BEIR (full), we empirically showed that Nano-sets reproduce the overall ranking at Spearman $0.975$ / $0.983$ / $0.973$, Pearson $0.969$ / $0.981$ / $0.974$, and max rank difference $4$ / $2$ / $3$ (\secref{sec:corr}).

Future work is threefold: (i) expanding the evaluated models to include public models over $1$B and commercial APIs; (ii) extending the real-data model-adoption use cases (\secref{sec:usecases}, \appref{app:F}) to full-corpus verification; and (iii) improving the down-sampling method at Nano-set construction. On (iii) in particular, the current shrinking fills the corpus of tasks without hard negatives with filler documents in original corpus order, making the retrieval space easy (\secref{sec:tasks}). Adding hard negatives collected by first-stage retrieval, or document sampling that considers difficulty and diversity, could raise the representativeness of absolute scores and the discriminative power between nearby models. This is an improvement that raises quality while preserving Nano-set rank reproducibility (\secref{sec:corr}), and better sampling design is an important task in refining the benchmark. We hope the benchmark contributes to the community as a lightweight evaluation infrastructure that compares multilingual, multi-domain retrieval models, rerankers, and efficiency settings under the same conditions.

\nocite{benabacha2019medquad,boteva2016nfcorpus,hoppe2021legalquad,li2023lecardv2,lu2024publichealthqa,manor2019consumercontracts,qiu2022dureader,roberts2021treccovid,wadden2020scifact,wojtasik2024beirpl,wrzalik2021gerdalir,zhang2018cmedqa}
\bibliographystyle{plainnat}
\bibliography{references}

\begin{thebibliography}{76}
\providecommand{\natexlab}[1]{#1}
\providecommand{\url}[1]{\texttt{#1}}
\expandafter\ifx\csname urlstyle\endcsname\relax
  \providecommand{\doi}[1]{doi: #1}\else
  \providecommand{\doi}{doi: \begingroup \urlstyle{rm}\Url}\fi

\bibitem[Aarsen(2024)]{aarsen2024nanobeir}
Tom Aarsen.
\newblock {NanoBEIR}: Lightweight {BEIR} subsets for iterative retrieval
  evaluation.
\newblock Hugging Face Hub Dataset Collection / Sentence Transformers, 2024.
\newblock URL
  \url{https://huggingface.co/collections/zeta-alpha-ai/nanobeir-66e1a0af21dfd93e620cd9f6}.

\bibitem[Akram et~al.(2026)Akram, Sturua, Havriushenko, Herreros, G{\"u}nther,
  Werk, and Xiao]{akram2026jinav5}
Mohammad~Kalim Akram, Saba Sturua, Nastia Havriushenko, Quentin Herreros,
  Michael G{\"u}nther, Maximilian Werk, and Han Xiao.
\newblock jina-embeddings-v5-text: Task-targeted embedding distillation.
\newblock \emph{arXiv preprint arXiv:2602.15547}, 2026.
\newblock URL \url{https://arxiv.org/abs/2602.15547}.

\bibitem[Ayaou et~al.(2026)Ayaou, Cavallucci, and Chibane]{ayaou2026dapfam}
Iliass Ayaou, Denis Cavallucci, and Hicham Chibane.
\newblock {DAPFAM}: A domain-aware family-level dataset to benchmark cross
  domain patent retrieval.
\newblock \emph{Array}, page 100720, 2026.
\newblock URL \url{https://doi.org/10.1016/j.array.2026.100720}.

\bibitem[Bajaj et~al.(2016)Bajaj, Campos, Craswell, Deng, Gao, Liu, Majumder,
  McNamara, Mitra, Nguyen, Rosenberg, Song, Stoica, Tiwary, and
  Wang]{bajaj2016msmarco}
Payal Bajaj, Daniel Campos, Nick Craswell, Li~Deng, Jianfeng Gao, Xiaodong Liu,
  Rangan Majumder, Andrew McNamara, Bhaskar Mitra, Tri Nguyen, Mir Rosenberg,
  Xia Song, Alina Stoica, Saurabh Tiwary, and Tong Wang.
\newblock {MS MARCO}: A human generated machine reading comprehension dataset.
\newblock \emph{arXiv preprint arXiv:1611.09268}, 2016.
\newblock URL \url{https://arxiv.org/abs/1611.09268}.

\bibitem[Banar et~al.(2025)Banar, Lotfi, Van~Nooten, Arhiliuc, Kliocaite, and
  Daelemans]{banar2025mtebnl}
Nikolay Banar, Ehsan Lotfi, Jens Van~Nooten, Cristina Arhiliuc, Marija
  Kliocaite, and Walter Daelemans.
\newblock {MTEB-NL} and {E5-NL}: Embedding benchmark and models for dutch.
\newblock \emph{arXiv preprint arXiv:2509.12340}, 2025.
\newblock URL \url{https://arxiv.org/abs/2509.12340}.

\bibitem[Ben~Abacha and Demner-Fushman(2019)]{benabacha2019medquad}
Asma Ben~Abacha and Dina Demner-Fushman.
\newblock A question-entailment approach to question answering.
\newblock \emph{BMC Bioinformatics}, 20:\penalty0 511, 2019.
\newblock URL
  \url{https://bmcbioinformatics.biomedcentral.com/articles/10.1186/s12859-019-3119-4}.

\bibitem[Bhattacharya et~al.(2019)Bhattacharya, Ghosh, Ghosh, Pal, Mehta,
  Bhattacharya, and Majumder]{bhattacharya2019aila}
Paheli Bhattacharya, Kripabandhu Ghosh, Saptarshi Ghosh, Arindam Pal, Parth
  Mehta, Arnab Bhattacharya, and Prasenjit Majumder.
\newblock Overview of the {FIRE} 2019 {AILA} track: Artificial intelligence for
  legal assistance.
\newblock In \emph{CEUR-WS Vol-2517}, 2019.
\newblock URL \url{https://ceur-ws.org/Vol-2517/T1-1.pdf}.

\bibitem[Boteva et~al.(2016)Boteva, Gholipour~Ghalandari, Sokolov, and
  Riezler]{boteva2016nfcorpus}
Vera Boteva, Demian Gholipour~Ghalandari, Artem Sokolov, and Stefan Riezler.
\newblock A full-text learning to rank dataset for medical information
  retrieval.
\newblock In \emph{ECIR 2016 (LNCS 9626)}, 2016.
\newblock URL \url{https://doi.org/10.1007/978-3-319-30671-1_58}.

\bibitem[C{\^a}mara(2024)]{camara2024nanobeir}
Arthur C{\^a}mara.
\newblock Fine-tuning an {LLM} for state-of-the-art retrieval: Zeta alpha's
  top-10 submission to the {MTEB} benchmark.
\newblock Zeta Alpha Blog, 2024.
\newblock URL
  \url{https://www.zeta-alpha.com/post/fine-tuning-an-llm-for-state-of-the-art-retrieval-zeta-alpha-s-top-10-submission-to-the-the-mteb-be}.

\bibitem[Chen et~al.(2024)Chen, Xiao, Zhang, Luo, Lian, and Liu]{chen2024bgem3}
Jianlv Chen, Shitao Xiao, Peitian Zhang, Kun Luo, Defu Lian, and Zheng Liu.
\newblock {M3-Embedding} ({BGE-M3}): Multi-linguality, multi-functionality,
  multi-granularity text embeddings through self-knowledge distillation.
\newblock In \emph{Findings of ACL 2024}, pages 2318--2335, 2024.
\newblock URL \url{https://aclanthology.org/2024.findings-acl.137/}.

\bibitem[Ciancone et~al.(2024)Ciancone, Kerboua, Schaeffer, and
  Siblini]{ciancone2024mtebfrench}
Mathieu Ciancone, Imene Kerboua, Marion Schaeffer, and Wissam Siblini.
\newblock {MTEB-French}: Resources for french sentence embedding evaluation and
  analysis.
\newblock \emph{arXiv preprint arXiv:2405.20468}, 2024.
\newblock URL \url{https://arxiv.org/abs/2405.20468}.

\bibitem[Dao(2023)]{dao2023flashattention2}
Tri Dao.
\newblock {FlashAttention-2}: Faster attention with better parallelism and work
  partitioning.
\newblock \emph{arXiv preprint arXiv:2307.08691}, 2023.
\newblock URL \url{https://arxiv.org/abs/2307.08691}.

\bibitem[Doddapaneni et~al.(2023)Doddapaneni, Aralikatte, Ramesh, Goyal,
  Khapra, Kunchukuttan, and Kumar]{doddapaneni2023indicxtreme}
Sumanth Doddapaneni, Rahul Aralikatte, Gowtham Ramesh, Shreya Goyal, Mitesh~M.
  Khapra, Anoop Kunchukuttan, and Pratyush Kumar.
\newblock Towards leaving no indic language behind: Building monolingual
  corpora, benchmark and models for indic languages ({IndicXTREME}).
\newblock In \emph{ACL 2023}, 2023.
\newblock URL \url{https://aclanthology.org/2023.acl-long.693/}.

\bibitem[Enevoldsen et~al.(2024)Enevoldsen, Kardos, Muennighoff, and
  Nielbo]{enevoldsen2024seb}
Kenneth Enevoldsen, M{\'a}rton Kardos, Niklas Muennighoff, and Kristoffer~L.
  Nielbo.
\newblock The scandinavian embedding benchmarks: Comprehensive assessment of
  multilingual and monolingual text embedding.
\newblock \emph{arXiv preprint arXiv:2406.02396}, 2024.
\newblock URL \url{https://arxiv.org/abs/2406.02396}.

\bibitem[Enevoldsen et~al.(2025)Enevoldsen, Chung, Kerboua, Kardos, Mathur,
  et~al.]{enevoldsen2025mmteb}
Kenneth Enevoldsen, Isaac Chung, Imene Kerboua, M{\'a}rton Kardos, Ashwin
  Mathur, et~al.
\newblock {MMTEB}: Massive multilingual text embedding benchmark.
\newblock \emph{arXiv preprint arXiv:2502.13595}, 2025.
\newblock URL \url{https://arxiv.org/abs/2502.13595}.

\bibitem[Formal et~al.(2021)Formal, Piwowarski, and
  Clinchant]{formal2021splade}
Thibault Formal, Benjamin Piwowarski, and St{\'e}phane Clinchant.
\newblock {SPLADE}: Sparse lexical and expansion model for first stage ranking.
\newblock In \emph{SIGIR 2021}, 2021.
\newblock URL \url{https://arxiv.org/abs/2107.05720}.

\bibitem[Gao and Long(2024)]{gao2024rabitq}
Jianyang Gao and Cheng Long.
\newblock {RaBitQ}: Quantizing high-dimensional vectors with a theoretical
  error bound for approximate nearest neighbor search.
\newblock \emph{Proceedings of the ACM on Management of Data (SIGMOD 2024)},
  2024.
\newblock URL \url{https://doi.org/10.1145/3654970}.

\bibitem[Ge et~al.(2013)Ge, He, Ke, and Sun]{ge2013opq}
Tiezheng Ge, Kaiming He, Qifa Ke, and Jian Sun.
\newblock Optimized product quantization for approximate nearest neighbor
  search.
\newblock In \emph{IEEE Conference on Computer Vision and Pattern Recognition
  (CVPR)}, 2013.
\newblock URL
  \url{https://openaccess.thecvf.com/content_cvpr_2013/html/Ge_Optimized_Product_Quantization_2013_CVPR_paper.html}.

\bibitem[Guha et~al.(2023)Guha, Nyarko, Ho, R{\'e}, Chilton,
  et~al.]{guha2023legalbench}
Neel Guha, Julian Nyarko, Daniel~E. Ho, Christopher R{\'e}, Adam Chilton,
  et~al.
\newblock {LegalBench}: A collaboratively built benchmark for measuring legal
  reasoning in large language models.
\newblock \emph{arXiv preprint arXiv:2308.11462}, 2023.
\newblock URL \url{https://arxiv.org/abs/2308.11462}.

\bibitem[Hoppe et~al.(2021)Hoppe, Pelkmann, Migenda, Hotte, and
  Schenck]{hoppe2021legalquad}
Christoph Hoppe, David Pelkmann, Nico Migenda, Daniel Hotte, and Wolfram
  Schenck.
\newblock Towards intelligent legal advisors for document retrieval and
  question-answering in german legal documents.
\newblock In \emph{AIKE 2021}, 2021.
\newblock URL \url{https://doi.org/10.1109/AIKE52691.2021.00011}.

\bibitem[J{\'e}gou et~al.(2011)J{\'e}gou, Douze, and Schmid]{jegou2011pq}
Herv{\'e} J{\'e}gou, Matthijs Douze, and Cordelia Schmid.
\newblock Product quantization for nearest neighbor search.
\newblock \emph{IEEE Transactions on Pattern Analysis and Machine
  Intelligence}, 33\penalty0 (1):\penalty0 117--128, 2011.
\newblock URL \url{https://doi.org/10.1109/TPAMI.2010.57}.

\bibitem[Karpukhin et~al.(2020)Karpukhin, O{\u{g}}uz, Min, Lewis, Wu, Edunov,
  Chen, and Yih]{karpukhin2020dpr}
Vladimir Karpukhin, Barlas O{\u{g}}uz, Sewon Min, Patrick Lewis, Ledell Wu,
  Sergey Edunov, Danqi Chen, and Wen-tau Yih.
\newblock Dense passage retrieval for open-domain question answering.
\newblock In \emph{EMNLP 2020}, 2020.
\newblock URL \url{https://arxiv.org/abs/2004.04906}.

\bibitem[Khattab and Zaharia(2020)]{khattab2020colbert}
Omar Khattab and Matei Zaharia.
\newblock {ColBERT}: Efficient and effective passage search via contextualized
  late interaction over {BERT}.
\newblock In \emph{SIGIR 2020}, 2020.
\newblock URL \url{https://arxiv.org/abs/2004.12832}.

\bibitem[Kusupati et~al.(2022)Kusupati, Bhatt, Rege, Wallingford, Sinha,
  Ramanujan, Howard-Snyder, Chen, Kakade, Jain, and
  Farhadi]{kusupati2022matryoshka}
Aditya Kusupati, Gantavya Bhatt, Aniket Rege, Matthew Wallingford, Aditya
  Sinha, Vivek Ramanujan, William Howard-Snyder, Kaifeng Chen, Sham Kakade,
  Prateek Jain, and Ali Farhadi.
\newblock Matryoshka representation learning.
\newblock In \emph{NeurIPS 2022}, 2022.
\newblock URL \url{https://arxiv.org/abs/2205.13147}.

\bibitem[Lassance et~al.(2024)Lassance, D{\'e}jean, Formal, and
  Clinchant]{lassance2024spladev3}
Carlos Lassance, Herv{\'e} D{\'e}jean, Thibault Formal, and St{\'e}phane
  Clinchant.
\newblock {SPLADE-v3}: New baselines for {SPLADE}.
\newblock \emph{arXiv preprint arXiv:2403.06789}, 2024.
\newblock URL \url{https://arxiv.org/abs/2403.06789}.

\bibitem[Li et~al.(2023)Li, Shao, Wu, Ai, Ma, and Liu]{li2023lecardv2}
Haitao Li, Yunqiu Shao, Yueyue Wu, Qingyao Ai, Yixiao Ma, and Yiqun Liu.
\newblock {LeCaRDv2}: A large-scale chinese legal case retrieval dataset.
\newblock \emph{arXiv preprint arXiv:2310.17609}, 2023.
\newblock URL \url{https://arxiv.org/abs/2310.17609}.

\bibitem[Li et~al.(2024)Li, Dong, Lee, Xia, Zhang, Dai, Wang, and
  Tang]{li2024coir}
Xiangyang Li, Kuicai Dong, Yi~Quan Lee, Wei Xia, Hao Zhang, Xinyi Dai, Yong
  Wang, and Ruiming Tang.
\newblock {CoIR}: A comprehensive benchmark for code information retrieval
  models.
\newblock \emph{arXiv preprint arXiv:2407.02883}, 2024.
\newblock URL \url{https://arxiv.org/abs/2407.02883}.

\bibitem[{Liquid AI}(2025)]{liquidai2025nanobeir}
{Liquid AI}.
\newblock {LiquidAI/nanobeir-multilingual-extended}.
\newblock Hugging Face Hub Dataset, 2025.
\newblock URL
  \url{https://huggingface.co/datasets/LiquidAI/nanobeir-multilingual-extended}.

\bibitem[Liu et~al.(2025)Liu, Enevoldsen, Solomatin, Chung, Aarsen, and
  F{\H{o}}di]{liu2025rteb}
Friso Liu, Kenneth Enevoldsen, Roman Solomatin, Isaac Chung, Tom Aarsen, and
  Zolt{\'a}n F{\H{o}}di.
\newblock Introducing {RTEB}: A new standard for retrieval evaluation.
\newblock Hugging Face Blog, 2025.
\newblock URL \url{https://huggingface.co/blog/rteb}.

\bibitem[Lu(2024)]{lu2024publichealthqa}
Xing~Han Lu.
\newblock publichealth-qa.
\newblock Hugging Face Hub Dataset, 2024.
\newblock URL \url{https://huggingface.co/datasets/xhluca/publichealth-qa}.

\bibitem[Manor and Li(2019)]{manor2019consumercontracts}
Laura Manor and Junyi~Jessy Li.
\newblock Plain english summarization of contracts.
\newblock In \emph{Proceedings of the Natural Legal Language Processing
  Workshop 2019}, pages 1--11, 2019.
\newblock URL \url{https://aclanthology.org/W19-2201/}.

\bibitem[Muennighoff et~al.(2023)Muennighoff, Tazi, Magne, and
  Reimers]{muennighoff2023mteb}
Niklas Muennighoff, Nouamane Tazi, Lo{\"i}c Magne, and Nils Reimers.
\newblock {MTEB}: Massive text embedding benchmark.
\newblock In \emph{EACL 2023}, 2023.
\newblock URL \url{https://arxiv.org/abs/2210.07316}.

\bibitem[Nogueira and Cho(2019)]{nogueira2019passrerank}
Rodrigo Nogueira and Kyunghyun Cho.
\newblock Passage re-ranking with {BERT}.
\newblock \emph{arXiv preprint arXiv:1901.04085}, 2019.
\newblock URL \url{https://arxiv.org/abs/1901.04085}.

\bibitem[Nussbaum et~al.(2024)Nussbaum, Morris, Duderstadt, and
  Mulyar]{nussbaum2024nomic}
Zach Nussbaum, John~X. Morris, Brandon Duderstadt, and Andriy Mulyar.
\newblock Nomic embed: Training a reproducible long context text embedder.
\newblock \emph{Transactions on Machine Learning Research (arXiv:2402.01613)},
  2024.
\newblock URL \url{https://arxiv.org/abs/2402.01613}.

\bibitem[Pham et~al.(2026)Pham, Luu, Vo, Nguyen, and Hoang]{pham2026vnmteb}
Long Pham, Tuan Luu, Thang Vo, Minh Nguyen, and Vu~Hoang.
\newblock {VN-MTEB}: Vietnamese massive text embedding benchmark.
\newblock In \emph{Findings of EACL 2026}, 2026.
\newblock URL \url{https://aclanthology.org/2026.findings-eacl.86/}.

\bibitem[Pijpelink(2026)]{pijpelink2026turboquant}
Arnaud Pijpelink.
\newblock Qdrant 1.18 --- {TurboQuant}.
\newblock Qdrant Blog, 2026.
\newblock URL \url{https://qdrant.tech/blog/qdrant-1.18.x/}.

\bibitem[Qiu et~al.(2022)Qiu, Li, Qu, Chen, She, Liu, Wu, and
  Wang]{qiu2022dureader}
Yifu Qiu, Hongyu Li, Yingqi Qu, Ying Chen, Qiaoqiao She, Jing Liu, Hua Wu, and
  Haifeng Wang.
\newblock {DuReader\_retrieval}: A large-scale chinese benchmark for passage
  retrieval from web search engine.
\newblock In \emph{EMNLP 2022}, pages 5326--5338, 2022.
\newblock URL \url{https://aclanthology.org/2022.emnlp-main.357/}.

\bibitem[Reimers and Gurevych(2019)]{reimers2019sbert}
Nils Reimers and Iryna Gurevych.
\newblock {Sentence-BERT}: Sentence embeddings using siamese {BERT}-networks.
\newblock In \emph{EMNLP-IJCNLP 2019}, 2019.
\newblock URL \url{https://arxiv.org/abs/1908.10084}.

\bibitem[Roberts et~al.(2021)Roberts, Alam, Bedrick, Demner-Fushman, Lo,
  Soboroff, Voorhees, Wang, and Hersh]{roberts2021treccovid}
Kirk Roberts, Tasmeer Alam, Steven Bedrick, Dina Demner-Fushman, Kyle Lo, Ian
  Soboroff, Ellen Voorhees, Lucy~Lu Wang, and William~R. Hersh.
\newblock Searching for scientific evidence in a pandemic: An overview of
  {TREC-COVID}.
\newblock \emph{arXiv preprint arXiv:2104.09632}, 2021.
\newblock URL \url{https://arxiv.org/abs/2104.09632}.

\bibitem[Robertson and Zaragoza(2009)]{robertson2009bm25}
Stephen Robertson and Hugo Zaragoza.
\newblock The probabilistic relevance framework: {BM25} and beyond.
\newblock \emph{Foundations and Trends in Information Retrieval}, 3\penalty0
  (4):\penalty0 333--389, 2009.
\newblock URL \url{https://doi.org/10.1561/1500000019}.

\bibitem[Santhanam et~al.(2021)Santhanam, Khattab, Saad-Falcon, Potts, and
  Zaharia]{santhanam2021colbertv2}
Keshav Santhanam, Omar Khattab, Jon Saad-Falcon, Christopher Potts, and Matei
  Zaharia.
\newblock {ColBERTv2}: Effective and efficient retrieval via lightweight late
  interaction.
\newblock \emph{arXiv preprint arXiv:2112.01488}, 2021.
\newblock URL \url{https://arxiv.org/abs/2112.01488}.

\bibitem[{Sentence Transformers}(2024)]{sentencetransformers2024nanobeirbm25}
{Sentence Transformers}.
\newblock {NanoBEIR} with {BM25} rankings.
\newblock Hugging Face Hub Collection, 2024.
\newblock URL
  \url{https://huggingface.co/collections/sentence-transformers/nanobeir-with-bm25-rankings}.

\bibitem[Shahinmoghadam and Motamedi(2025)]{shahinmoghadam2025builtbench}
Mehrzad Shahinmoghadam and Ali Motamedi.
\newblock Benchmarking pre-trained text embedding models in aligning built
  asset information.
\newblock \emph{Scientific Reports}, 15, 2025.
\newblock URL \url{https://www.nature.com/articles/s41598-025-09052-5}.

\bibitem[Shakir et~al.(2024)Shakir, Aarsen, and
  {SeanLee}]{shakir2024quantization}
Aamir Shakir, Tom Aarsen, and {SeanLee}.
\newblock Binary and scalar embedding quantization for significantly faster and
  cheaper retrieval.
\newblock Hugging Face Blog, 2024.
\newblock URL \url{https://huggingface.co/blog/embedding-quantization}.

\bibitem[Sheikh et~al.(2025)Sheikh, Buades~Marcos, Jousse, Oladipo, Rousseau,
  and Lin]{sheikh2025cure}
Nadia~Amin Sheikh, David Buades~Marcos, Anne-Laure Jousse, Akintunde Oladipo,
  Olivier Rousseau, and Jimmy Lin.
\newblock {CURE}: A dataset for clinical understanding \& retrieval evaluation.
\newblock In \emph{SIGIR 2025}, 2025.
\newblock URL \url{https://doi.org/10.1145/3711896.3737435}.

\bibitem[Shiraee~Kasmaee et~al.(2024)Shiraee~Kasmaee, Khodadad, Saloot, Sherck,
  Dokas, Mahyar, and Samiee]{shiraee2024chemteb}
Ali Shiraee~Kasmaee, Mohammad Khodadad, Mohammad~Arshi Saloot, Nick Sherck,
  Stephen Dokas, Hamidreza Mahyar, and Soheila Samiee.
\newblock {ChemTEB}: Chemical text embedding benchmark.
\newblock In \emph{Proceedings of the 4th NeurIPS Efficient Natural Language
  and Speech Processing Workshop, PMLR 262}, pages 512--531, 2024.
\newblock URL \url{https://arxiv.org/abs/2412.00532}.

\bibitem[{Sionic AI}(2025)]{sionic2025nanobeir}
{Sionic AI}.
\newblock Nano-{BEIR}: A multilingual information retrieval benchmark with
  quality-enhanced queries.
\newblock Hugging Face Blog, 2025.
\newblock URL
  \url{https://huggingface.co/blog/sionic-ai/eval-sionic-nano-beir}.

\bibitem[Snegirev et~al.(2025)Snegirev, Tikhonova, Maksimova, Fenogenova, and
  Abramov]{snegirev2025rumteb}
Artem Snegirev, Maria Tikhonova, Anna Maksimova, Alena Fenogenova, and
  Alexander Abramov.
\newblock The russian-focused embedders' exploration: {ruMTEB} benchmark and
  russian embedding model design.
\newblock In \emph{NAACL 2025}, 2025.
\newblock URL \url{https://aclanthology.org/2025.naacl-long.12/}.

\bibitem[Song et~al.(2025)Song, Gan, Shang, and Zhao]{song2025ifir}
Tingyu Song, Guo Gan, Mingsheng Shang, and Yilun Zhao.
\newblock {IFIR}: A comprehensive benchmark for evaluating
  instruction-following in expert-domain information retrieval.
\newblock In \emph{NAACL 2025}, 2025.
\newblock URL \url{https://aclanthology.org/2025.naacl-long.511/}.

\bibitem[Sourty(2025)]{sourty2025nanobeir}
Rapha{\"e}l Sourty.
\newblock {lightonai/nanobeir-multilingual}.
\newblock Hugging Face Hub Dataset, 2025.
\newblock URL
  \url{https://huggingface.co/datasets/lightonai/nanobeir-multilingual}.

\bibitem[Steinberger et~al.(2014)Steinberger, Ebrahim, Poulis,
  Carrasco-Benitez, Schl{\"u}ter, Przybyszewski, and
  Gilbro]{steinberger2014dgt}
Ralf Steinberger, Mohamed Ebrahim, Alexandros Poulis, Manuel Carrasco-Benitez,
  Patrick Schl{\"u}ter, Marek Przybyszewski, and Signe Gilbro.
\newblock An overview of the european union's highly multilingual parallel
  corpora.
\newblock \emph{Language Resources and Evaluation}, 48\penalty0 (4):\penalty0
  679--707, 2014.
\newblock URL \url{https://doi.org/10.1007/s10579-014-9277-0}.

\bibitem[Su et~al.(2024)Su, Yen, Xia, Shi, Muennighoff, Wang, Liu, Shi, Siegel,
  Tang, Sun, Yoon, Arik, Chen, and Yu]{su2024bright}
Hongjin Su, Howard Yen, Mengzhou Xia, Weijia Shi, Niklas Muennighoff, Han-yu
  Wang, Haisu Liu, Quan Shi, Zachary~S. Siegel, Michael Tang, Ruoxi Sun,
  Jinsung Yoon, Sercan~O. Arik, Danqi Chen, and Tao Yu.
\newblock {BRIGHT}: A realistic and challenging benchmark for
  reasoning-intensive retrieval.
\newblock \emph{arXiv preprint arXiv:2407.12883}, 2024.
\newblock URL \url{https://arxiv.org/abs/2407.12883}.

\bibitem[Thakur et~al.(2021)Thakur, Reimers, R{\"u}ckl{\'e}, Srivastava, and
  Gurevych]{thakur2021beir}
Nandan Thakur, Nils Reimers, Andreas R{\"u}ckl{\'e}, Abhishek Srivastava, and
  Iryna Gurevych.
\newblock {BEIR}: A heterogeneous benchmark for zero-shot evaluation of
  information retrieval models.
\newblock In \emph{NeurIPS Datasets and Benchmarks 2021}, 2021.
\newblock URL \url{https://arxiv.org/abs/2104.08663}.

\bibitem[Thoresen(2026)]{thoresen2026vespa}
Thomas~Hjelde Thoresen.
\newblock Embedding tradeoffs, quantified.
\newblock Vespa Blog, 2026.
\newblock URL \url{https://blog.vespa.ai/embedding-tradeoffs-quantified/}.

\bibitem[Trent(2024)]{trent2024bbq}
Benjamin Trent.
\newblock Better binary quantization ({BBQ}) in lucene and elasticsearch.
\newblock Elasticsearch Labs Blog, 2024.
\newblock URL
  \url{https://www.elastic.co/search-labs/blog/better-binary-quantization-lucene-elasticsearch}.

\bibitem[Tsukagoshi and Sasano(2024)]{tsukagoshi2024ruri}
Hayato Tsukagoshi and Ryohei Sasano.
\newblock Ruri: Japanese general text embeddings.
\newblock \emph{arXiv preprint arXiv:2409.07737}, 2024.
\newblock URL \url{https://arxiv.org/abs/2409.07737}.

\bibitem[Veasey(2026)]{veasey2026bbqturbo}
Thomas Veasey.
\newblock Elasticsearch's {BBQ} vs. {TurboQuant}: 10--40x faster on cpu and
  lower ranking noise.
\newblock Elasticsearch Labs Blog, 2026.
\newblock URL
  \url{https://www.elastic.co/search-labs/blog/elasticsearch-bbq-osq-vs-turbo}.

\bibitem[Voorhees and Harman(2005)]{voorhees2005trec}
Ellen~M. Voorhees and Donna~K. Harman.
\newblock \emph{{TREC}: Experiment and Evaluation in Information Retrieval}.
\newblock MIT Press, 2005.
\newblock ISBN 9780262220736.

\bibitem[Wadden et~al.(2020)Wadden, Lin, Lo, Wang, van Zuylen, Cohan, and
  Hajishirzi]{wadden2020scifact}
David Wadden, Shanchuan Lin, Kyle Lo, Lucy~Lu Wang, Madeleine van Zuylen, Arman
  Cohan, and Hannaneh Hajishirzi.
\newblock Fact or fiction: Verifying scientific claims.
\newblock In \emph{EMNLP 2020}, pages 7534--7550, 2020.
\newblock URL \url{https://aclanthology.org/2020.emnlp-main.609/}.

\bibitem[Wang et~al.(2024{\natexlab{a}})Wang, Yang, Huang, Yang, Majumder, and
  Wei]{wang2024me5}
Liang Wang, Nan Yang, Xiaolong Huang, Linjun Yang, Rangan Majumder, and Furu
  Wei.
\newblock Multilingual {E5} text embeddings: A technical report.
\newblock \emph{arXiv preprint arXiv:2402.05672}, 2024{\natexlab{a}}.
\newblock URL \url{https://arxiv.org/abs/2402.05672}.

\bibitem[Wang et~al.(2024{\natexlab{b}})Wang, Wang, Cao, Wang, Paturi, and
  Bergen]{wang2024birco}
Xiaoyue Wang, Jianyou Wang, Weili Cao, Kaicheng Wang, Ramamohan Paturi, and
  Leon Bergen.
\newblock {BIRCO}: A benchmark of information retrieval tasks with complex
  objectives.
\newblock \emph{arXiv preprint arXiv:2402.14151}, 2024{\natexlab{b}}.
\newblock URL \url{https://arxiv.org/abs/2402.14151}.

\bibitem[Wang et~al.(2025)Wang, Asai, Yu, Xu, Xie, Neubig, and
  Fried]{wang2025coderag}
Zora~Zhiruo Wang, Akari Asai, Xinyan~Velocity Yu, Frank~F. Xu, Yiqing Xie,
  Graham Neubig, and Daniel Fried.
\newblock {CodeRAG-Bench}: Can retrieval augment code generation?
\newblock In \emph{Findings of NAACL 2025}, pages 3199--3214, 2025.
\newblock URL \url{https://arxiv.org/abs/2406.14497}.

\bibitem[Weller et~al.(2024)Weller, Chang, MacAvaney, Lo, Cohan, Van~Durme,
  Lawrie, and Soldaini]{weller2024followir}
Orion Weller, Benjamin Chang, Sean MacAvaney, Kyle Lo, Arman Cohan, Benjamin
  Van~Durme, Dawn Lawrie, and Luca Soldaini.
\newblock {FollowIR}: Evaluating and teaching information retrieval models to
  follow instructions.
\newblock \emph{arXiv preprint arXiv:2403.15246}, 2024.
\newblock URL \url{https://arxiv.org/abs/2403.15246}.

\bibitem[Weller et~al.(2025)Weller, Ricci, Marone, Chaffin, Lawrie, and
  Van~Durme]{weller2025ettin}
Orion Weller, Kathryn Ricci, Marc Marone, Antoine Chaffin, Dawn Lawrie, and
  Benjamin Van~Durme.
\newblock Seq vs seq: An open suite of paired encoders and decoders.
\newblock In \emph{ICLR 2026 (arXiv:2507.11412)}, 2025.
\newblock URL \url{https://arxiv.org/abs/2507.11412}.

\bibitem[Wojtasik et~al.(2024)Wojtasik, Wo{\l}owiec, Shishkin, Janz, and
  Piasecki]{wojtasik2024beirpl}
Konrad Wojtasik, Kacper Wo{\l}owiec, Vadim Shishkin, Arkadiusz Janz, and Maciej
  Piasecki.
\newblock {BEIR-PL}: Zero shot information retrieval benchmark for the polish
  language.
\newblock In \emph{LREC-COLING 2024}, 2024.
\newblock URL \url{https://aclanthology.org/2024.lrec-main.194/}.

\bibitem[Wrzalik and Krechel(2021)]{wrzalik2021gerdalir}
Marco Wrzalik and Dirk Krechel.
\newblock {GerDaLIR}: A german dataset for legal information retrieval.
\newblock In \emph{Proceedings of the Natural Legal Language Processing
  Workshop 2021}, 2021.
\newblock URL \url{https://aclanthology.org/2021.nllp-1.13/}.

\bibitem[Xiao et~al.(2024{\natexlab{a}})Xiao, Hudson, and
  Al~Moubayed]{xiao2024rarb}
Chenghao Xiao, G.~Thomas Hudson, and Noura Al~Moubayed.
\newblock {RAR-b}: Reasoning as retrieval benchmark.
\newblock \emph{arXiv preprint arXiv:2404.06347}, 2024{\natexlab{a}}.
\newblock URL \url{https://arxiv.org/abs/2404.06347}.

\bibitem[Xiao et~al.(2024{\natexlab{b}})Xiao, Liu, Zhang, Muennighoff, Lian,
  and Nie]{xiao2024cpack}
Shitao Xiao, Zheng Liu, Peitian Zhang, Niklas Muennighoff, Defu Lian, and
  Jian-Yun Nie.
\newblock {C-Pack}: Packed resources for general chinese embeddings.
\newblock In \emph{SIGIR 2024}, 2024{\natexlab{b}}.
\newblock URL \url{https://doi.org/10.1145/3626772.3657878}.

\bibitem[Xu et~al.(2024)Xu, Guan, Greene, and Kechadi]{xu2024contamination}
Cheng Xu, Shuhao Guan, Derek Greene, and M-Tahar Kechadi.
\newblock Benchmark data contamination of large language models: A survey.
\newblock \emph{arXiv preprint arXiv:2406.04244}, 2024.
\newblock URL \url{https://arxiv.org/abs/2406.04244}.

\bibitem[Yamada et~al.(2021)Yamada, Asai, and Hajishirzi]{yamada2021bpr}
Ikuya Yamada, Akari Asai, and Hannaneh Hajishirzi.
\newblock Efficient passage retrieval with hashing for open-domain question
  answering.
\newblock In \emph{ACL 2021}, 2021.
\newblock URL \url{https://arxiv.org/abs/2106.00882}.

\bibitem[Zhang et~al.(2018)Zhang, Zhang, Wang, Guo, and Liu]{zhang2018cmedqa}
Sheng Zhang, Xin Zhang, Hui Wang, Lixiang Guo, and Shanshan Liu.
\newblock Multi-scale attentive interaction networks for chinese medical
  question answer selection.
\newblock \emph{IEEE Access}, 6:\penalty0 74061--74071, 2018.
\newblock URL \url{https://doi.org/10.1109/ACCESS.2018.2883637}.

\bibitem[Zhang et~al.(2024)Zhang, Zhang, Long, Xie, Dai, Tang, Lin, Yang, Xie,
  Huang, Zhang, Li, and Zhang]{zhang2024mgte}
Xin Zhang, Yanzhao Zhang, Dingkun Long, Wen Xie, Ziqi Dai, Jialong Tang, Huan
  Lin, Baosong Yang, Pengjun Xie, Fei Huang, Meishan Zhang, Wenjie Li, and Min
  Zhang.
\newblock {mGTE}: Generalized long-context text representation and reranking
  models for multilingual text retrieval.
\newblock In \emph{EMNLP 2024 (Industry Track)}, 2024.
\newblock URL \url{https://arxiv.org/abs/2407.19669}.

\bibitem[Zhang et~al.(2025{\natexlab{a}})Zhang, Li, Zhou, and
  Liu]{zhangx2025r2med}
Xin Zhang, Lei Li, Xiaohan Zhou, and Zheng Liu.
\newblock {R2MED}: A benchmark for reasoning-driven medical retrieval.
\newblock \emph{arXiv preprint arXiv:2505.14558}, 2025{\natexlab{a}}.
\newblock URL \url{https://arxiv.org/abs/2505.14558}.

\bibitem[Zhang et~al.(2023)Zhang, Thakur, Ogundepo, Kamalloo, Alfonso-Hermelo,
  Li, Liu, Rezagholizadeh, and Lin]{zhang2023miracl}
Xinyu Zhang, Nandan Thakur, Odunayo Ogundepo, Ehsan Kamalloo, David
  Alfonso-Hermelo, Xiaoguang Li, Qun Liu, Mehdi Rezagholizadeh, and Jimmy Lin.
\newblock Making a {MIRACL}: Multilingual information retrieval across a
  continuum of languages.
\newblock In \emph{WSDM 2023}, 2023.
\newblock URL \url{https://arxiv.org/abs/2210.09984}.

\bibitem[Zhang et~al.(2025{\natexlab{b}})Zhang, Li, Long, Zhang, Lin, Yang,
  Xie, Yang, Liu, Lin, Huang, and Zhou]{zhangy2025qwen3}
Yanzhao Zhang, Mingxin Li, Dingkun Long, Xin Zhang, Huan Lin, Baosong Yang,
  Pengjun Xie, An~Yang, Dayiheng Liu, Junyang Lin, Fei Huang, and Jingren Zhou.
\newblock {Qwen3} embedding: Advancing text embedding and reranking through
  foundation models.
\newblock \emph{arXiv preprint arXiv:2506.05176}, 2025{\natexlab{b}}.
\newblock URL \url{https://arxiv.org/abs/2506.05176}.

\bibitem[Zhu et~al.(2024)Zhu, Wang, Yang, Song, Wu, Wei, and
  Li]{zhu2024longembed}
Dawei Zhu, Liang Wang, Nan Yang, Yifan Song, Wenhao Wu, Furu Wei, and Sujian
  Li.
\newblock {LongEmbed}: Extending embedding models for long context retrieval.
\newblock \emph{arXiv preprint arXiv:2404.12096}, 2024.
\newblock URL \url{https://arxiv.org/abs/2404.12096}.

\end{thebibliography}

\clearpage
\appendix
\addtocontents{toc}{\protect\setcounter{tocdepth}{2}}
\renewcommand{\contentsname}{Appendix Table of Contents}
\begingroup
\hypersetup{linkcolor=black}
\tableofcontents
\endgroup
\clearpage

\section{Nano-set construction and dataset list}
\label{app:A}

\subsection{Benchmark/task list}
\label{app:A1}
The $35$ benchmarks and $551$ retrieval tasks are classified into the five families of \secref{sec:tasks}. Of the $551$ tasks, $526$ are natural-language and $25$ are code; code tasks are distributed over $5$ benchmarks (NanoCoIR and NanoCodeRAG are code-only; NanoBRIGHT, NanoRTEB, NanoRARb are mixed with natural language). Natural-language tasks cover $43$ languages in total (\secref{sec:dist}). Table~\ref{tab:a1} shows each benchmark's natural-language task count, code task count, language count, and the main source benchmark its Nano-set references (task/language counts are machine-generated from the evaluation results; sources are based on each dataset spec's citation metadata). Each source benchmark, and the individual datasets composing the composite tasks NanoLaw and NanoMedical, appear as formal entries in the reference list. Note that MNanoBEIR's task count $182$ corresponds to $13$ BEIR datasets $\times$ $14$ language editions (the ``Langs'' column counts the distinct actual languages tagged on each task, so for MNanoBEIR, which contains a \texttt{multilingual} edition spanning several languages, it is $19$, larger than the $14$ language editions). The task count $182$ stands out in the table, but in the macro aggregation that is our primary basis it is grouped hierarchically by language and dataset as in \secref{sec:metrics} and contributes as one benchmark ($1/35$ in macro), like the others (in micro aggregation the weight grows in proportion to task count).

{\footnotesize\setlength{\tabcolsep}{4pt}\renewcommand{\arraystretch}{1.08}
\rowcolors{2}{rowgray}{white}
\begin{longtable}{@{}lrrr>{\raggedright\arraybackslash}p{6.2cm}@{}}
\caption{Composition of the $35$ benchmarks (an empty ``Code'' column means natural language only).}
\label{tab:a1}\\
\toprule
\rowcolor{headcol} \hdr{Benchmark} & \hdr{NL} & \hdr{Code} & \hdr{Langs} & \hdr{Main source} \\
\midrule
\endfirsthead
\rowcolor{headcol} \hdr{Benchmark} & \hdr{NL} & \hdr{Code} & \hdr{Langs} & \hdr{Main source} \\
\midrule
\endhead
\bottomrule
\endfoot
MNanoBEIR & 182 & --- & 19 & BEIR / NanoBEIR \citep{thakur2021beir,aarsen2024nanobeir} \\
NanoBIRCO & 5 & --- & 1 & BIRCO \citep{wang2024birco} \\
NanoBRIGHT & 15 & 5 & 1 & BRIGHT \citep{su2024bright} \\
NanoBuiltBench & 2 & --- & 1 & BuiltBench \citep{shahinmoghadam2025builtbench} \\
NanoCMTEB & 8 & --- & 2 & C-MTEB (C-Pack) \citep{xiao2024cpack} \\
NanoChemTEB & 3 & --- & 1 & ChemTEB \citep{shiraee2024chemteb} \\
NanoCoIR & --- & 10 & 1 & CoIR \citep{li2024coir} \\
NanoCodeRAG & --- & 4 & 1 & CodeRAG-Bench \citep{wang2025coderag} \\
NanoDAPFAM & 12 & --- & 1 & DAPFAM \citep{ayaou2026dapfam} \\
NanoFaMTEB-v2 & 17 & --- & 1 & MMTEB \citep{enevoldsen2025mmteb} \\
NanoIFIR & 7 & --- & 1 & IFIR \citep{song2025ifir} \\
NanoIndicQA & 11 & --- & 11 & IndicQA (IndicXTREME) \citep{doddapaneni2023indicxtreme} \\
NanoJMTEB-v2 & 11 & --- & 1 & MMTEB \citep{enevoldsen2025mmteb} \\
NanoLaw & 8 & --- & 3 & Legal IR composite (AILA, etc.) \citep{bhattacharya2019aila,guha2023legalbench} \\
NanoLongEmbed & 6 & --- & 1 & LongEmbed \citep{zhu2024longembed} \\
NanoMIRACL & 18 & --- & 18 & MIRACL \citep{zhang2023miracl} \\
NanoMLDR & 13 & --- & 13 & MLDR (BGE-M3) \citep{chen2024bgem3} \\
NanoMMTEB-v2 & 18 & --- & 10 & MMTEB \citep{enevoldsen2025mmteb} \\
NanoMTEB-Dutch & 27 & --- & 2 & MTEB-NL \citep{banar2025mtebnl} \\
NanoMTEB-French & 8 & --- & 2 & MTEB-French \citep{ciancone2024mtebfrench} \\
NanoMTEB-German & 5 & --- & 2 & MTEB \citep{muennighoff2023mteb} \\
NanoMTEB-Korean & 5 & --- & 1 & MTEB \citep{muennighoff2023mteb} \\
NanoMTEB-Misc & 12 & --- & 8 & MTEB \citep{muennighoff2023mteb} \\
NanoMTEB-Polish & 14 & --- & 1 & MTEB \citep{muennighoff2023mteb} \\
NanoMTEB-Scandinavian & 7 & --- & 5 & SEB \citep{enevoldsen2024seb} \\
NanoMTEB-Spanish & 7 & --- & 2 & MTEB \citep{muennighoff2023mteb} \\
NanoMTEB-Thai & 9 & --- & 2 & MTEB \citep{muennighoff2023mteb} \\
NanoMTEB-v2 & 10 & --- & 1 & MTEB \citep{muennighoff2023mteb} \\
NanoMedical & 10 & --- & 4 & Medical IR composite (CURE, etc.) \citep{sheikh2025cure} \\
NanoMuPLeR & 14 & --- & 14 & MuPLeR (EU DGT-Acquis; MTEB) \citep{steinberger2014dgt} \\
NanoR2MED & 8 & --- & 1 & R2MED \citep{zhangx2025r2med} \\
NanoRARb & 16 & 1 & 2 & RAR-b \citep{xiao2024rarb} \\
NanoRTEB & 9 & 5 & 1 & RTEB \citep{liu2025rteb} \\
NanoRuMTEB & 3 & --- & 1 & ruMTEB \citep{snegirev2025rumteb} \\
NanoVNMTEB & 26 & --- & 2 & VN-MTEB \citep{pham2026vnmteb} \\
\end{longtable}}

\subsection{Dataset versions and sources}
\label{app:A2}
In evaluation, the dataset version (commit SHA) can be specified explicitly. The resolved SHA is always recorded in the result file, so even when a dataset's contents are updated, past and new numbers can be distinguished by version. Rather than re-implementing each Nano-set's shrinking procedure, the benchmark references already-published Nano-family datasets on the Hugging Face Hub by name and version. Each dataset's shrinking procedure follows the original paper or distributor's description.

\subsection{Known differences from Nano-set construction}
\label{app:A3}
Nano-set construction introduces the following differences from the original benchmark. (i) Due to sampling and retrieval-space reconstruction, absolute scores and variance do not necessarily match the original; in particular, tasks that compress multiple subsets into a $10$K-document combined corpus (e.g., \texttt{belebele}, \texttt{mlqa} in NanoMMTEB-v2) have a compressed score range (\appref{app:D3}). (ii) Because the query count is limited, the standard error of the evaluation values is larger than the original. (iii) Because hard negatives are not necessarily preserved, scores may saturate for tasks from large corpora. (iv) The fixed candidate set aids reproducibility but reflects candidate-generation language/domain bias as-is, so candidate coverage may drop for low-resource languages or instruction-following tasks (\secref{sec:rerankanalysis}). (v) Because query/document counts differ greatly across tasks, the simple (micro) average is pulled by large benchmarks; the benchmark co-reports the macro average to address this (\secref{sec:metrics}). Neither the micro average (the leaderboard/viewer default) nor the macro average is ``correct''; use macro to suppress scale skew and micro for equal-weight over all tasks, switchable in the leaderboard/viewer (\secref{sec:metrics}, \ref{app:B2}). This paper uses macro as primary to suppress scale skew. (vi) Duplicate tasks deriving from the same original dataset coexist across families (e.g., Nano-sets from NanoBEIR and from official MTEB / MMTEB retrieval). The benchmark keeps duplicates, respecting sampling differences (\secref{sec:tasks}), so the micro average may double-count duplicates; the macro average that is our primary basis mitigates this (\secref{sec:metrics}). Table~\ref{tab:a3} lists task names appearing in multiple benchmarks (machine-generated). $23$ task names appear in $2$--$4$ benchmarks. For families that use language codes (\texttt{en}, \texttt{fr}, etc.) in task names, the same name is a different corpus, so these $17$ name collisions were excluded from the duplicate list (only \texttt{nq}, despite being language-code length, is a truly shared original dataset and is included). Same-named tasks include both the same English original re-sampled by different Nano families (\texttt{scidocs}, \texttt{treccovid}, etc.) and language-translated versions of the same original (Dutch/Polish versions of \texttt{cqadupstack\_*}, etc.); both are kept as different evaluation surfaces with different sampling/language (\secref{sec:tasks}).

{\footnotesize\setlength{\tabcolsep}{4pt}\renewcommand{\arraystretch}{1.08}
\rowcolors{2}{rowgray}{white}
\begin{longtable}{@{}>{\raggedright\arraybackslash}p{3.5cm}>{\raggedright\arraybackslash}p{4.7cm}>{\raggedright\arraybackslash}p{5.8cm}@{}}
\caption{Tasks deriving from the same original dataset across families.}
\label{tab:a3}\\
\toprule
\rowcolor{headcol} \hdr{Task name} & \hdr{Benchmarks} & \hdr{Languages} \\
\midrule
\endfirsthead
\rowcolor{headcol} \hdr{Task name} & \hdr{Benchmarks} & \hdr{Languages} \\
\midrule
\endhead
\bottomrule
\endfoot
\texttt{fever} & MNanoBEIR, NanoMTEB-Dutch, NanoMTEB-v2 & ar, de, en, es, fr, it, ja, ko, multi, nl, pt, sv, th, vi \\
\texttt{nq} & MNanoBEIR, NanoMTEB-Dutch, NanoMTEB-Polish & ar, de, en, es, fr, it, ja, ko, multi, nl, pl, pt, sv, th, vi \\
\texttt{scidocs} & MNanoBEIR, NanoMMTEB-v2, NanoMTEB-v2 & en, ja, ko, multi, vi \\
\texttt{NanoAILACasedocs} & NanoLaw, NanoRTEB & en \\
\texttt{NanoAILAStatutes} & NanoLaw, NanoRTEB & en \\
\texttt{NanoApps} & NanoCoIR, NanoRTEB & en \\
\texttt{NanoCUREv1} & NanoMedical, NanoRTEB & en \\
\texttt{NanoLegalSummarization} & NanoLaw, NanoRTEB & en \\
\texttt{argu\_ana} & NanoMMTEB-v2, NanoMTEB-v2 & en \\
\texttt{covid} & NanoCMTEB, NanoMMTEB-v2 & zh \\
\texttt{cqadupstack\_android} & NanoMTEB-Dutch, NanoMTEB-Polish & nl, pl \\
\texttt{cqadupstack\_english} & NanoMTEB-Dutch, NanoMTEB-Polish & nl, pl \\
\texttt{cqadupstack\_gis} & NanoMTEB-Dutch, NanoMTEB-Polish & nl, pl \\
\texttt{cqadupstack\_mathematica} & NanoMTEB-Dutch, NanoMTEB-Polish & nl, pl \\
\texttt{cqadupstack\_physics} & NanoMTEB-Dutch, NanoMTEB-Polish & nl, pl \\
\texttt{cqadupstack\_programmers} & NanoMTEB-Dutch, NanoMTEB-Polish & nl, pl \\
\texttt{cqadupstack\_stats} & NanoMTEB-Dutch, NanoMTEB-Polish & nl, pl \\
\texttt{cqadupstack\_tex} & NanoMTEB-Dutch, NanoMTEB-Polish & nl, pl \\
\texttt{cqadupstack\_webmasters} & NanoMTEB-Dutch, NanoMTEB-Polish & nl, pl \\
\texttt{cqadupstack\_wordpress} & NanoMTEB-Dutch, NanoMTEB-Polish & nl, pl \\
\texttt{quora} & NanoMTEB-Dutch, NanoMTEB-Polish & nl, pl \\
\texttt{treccovid} & NanoMMTEB-v2, NanoMTEB-v2 & en \\
\texttt{twitter\_hjerne} & NanoMMTEB-v2, NanoMTEB-Scandinavian & da \\
\end{longtable}}

\section{Evaluation protocol details}
\label{app:B}

\subsection{Metric definitions}
\label{app:B1}
The metrics computed and stored at evaluation time are \nDCG{} and accuracy@$100$, and each task's result also stores the top $100$ ranking as an artifact. When building the leaderboard (DuckDB warehouse), nDCG@$100$, recall@$\{10,100\}$, accuracy@$\{1,10,100\}$, MRR@$10$, and MAP@$100$ are recomputed from this stored ranking. This design avoids bloating result files with redundant metric columns while allowing needed metrics to be recomputed downstream. \nDCG{} follows the standard definition, normalizing $\mathrm{DCG} = \sum_i \mathrm{rel}_i / \log_2(i+2)$ by IDCG (relevance labels are binary in this benchmark). Unlike the full $k$ list MTEB adopts ($1,3,5,10,20,100,1000$), $k$ is restricted to $\{1,10,100\}$ to match the design of handling top-$100$ candidate sets/rankings. Reliability metrics (e.g., nAUC) are not adopted; evaluation-noise quantification is done by task bootstrap (\secref{sec:noise}).

\subsection{Aggregation method}
\label{app:B2}
Per benchmark, the simple average over tasks ($\times 100$) is displayed, and task rank uses competition rank (ties share a rank, skipping the next). For cross-benchmark aggregation, the equal-weight micro average over all tasks and the per-benchmark equal-weight macro average are co-reported. The leaderboard/viewer default is the micro average (simpler interpretation when combined with filters/Nano-set narrowing), with macro equally switchable. This paper uses macro as the primary reporting basis to suppress scale skew (\secref{sec:metrics}). In macro aggregation, benchmarks bundling multiple-language derivatives are first averaged per derivative (language/task) before being combined into one benchmark score. Concretely, MNanoBEIR is grouped by BEIR dataset name (\texttt{arguana}, \texttt{fever}, \texttt{scidocs}, etc., $13$ kinds), each averaged over $14$ languages, and the $13$ dataset averages averaged again. Thus even if one BEIR dataset has more languages than another, the per-dataset contribution is equal. In the rank-correlation analysis (\secref{sec:corr}), the Borda score $100 \times (N - \mathrm{rank})/(N-1)$ ($N$ = number of models) is used per task, and a model's Borda value is its average; this is a task-scale-independent ranking metric.

\subsection{Handling missing tasks}
\label{app:B3}
Ranking targets only models that have the entire expected task set within the selected display range (complete-model rule). When displaying efficiency variants, completeness is judged per variant for the same model--variant pair. Models with un-evaluable tasks are excluded from that ranking table, aligning the comparison conditions.

\section{Models, prompts, and execution environment}
\label{app:C}

\subsection{List of evaluated models}
\label{app:C1}
The evaluated models split into candidate-generation methods (dense $33$, learned sparse $4$, late interaction $6$, lexical-baseline BM25 $1$) and rerankers ($10$ cross-encoders, $1$ LLM-style) that take the top candidate set as input. Note that the fixed DuckDB snapshot itself contains $57$ models, but our analysis excludes two unreleased (dense) models from all pools, aggregations, and figures/tables, so the evaluation target is $55$ models.

Individual model IDs, resolved Hugging Face revisions, parameter counts, and embedding dimensions are machine-readably available from the public leaderboard (\texttt{hakari-bench/leaderboard}) and the result dataset (\texttt{hakari-bench/leaderboard\_database}), so a complete list is not included here. The models appearing in the tables/figures of the main text are representative of each method. All models have base rows for all $551$ tasks.

\subsection{Prompt settings}
\label{app:C2}
We perform no fine instruction control over the query/document transformation. Following each model's official documentation or the SentenceTransformers standard prompt format, for models that officially define query/document prompts (or prompt names / encode-task specifications), we applied them \textbf{uniformly to all $551$ tasks}. That is, at most one fixed setting each for query and document per model, with no per-task prompt switching (uniqueness per model in the result table was verified mechanically). As a guide, of the $55$ models, $20$ officially define a query-side prompt (or prompt name / encode-task spec) and $16$ also define a document-side one (the remaining $35$ have no prompt specification and are evaluated with the model's default transformation only; $4$ models without a document side specify only the query side). The applied format differs per model; e.g., E5 family uses \texttt{query:} / \texttt{passage:}, Qwen3 / Jina v5 family uses prompt names \texttt{query} / \texttt{document}, and ruri uses Japanese \texttt{kensaku kueri:} / \texttt{kensaku bunsho:} (search query / search document). All are applied uniformly to all $551$ tasks, with no per-task switching.

This uniform application has a fairness limitation. For models designed to switch prompts (instructions) finely per task type, uniform application of a single retrieval prompt may not draw out their true performance, making the comparison unfair to them. Conversely, against prompt-agnostic models, not doing per-task prompt search works to align conditions. Our numbers should be read as ``performance under uniform application of the official base prompt''; comparison including per-task instruction optimization is future work.

\subsection{Execution environment}
\label{app:C3}
The dense evaluation data type is basically \texttt{bf16}; for models that cannot be loaded in bf16 or suffer large score degradation, \texttt{fp32} / \texttt{fp16} is used. The similarity function is evaluated with both cosine and dot (and any function the model specifies); for each model--task pair we report whichever similarity maximizes the task \nDCG{}. This is therefore a per-task best-of-similarity \emph{upper bound} (an oracle over the similarity choice), not a single similarity fixed in advance; the same procedure is applied uniformly to all dense models, so it does not advantage any particular model. Score computation and top-$k$ extraction are, in principle, done on the same device (CPU or GPU) as the base embeddings. When scoring int8/binary quantization on GPU, values are cast to float32 for matrix multiplication; this cast is numerically equivalent and does not affect quality. The execution environment uses PyTorch, Transformers, and Sentence-Transformers for dense/sparse/reranker evaluation, and PyLate for late interaction (ColBERT family) (specific versions are recorded per result row and machine-aggregatable). Because the evaluation period spanned library updates, versions differ; for models confirmed to suffer large score degradation without a specific version, the version is fixed. Late interaction (ColBERT family) is evaluated in fp32 due to implementation requirements. For attention, flash\_attention\_2 is used where supported, otherwise sdpa (or each implementation's default).

\section{Rank correlation and reliability details}
\label{app:D}
So that readers can independently verify the rank correlation shown in \secref{sec:corr}, this appendix shows the common model set, per-model ranks and scores, per-task mean/variance differences, and the figure of the \secref{sec:noise} bootstrap confidence intervals.

\subsection{Common model set}
\label{app:D1}
Both pairs use the official \texttt{mteb/results} commit \texttt{1e8ab5d} (reflected up to 2026-06-08) as reference and compare against the base rows (excluding efficiency variants) of the local DuckDB (same results as \secref{sec:results}). The $95\%$ confidence intervals for Spearman from bootstrap ($10{,}000$ resamples with replacement) over the common model set, for the three comparisons (MMTEB / MTEB-v2 / BEIR-en), are $[0.915, 0.995]$, $[0.912, 0.998]$, and $[0.882, 0.997]$ (same procedure as the \secref{sec:noise} reliability analysis). A rank is assigned by descending score within each task, with ties as the average rank. For the official \texttt{mteb/results} JSON, \texttt{main\_score} is used where present, otherwise the retrieval primary metric \texttt{ndcg\_at\_10} converted to a 0--100 scale. Models with results only on the Nano side, or for which the official side does not have all tasks, are excluded as they could distort the correlation (common model set: MMTEB $24$ / MTEB-v2 $18$ / BEIR-en $19$). Note that in the official \texttt{mteb/results}, a model's results are stored in per-revision (per-measurement) directories, and even the same model can have both official-pipeline revisions and external submissions (\texttt{external}), so no single revision may have all target tasks. We do not stitch results from different-condition revisions to complete all tasks; only models with all target tasks in a single revision are compared. This single-revision-completeness condition prevents the correlation from being distorted by mixing results with different evaluation settings.

In every comparison, the common set is only models whose original tasks corresponding to the Nano-side tasks are present in a single revision of the official \texttt{mteb/results}, and ranking is over the intersection of tasks common to both.
\begin{itemize}[leftmargin=*]
\item \textbf{MMTEB v2 retrieval vs NanoMMTEB-v2}: common $24$ models $\times$ $18$ tasks.
\item \textbf{MTEB retrieval v2 vs NanoMTEB-v2}: common $18$ models $\times$ $10$ tasks.
\item \textbf{BEIR (full) vs NanoBEIR-en} (\secref{sec:validity}): common $19$ models $\times$ $13$ tasks. NanoBEIR-en matches the English tasks of MNanoBEIR against the original BEIR version (not the HardNegatives / .v3 version used by MTEB-v2). Spearman $0.973$, Borda Pearson $0.974$, max rank difference $3$.
\end{itemize}

\subsection{Per-model ranking tables}
\label{app:D2}
Table~\ref{tab:d1} lists the common $24$ models in MMTEB rank order. $\Delta$rank $=$ Nano rank $-$ MMTEB rank; negative means Nano ranks higher than official, positive the reverse. The Borda columns are the $18$-task average Borda, and the mean columns the $18$-task average \nDCG{} ($\times 100$).

{\footnotesize\setlength{\tabcolsep}{5pt}\renewcommand{\arraystretch}{1.1}
\rowcolors{2}{rowgray}{white}
\begin{longtable}{@{}>{\ttfamily\raggedright\arraybackslash}p{3.7cm}rrrrrrr@{}}
\caption{MMTEB / Nano ranks and scores for the common $24$ models. Model IDs are abbreviated (Hugging Face org prefix omitted).}
\label{tab:d1}\\
\toprule
\rowcolor{headcol} \hdr{\normalfont model} & \hdr{MMTEB} & \hdr{Nano} & \hdr{$\Delta$rk} & \hdr{M.Borda} & \hdr{N.Borda} & \hdr{M.mean} & \hdr{N.mean} \\
\midrule
\endfirsthead
\rowcolor{headcol} \hdr{\normalfont model} & \hdr{MMTEB} & \hdr{Nano} & \hdr{$\Delta$rk} & \hdr{M.Borda} & \hdr{N.Borda} & \hdr{M.mean} & \hdr{N.mean} \\
\midrule
\endhead
\bottomrule
\endfoot
harrier-0.6b & 1.0 & 1.0 & 0.0 & 85.14 & 82.85 & 71.74 & 56.62 \\
pplx-embed-0.6b & 2.0 & 4.0 & $+2.0$ & 81.04 & 76.69 & 66.34 & 53.79 \\
jina-v5-small & 3.0 & 2.0 & $-1.0$ & 79.95 & 82.25 & 65.71 & 55.90 \\
embeddinggemma-300m & 4.0 & 5.0 & $+1.0$ & 76.33 & 72.46 & 64.37 & 51.72 \\
jina-v5-nano & 5.0 & 3.0 & $-2.0$ & 75.12 & 79.35 & 64.06 & 53.66 \\
harrier-270m & 6.0 & 9.0 & $+3.0$ & 74.88 & 57.01 & 66.89 & 52.22 \\
Qwen3-Embed-0.6B & 7.0 & 6.0 & $-1.0$ & 71.26 & 72.22 & 65.14 & 55.80 \\
granite-311m-ml-r2 & 8.0 & 7.0 & $-1.0$ & 65.70 & 71.38 & 64.81 & 57.73 \\
arctic-embed-l-v2.0 & 9.0 & 8.0 & $-1.0$ & 58.94 & 62.80 & 59.26 & 50.23 \\
F2LLM-v2-330M & 10.0 & 10.5 & $+0.5$ & 53.87 & 52.41 & 57.50 & 48.69 \\
jina-v3 & 11.0 & 12.0 & $+1.0$ & 51.81 & 47.83 & 56.67 & 47.23 \\
bge-m3 & 12.0 & 13.0 & $+1.0$ & 51.21 & 47.70 & 56.55 & 48.46 \\
granite-97m-ml-r2 & 13.0 & 10.5 & $-2.5$ & 50.00 & 52.41 & 60.58 & 53.15 \\
F2LLM-v2-160M & 14.0 & 14.0 & 0.0 & 43.24 & 46.01 & 55.34 & 48.02 \\
granite-278m-ml & 15.0 & 15.0 & 0.0 & 41.30 & 45.05 & 55.13 & 46.87 \\
mE5-base & 16.0 & 16.0 & 0.0 & 39.01 & 41.30 & 53.56 & 46.93 \\
granite-107m-ml & 17.0 & 19.0 & $+2.0$ & 32.37 & 31.88 & 51.57 & 43.98 \\
F2LLM-v2-80M & 18.0 & 17.0 & $-1.0$ & 31.16 & 34.78 & 51.91 & 44.74 \\
mE5-small & 19.0 & 20.0 & $+1.0$ & 31.04 & 31.40 & 53.82 & 44.55 \\
bge-small-en-v1.5 & 20.0 & 21.0 & $+1.0$ & 28.50 & 28.26 & 36.64 & 36.83 \\
all-MiniLM-L6-v2 & 21.0 & 22.0 & $+1.0$ & 22.83 & 20.29 & 33.15 & 34.01 \\
nomic-embed-text-v1.5 & 22.0 & 18.0 & $-4.0$ & 20.65 & 33.21 & 33.35 & 39.54 \\
static-sim-mrl-ml-v1 & 23.0 & 24.0 & $+1.0$ & 17.63 & 14.73 & 40.58 & 36.08 \\
paraphrase-ml-MiniLM-L12 & 24.0 & 23.0 & $-1.0$ & 17.03 & 15.70 & 36.59 & 34.93 \\
\end{longtable}}

Of the $24$ models, $\Delta$rank $=0$ for $4$, $|\Delta\mathrm{rank}| \le 1$ for $18$ in total, and $|\Delta\mathrm{rank}| \le 2$ for $21$. The largest difference is $-4$ for \model{nomic-ai/nomic-embed-text-v1.5} \citep[Nomic Embed;][]{nussbaum2024nomic}. It ranks higher on the Nano side because the original large corpus shrinks greatly under Nano-ization into a retrieval space favorable to this model, strong on English BEIR; this should be read as a rank swap, not an absolute score (\appref{app:D3}). That the top \model{microsoft/harrier-oss-v1-0.6b} achieves rank $1$ on both official and Nano ($\Delta$rank $=0$) is a representative agreement on top-band rank preservation. The mean column is lower on the Nano side for almost all models, by about $7$ points on average, mainly because NanoMMTEB-v2 contains tasks compressing multiple subsets into a $10$K-document combined corpus, giving a retrieval space different from official with a compressed score range. Importantly, even when absolute values drop, the relative ranking among models is largely preserved.

Table~\ref{tab:d2} lists the common $18$ models in MTEB-v2 rank order.

{\footnotesize\setlength{\tabcolsep}{5pt}\renewcommand{\arraystretch}{1.1}
\rowcolors{2}{rowgray}{white}
\begin{longtable}{@{}>{\ttfamily\raggedright\arraybackslash}p{3.7cm}rrrrrrr@{}}
\caption{MTEB-v2 / Nano ranks and scores for the common $18$ models. Model IDs are abbreviated (Hugging Face org prefix omitted).}
\label{tab:d2}\\
\toprule
\rowcolor{headcol} \hdr{\normalfont model} & \hdr{MTEB-v2} & \hdr{Nano} & \hdr{$\Delta$rk} & \hdr{M.Borda} & \hdr{N.Borda} & \hdr{M.mean} & \hdr{N.mean} \\
\midrule
\endfirsthead
\rowcolor{headcol} \hdr{\normalfont model} & \hdr{MTEB-v2} & \hdr{Nano} & \hdr{$\Delta$rk} & \hdr{M.Borda} & \hdr{N.Borda} & \hdr{M.mean} & \hdr{N.mean} \\
\midrule
\endhead
\bottomrule
\endfoot
jina-v5-small & 1.0 & 1.0 & 0.0 & 91.18 & 93.53 & 60.07 & 64.50 \\
Qwen3-Embed-0.6B & 2.0 & 3.0 & $+1.0$ & 91.18 & 85.29 & 61.83 & 63.72 \\
jina-v5-nano & 3.0 & 2.0 & $-1.0$ & 84.71 & 91.18 & 58.80 & 63.86 \\
arctic-embed-l-v2.0 & 4.0 & 4.0 & 0.0 & 83.53 & 75.29 & 58.56 & 61.90 \\
embeddinggemma-300m & 5.0 & 5.0 & 0.0 & 69.41 & 68.82 & 55.69 & 60.92 \\
F2LLM-v2-330M & 6.0 & 7.0 & $+1.0$ & 64.71 & 58.23 & 53.34 & 58.40 \\
granite-311m-ml-r2 & 7.0 & 6.0 & $-1.0$ & 57.65 & 64.12 & 52.55 & 60.02 \\
granite-278m-ml & 8.0 & 8.5 & $+0.5$ & 50.59 & 56.47 & 51.45 & 58.51 \\
mE5-large & 9.0 & 8.5 & $-0.5$ & 49.41 & 56.47 & 51.53 & 58.44 \\
granite-97m-ml-r2 & 10.0 & 11.0 & $+1.0$ & 45.29 & 41.77 & 50.09 & 57.15 \\
F2LLM-v2-160M & 11.0 & 12.0 & $+1.0$ & 42.94 & 38.82 & 49.33 & 55.67 \\
mE5-base & 12.0 & 10.0 & $-2.0$ & 38.82 & 42.35 & 48.98 & 55.95 \\
granite-107m-ml & 13.0 & 13.0 & 0.0 & 33.53 & 37.06 & 47.91 & 55.83 \\
F2LLM-v2-80M & 14.0 & 16.0 & $+2.0$ & 32.35 & 23.53 & 47.54 & 53.06 \\
all-MiniLM-L6-v2 & 15.0 & 14.0 & $-1.0$ & 29.41 & 34.12 & 42.92 & 55.14 \\
mE5-small & 16.0 & 15.0 & $-1.0$ & 27.06 & 24.12 & 46.43 & 53.48 \\
paraphrase-ml-MiniLM-L12 & 17.0 & 17.0 & 0.0 & 7.65 & 7.65 & 35.93 & 47.29 \\
static-sim-mrl-ml-v1 & 18.0 & 18.0 & 0.0 & 0.59 & 1.18 & 28.81 & 41.45 \\
\end{longtable}}

Of the $18$ models, $\Delta$rank $=0$ for $6$, $|\Delta\mathrm{rank}| \le 1$ for $16$, and $|\Delta\mathrm{rank}| \le 2$ covers all. The max rank difference is $2$. The top \model{jinaai/jina-embeddings-v5-text-small} is rank $1$ on both official and Nano ($\Delta$rank $=0$). The mean column, opposite to MMTEB, is higher on the Nano side, by about $7$ points on average; in NanoMTEB-v2, extreme low-score regions are compressed on the Nano side by the hard-negative pool and corpus cap (e.g., \texttt{fever\_hard\_negatives}, \texttt{touche2020\_v3}), pushing up the mean. As a ranking proxy this is also not a problem.

\subsection{Per-task mean/variance differences}
\label{app:D3}
That rank correlation is high while per-task mean scores and standard deviations shift is because the monotone superiority among models is well preserved, whereas each task's score scale is sensitive to query/document counts, qrels density, candidate-pool difficulty, and subset mixing. The observed differences fall into five types (the specific per-task std values below are representative examples from an earlier snapshot with more detailed std analysis; the type-level trends reproduce regardless of snapshot).
\begin{itemize}[leftmargin=*]
\item \textbf{(i) Scale change from combined subsets:} \texttt{belebele} ($122$ languages) and \texttt{mlqa} (multilingual QA) do many small retrievals per language pair officially, whereas Nano-sets combine multiple pairs/subsets into one $10$K-document corpus. The retrieval space thus differs greatly from official and the mean drops (official means belebele $64.2$ / mlqa $65.5$; Nano means $17.1$ / $13.2$). Variance behavior splits by task: \texttt{mlqa}'s std is compressed to $27\%$ of official, while \texttt{belebele}'s widens to $1.11\times$. Both are examples to read as a rank proxy, not absolute score.
\item \textbf{(ii) Compression by hard-negative pool and corpus cap:} \texttt{fever\_hard\_negatives}, \texttt{treccovid}, \texttt{touche2020\_v3}, \texttt{mlqa} have std compressed to $27$--$51\%$ on the Nano side. For example, \texttt{fever\_hard\_negatives}'s score range shifts to the high side, official $27.5$--$92.9$ vs Nano $74.1$--$99.1$. The task did not become ``easy''; rather, the candidate pool and qrels design sometimes do not represent the full set's wide difficulty range.
\item \textbf{(iii) Ceiling effect and instability in the low-score region:} \texttt{hagrid} has std ratio $2.26$ but its score range ($97.3$--$98.9$ official, $95.7$--$99.3$ Nano) is in the saturated band; conversely \texttt{temp\_reason\_l1} is in a low band (mean $3.89$ / $3.37$) with high CV. Because a few successes dominate variance, bootstrap CIs or query-level success-rate distributions are more appropriate than the std ratio.
\item \textbf{(iv) Difference in evaluation policy:} \texttt{lembpasskey} has std ratio $0.86$ but task Spearman $0.662$ and max rank difference $20$. MTEB's $8$ context-length splits are stratified-compressed by Nano into $100$ queries $\times$ $100$ documents, so the official ``success rate per length condition'' and Nano's ``length-balanced small mixed retrieval'' are not the same scale; read as a difference in evaluation policy.
\item \textbf{(v) Sample bias of domain-specific queries:} \texttt{cqadupstack\_gaming}, \texttt{cqadupstack\_unix}, \texttt{fiqa2018}, \texttt{scidocs} have slightly larger Nano-side std ($1.05$--$1.18$). Domain-specific vocabulary sample bias emphasizes model-family strengths/weaknesses, but the ratio stays below $1.18$ and does not affect the overall Spearman $0.983$ (MTEB-v2).
\end{itemize}
These observations support that Nano-sets are designed as a rank proxy, not an absolute-score substitute. The overall rank swap in the common model set is at most $4$ (MMTEB) / $2$ (MTEB-v2). This appendix is relative ranking restricted to the common model set; any model without all tasks as a single-revision single measurement in the official \texttt{mteb/results} is excluded.

\subsection{Bootstrap confidence intervals of macro ranking}
\label{app:D4}
\begin{figure}[t]
\centering
\includegraphics[width=0.85\linewidth]{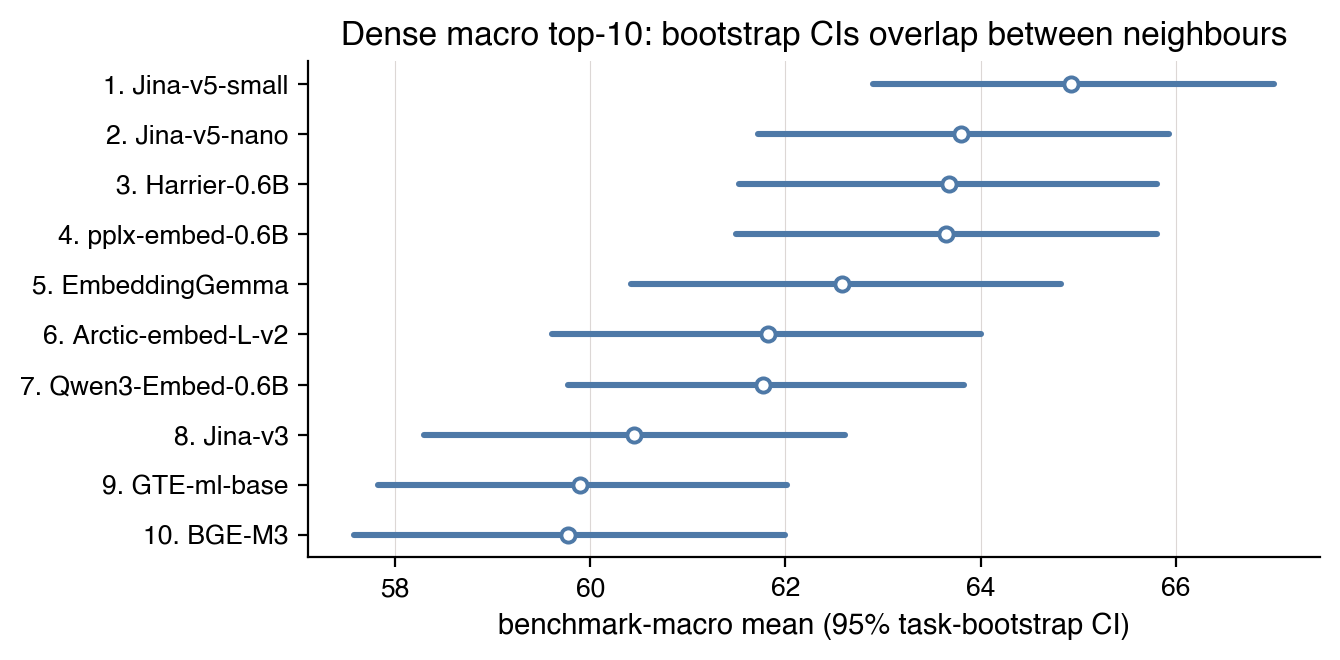}
\caption{Task-bootstrap $95\%$ confidence intervals for the top $10$ dense macro models.}
\label{fig:macroci}
\end{figure}

To show the magnitude of evaluation noise, Figure~\ref{fig:macroci} shows the $95\%$ confidence intervals for the top $10$ dense macro models from the \secref{sec:noise} task bootstrap (recomputing the macro average $2{,}000$ times by resampling tasks with replacement within each benchmark). The CI half-width averages $\pm 2.1$ points (max $\pm 2.3$), and neighboring models' intervals overlap greatly. Hence rank stability is determined by the score gap between models: pairs differing by about $1$ point or more virtually never swap (the rank-$1$ model never falls to rank $2$ over $2{,}000$ resamples), while pairs differing by around $0.1$ point swap with probability $30$--$40\%$. That is, a macro-average difference of under $1$ point should not be read as a rank.

\section{Efficiency settings and reranking details}
\label{app:E}

\subsection{Variant list}
\label{app:E1}
The dense efficiency variants are base, truncate (leading-dimension-preserving dimensionality reduction), int8, binary, rescore (rescoring the quantized-search top $100$ with float embeddings), and their cross product. The standard dense run, when no explicit variant is given, auto-computes int8, binary, and their rescore variants. Specifying explicit variants disables auto-derivation, so Matryoshka comparisons explicitly enumerate the standard dimension, dimensionality reduction, quantization, and their combinations. int8 is a per-dimension affine scalar quantization, not a simple cast to \texttt{float16}: from the corpus-side embeddings, take $\min_d$ / $\max_d$ per dimension $d$, set the step $\mathrm{step}_d = (\max_d - \min_d)/255$, map each value $x$ to a bucket by $(x - \min_d)/\mathrm{step}_d$, shift by $-128$, clip to $[-128, 127]$, and reduce to an 8-bit integer ($256$ levels; truncate the fractional part). Calibration is on the distribution-stable corpus side only; to avoid fitting buckets to evaluation queries, the query side uses the same corpus range and clips out-of-range values (no recalibration on query statistics). No separate calibration samples or calibration training are used (same family as the embedding quantization of \citealp{shakir2024quantization}, with calibration fixed to the corpus). Binary 1-bits each dimension by sign ($x > 0$) and packs the bits. The \texttt{*\_rescore} variants are the simplest two-stage retrieval, rescoring the quantized-search top $100$ with the float embeddings retained before quantization.

\subsection{Sparse pruning settings}
\label{app:E2}
Sparse-model pruning consists of single variants independently specifying the query-side and document-side \texttt{max active dims}, plus their cross-product variants. The query-side value determines the number of non-zero dimensions at search time, directly tied to search latency. The document-side value, in addition to latency, is directly tied to the inverted-index and embedding-matrix size, i.e., the production-time memory/disk footprint. For the SPLADE family we measured performance before/after pruning for combinations of query-side $q \in \{8,16,24,32\}$ $\times$ document-side $d \in \{64,128,256,512\}$. For \model{naver/splade-v3}, the average score decreases monotonically from $34.16$ at $q{=}32,d{=}512$ to $29.31$ at $q{=}8,d{=}64$; the document side nearly saturates beyond $256$, while cutting the query side below $16$ causes a large quality drop (\secref{sec:sparse}). The $q\times d$ grid figure in base ratio and the operating envelope are in \appref{app:F6} (Table~\ref{tab:f2}). Latency/memory measurement is out of scope per the \secref{sec:speed} policy; efficiency is read via active dims as a proxy.

\subsection{Candidate-set construction}
\label{app:E3}
The default candidate set a reranker re-orders is a hybrid candidate set that takes the tops from two first-stage retrievals and fuses them with RRF (Reciprocal Rank Fusion). Concretely, we build two candidate lists, (1) \textbf{BM25} (per-language tokenizer; top $500$ per query) and (2) \textbf{dense} (retrieval with \model{microsoft/harrier-oss-v1-270m}, with the \texttt{web\_search\_query} prompt on the query and cosine similarity of normalized embeddings; top $500$ per query), and fuse them with RRF (\texttt{rrf\_k=100}) to take the RRF top $100$ as the candidate set. RRF is a rank-based fusion that, for rank $r$ in each list, adds $1/(k + r)$ (where $k$ is \texttt{rrf\_k}, $100$ here). Combining BM25 and dense brings both lexical match and semantic proximity into the candidates, reducing first-stage misses.

This candidate set is \textbf{built once per task and fixed-stored on the dataset side, reused identically for the reranking of every evaluated model}. This establishes a fair comparison where all models compete only on ranking accuracy over the same candidate set (\secref{sec:reranker-method}). Only for queries with no positive in the RRF top $100$ do we apply the \textbf{safeguard} of appending one positive at the tail (rank $101$), ensuring every query has at least one relevant document (query coverage $100\%$). This does not guarantee inclusion of all relevant documents (about $87\%$ on dense average; \secref{sec:rerankanalysis}). A BM25-only candidate set (top $100$) is also stored as a lexical baseline, switchable as needed. The without-safeguard candidate-set metric (\texttt{reranking\_without\_safeguard}) is also co-computed to isolate the safeguard's uplift.

The per-language tokenizer breakdown for BM25 is as follows. Japanese, Chinese, Korean, Thai, and Vietnamese use the corresponding morphological analyzers (word segmenters); other languages use a regular-expression tokenizer based on Unicode word boundaries (Arabic, German, Spanish, French, Russian, etc.\ also use Snowball-family stemming). This prevents CJK/Thai/Vietnamese BM25 from being underestimated by naive whitespace splitting.

\subsection{Candidate coverage and reranker / dense comparison}
\label{app:E4}
Each task's diagnostic records include query coverage (fraction of queries with at least one relevant document), relevant-document coverage (fraction of relevant documents in the top candidates), the base and reranker scores and improvement, the candidate-set origin, and the runtime breakdown. On the $33$-dense-model average, query coverage was $100.0\%$ (by the safeguard) and relevant-document coverage $86.6\%$ (\secref{sec:rerankanalysis}). The improvement when a dense model re-evaluates itself over the hybrid candidate set averages $+1.9$ points ($+1.5$ without the safeguard metric), small (because the candidate set contains the dense top; \secref{sec:rerankanalysis}), differing in nature from the performance when a reranker re-orders the candidate set. Reading the two separately over the same candidate set lets candidate-generation performance and reranker ranking accuracy be analyzed apart. Benchmarks with low relevant-document coverage (e.g., NanoDAPFAM about $48\%$, NanoMTEB-Polish about $68\%$, NanoR2MED about $72\%$) are hard cases where candidate generation tends to miss positives, a signal to read independently of reranker ranking accuracy.

\paragraph{Decomposing reranker vs dense by type, scope, and query type.}
To make the comparison intuitive, based on the distribution of all models scored for reranking on each task ($52$ models; the old small multilingual cross-encoder \model{mmarco-mMiniLMv2} and the Japanese-specialized \model{japanese-reranker-xsmall-v2} are excluded from the field/tables/figures as extreme outliers obscuring the comparison), we express each model by its \textbf{$z$-score} ($z = (s - \mu)/\sigma$, where $s$ is the score and $\mu$, $\sigma$ are the mean and standard deviation of all models on that task; i.e., ``how many standard deviations above the field mean''). Normalizing per task reads relative strength against the whole field more stably than absolute scores (different scales across tasks) or the selection-biased ``difference from best dense.'' Rerankers split into two types: \textbf{cross-encoder} (encoder of BERT / XLM-R / ModernBERT / MiniLM, concatenating query and document to directly regress relevance; BGE / GTE / Jina / ettin, etc.) and \textbf{LLM-style reranker} (based on a large language model, using the predicted logit of the ``yes / no'' token following the query and document as the relevance score; here \model{Qwen/Qwen3-Reranker-0.6B}). Dense embedding models are the reference, scored as rerankers over the same candidate set.

Table~\ref{tab:e1} shows $z$-scores for $4$ representative dense models and $5$ rerankers, by all tasks / multilingual / English / short tasks (query $<70$ chars and document $<1000$ chars) / long tasks (query $>200$ chars or document $>3000$ chars).

\begin{table}[t]
\centering
\caption{Per-model $z$-scores ($\sigma$; reranking over the same candidate set, against the $52$-model field; larger means above the field mean).}
\label{tab:e1}
\footnotesize
\setlength{\tabcolsep}{5pt}
\renewcommand{\arraystretch}{1.12}
\rowcolors{2}{rowgray}{white}
\begin{tabular}{@{}>{\ttfamily}llrrrrr@{}}
\toprule
\rowcolor{headcol} \hdr{\normalfont Model} & \hdr{Type} & \hdr{All} & \hdr{ML} & \hdr{EN} & \hdr{Short} & \hdr{Long} \\
\midrule
Qwen3-Reranker-0.6B & \normalfont LLM reranker & \best{+1.17} & +1.11 & \best{+1.32} & +0.95 & \best{+1.62} \\
jina-v5-small & \normalfont dense & +1.10 & +1.09 & +1.11 & +1.06 & +1.17 \\
bge-reranker-v2-m3 & \normalfont cross-encoder & +0.83 & \best{+1.21} & $-0.16$ & +1.10 & +0.06 \\
jina-reranker-v2-ml & \normalfont cross-encoder & +0.81 & +0.93 & +0.48 & +0.92 & +0.31 \\
Qwen3-Embed-0.6B & \normalfont dense & +0.70 & +0.65 & +0.82 & +0.61 & +0.99 \\
ettin-reranker-400m & \normalfont CE (English-only) & +0.69 & +0.61 & +0.89 & +0.61 & +0.66 \\
gte-ml-reranker & \normalfont cross-encoder & +0.64 & +0.77 & +0.30 & +0.71 & +0.42 \\
bge-m3 & \normalfont dense & +0.54 & +0.78 & $-0.08$ & +0.77 & +0.24 \\
mE5-large & \normalfont dense & +0.51 & +0.73 & $-0.03$ & +0.79 & +0.00 \\
\bottomrule
\end{tabular}
\end{table}

\begin{figure}[t]
\centering
\includegraphics[width=\linewidth]{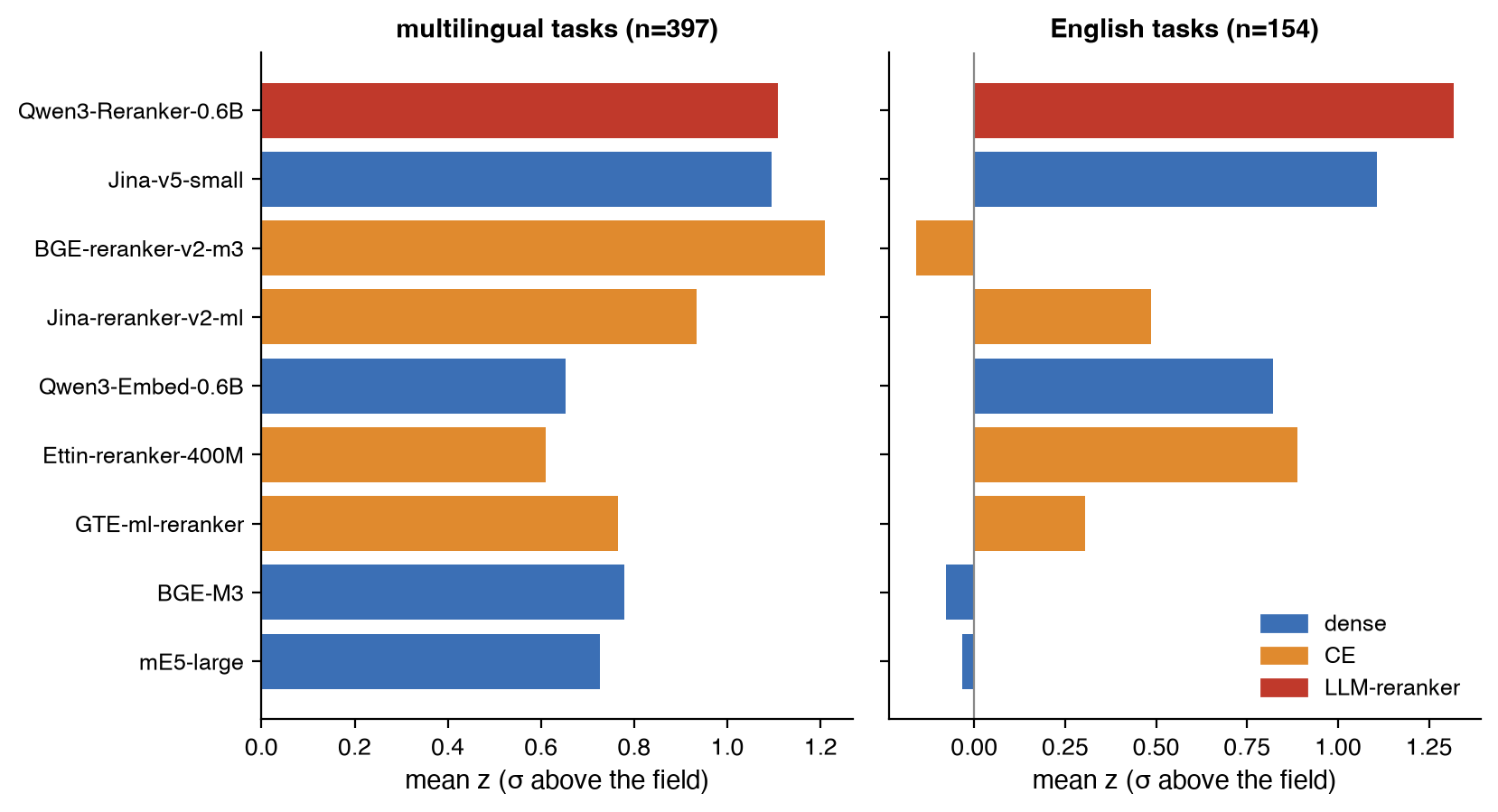}
\caption{$z$-scores of dense, cross-encoder, and LLM-style rerankers (left: multilingual tasks; right: English tasks).}
\label{fig:zscore}
\end{figure}

\paragraph{By scope (Figure~\ref{fig:zscore}).}
On multilingual tasks, the multilingual cross-encoder \model{BAAI/bge-reranker-v2-m3} ($+1.21$) tops the best dense \model{jinaai/jina-embeddings-v5-text-small} ($+1.09$), with \model{jina-reranker-v2} and \model{gte-multilingual-reranker} on par with the top dense group. That is, on the most common scope of multilingual semantic search, cross-encoder rerankers often beat the best dense. On English tasks the picture changes: the LLM-style \model{Qwen3-Reranker-0.6B} ($+1.32$) is first, then the best dense ($+1.11$), then the English-only cross-encoder \model{ettin-reranker-400m-v1} ($+0.89$). Multilingual cross-encoders sink greatly on English (bge-reranker $-0.16$, gte $+0.30$). On the overall macro the best dense beats the best multilingual cross-encoder (\secref{sec:rerankanalysis}), but this does not mean rerankers are useless: on the common task of ``quickly finding documents matching a short query,'' cross-encoders often rank higher.

\begin{figure}[t]
\centering
\includegraphics[width=0.8\linewidth]{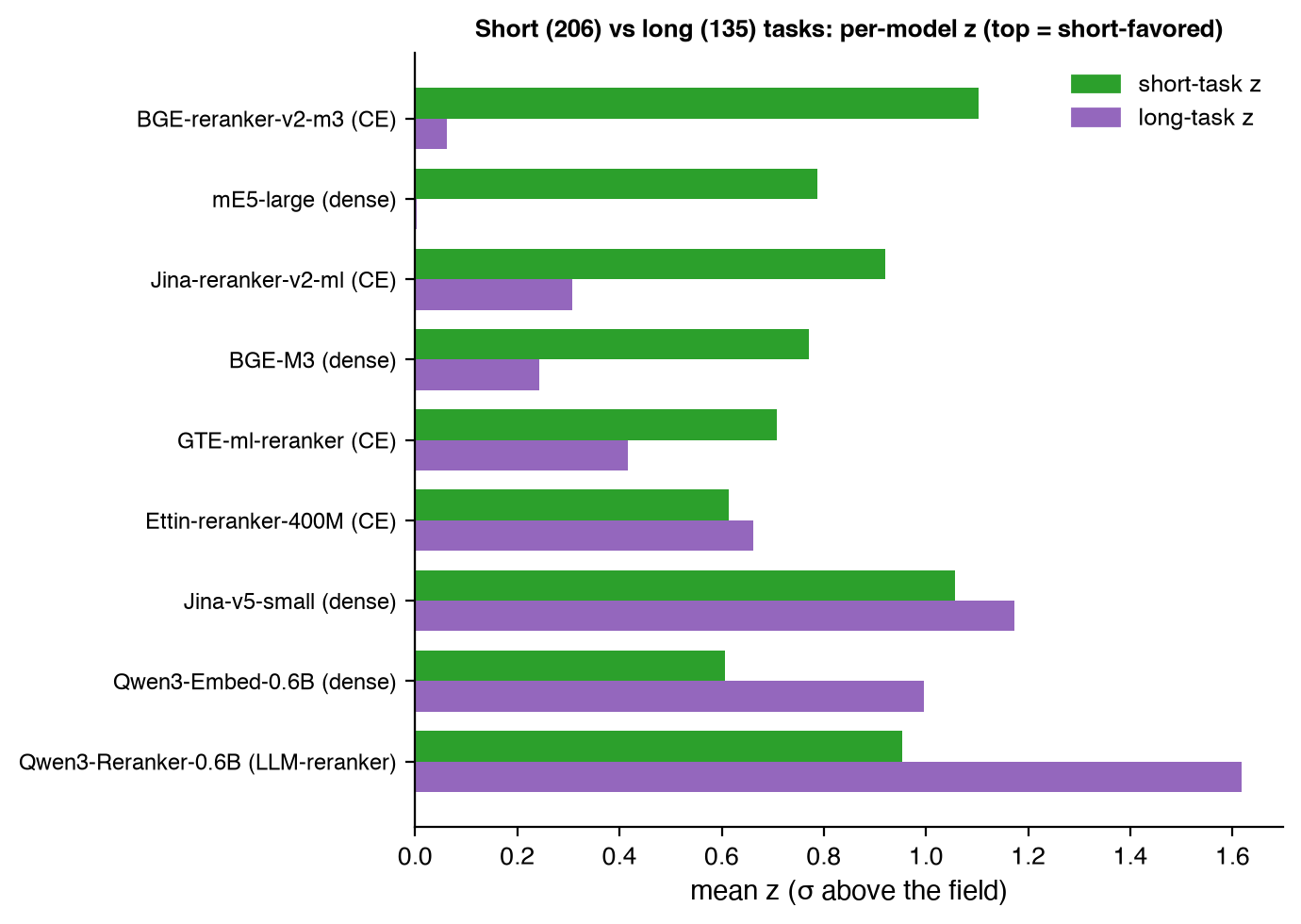}
\caption{Comparison of each model's short-task $z$ (query/document both short) and long-task $z$ (query or document long) as two bars (green = short-task $z$, purple = long-task $z$). Sorted by descending short $z$ $-$ long $z$; higher = short-favored, lower = long-favored. The type (dense / cross-encoder (CE) / LLM-reranker) is in parentheses after the model name.}
\label{fig:shortlong}
\end{figure}

\paragraph{By query/document type (Figure~\ref{fig:shortlong}).}
Comparing each model's short-task $z$ and long-task $z$ as two bars, clear trends emerge by type. Cross-encoders have short-task $z$ above long-task $z$ (short-favored); \model{bge-reranker-v2-m3} is short $+1.10$ / long $+0.06$, beating the best dense on short factual queries but collapsing on long queries/documents. This is because cross-encoders are trained on short-query/passage relevance (common retrieval data such as MS MARCO). By contrast, the LLM-style \model{Qwen3-Reranker-0.6B} has long-task $z$ above short-task $z$ (long-favored), at long $+1.62$, standing out on scopes with reasoning, instructions, and long text; its broad instruction-following capability from the LLM keeps it from collapsing on long discursive queries. Dense models have a small short--long gap and are moderate, not biased to a particular query type.

Looking more finely at query-length dependence, multilingual cross-encoders show short-query specialization. Stratifying \model{bge-reranker-v2-m3}'s advantage over dense by query length, it is consistently positive below $300$ chars (win rate $50$--$67\%$) but drops sharply to mean $-12.5$ points / win rate $24\%$ at $300$+ chars. By contrast, \model{Qwen3-Reranker-0.6B} is flat-to-slightly-positive across all lengths (win rate $57\%$ even at $300$+ chars), with no length penalty. This short-query specialization is not bge-specific but common to multilingual cross-encoders; restricting \model{gte-multilingual-reranker} to English tasks shows the same shape (parity on short, collapse on long). Note that \model{bge-reranker-v2-m3} falls below the best dense overall on English (English $z$ $-0.16$) not because it is weak at ``English itself'' but because this benchmark's English task set contains many code-retrieval and long-query reasoning/legal tasks outside bge's training distribution (MS MARCO short natural-language queries). Decomposing the advantage over dense on the $154$ English tasks: code ($25$ tasks) averages $-27.7$ points and queries $\ge 300$ chars average $-16.8$ points, while short English natural-language tasks (query $<70$ chars, $37$ tasks) average $-0.6$ points, essentially even. Thus the English disadvantage stems not from the language itself but from out-of-distribution task types (code, long reasoning) that are over-represented in English.

\paragraph{Real query examples.}
We give real examples of short factual queries favorable to rerankers and long reasoning queries favorable to dense.
\begin{itemize}[leftmargin=*]
\item Short factual (reranker-favored): English NanoMIRACL (avg $40$ chars) ``When did Marxism develop?'', ``Why is it called guerrilla?''; Japanese (avg $17$ chars) ``Where is Akiko Morigami from?''; Korean (avg $22$ chars) ``What is the capital of Luxembourg?''. All are exactly the short-query/passage distribution multilingual cross-encoders are trained on.
\item Long reasoning/instruction (dense / LLM-reranker-favored): NanoBRIGHT psychology (avg $693$ chars) paragraph-length reasoning questions like ``Can our beliefs change without reassessment or new evidence? \ldots''; NanoR2MED clinical (avg $2584$ chars) where the entire case record (\texttt{[Chief Complaint] \ldots [Current Medical History] \ldots}) is the query. Their ``relevant documents'' share underlying reasoning or techniques rather than surface word overlap, giving cross-encoders few cues.
\end{itemize}

\paragraph{Supplement: two axes.}
(a) Document length favors rerankers (single-vector bottleneck). On long-document multilingual retrieval NanoMLDR, cross-encoders and LLM rerankers both greatly beat dense (dense must compress thousand-character documents into one vector, while a cross-encoder reads the query and document jointly within its input window, subject to truncation at the reranker's maximum input length). (b) Similarity-type tasks favor dense. The scidocs family (citation prediction; query is a paper title/abstract, positive is a cited paper) leans dense across language versions, because it rewards broad topical similarity rather than query-to-answer relevance, the dense-embedding sweet spot. Note the English-only cross-encoder ettin (ModernBERT) strengthens monotonically with size ($17$M $\to$ $400$M, English $z$ up to $+0.89$), but its collapse axis is document length, not query length, degrading on long documents (in contrast to multilingual cross-encoders winning on long documents).

In summary, whether to adopt a reranker can be judged not by ``rerankers in general'' but by the task's query type and document length and the reranker type. On short factual queries multilingual cross-encoders tend to beat the best dense, on long reasoning/instruction queries or long documents LLM-style rerankers prevail, and on similarity-type tasks dense prevails. That this benchmark can measure reranker generalization itself over $500+$ diverse tasks can be a signal for developing more general-purpose rerankers. Note all $z$-scores here are candidate-set ranking accuracy under the candidate-set cap (\secref{sec:cands-limit}) and safeguard (\secref{sec:cands}); the absolute-value optimism from Nano-ization (\secref{sec:nano-diff}) remains---the claims concern relative superiority by type and query type.

\section{Real-data use cases (details)}
\label{app:F}
We detail, as six use cases, the model-adoption points summarized as three questions in \secref{sec:usecases}. Each example answers a practitioner's concrete question with measured values from these results, showing that decision material unavailable from a single overall leaderboard emerges only from same-harness measurement over many models $\times$ tasks $\times$ architectures $\times$ efficiency settings. All numbers are reproducible from the same results as \secref{sec:results} (queries and derivation scripts are in the repository under \texttt{facts/hakari-bench-results/12-usage-examples/}); aggregation is unified to the per-benchmark macro average that is our primary basis (\secref{sec:metrics}; the only exception is F.2 with $13$ tasks, using micro). As in \secref{sec:metrics}, the leaderboard default display is micro, and the numbers here correspond to switching to macro. The Nano-set caveats (\secref{sec:nano-diff}) apply throughout: each number is a proxy for rank/behavior, not the full-corpus absolute score.

\subsection{Retrieval: the best model and architecture depend on the scope}
\label{app:F1}
\begin{figure}[t]
\centering
\includegraphics[width=\linewidth]{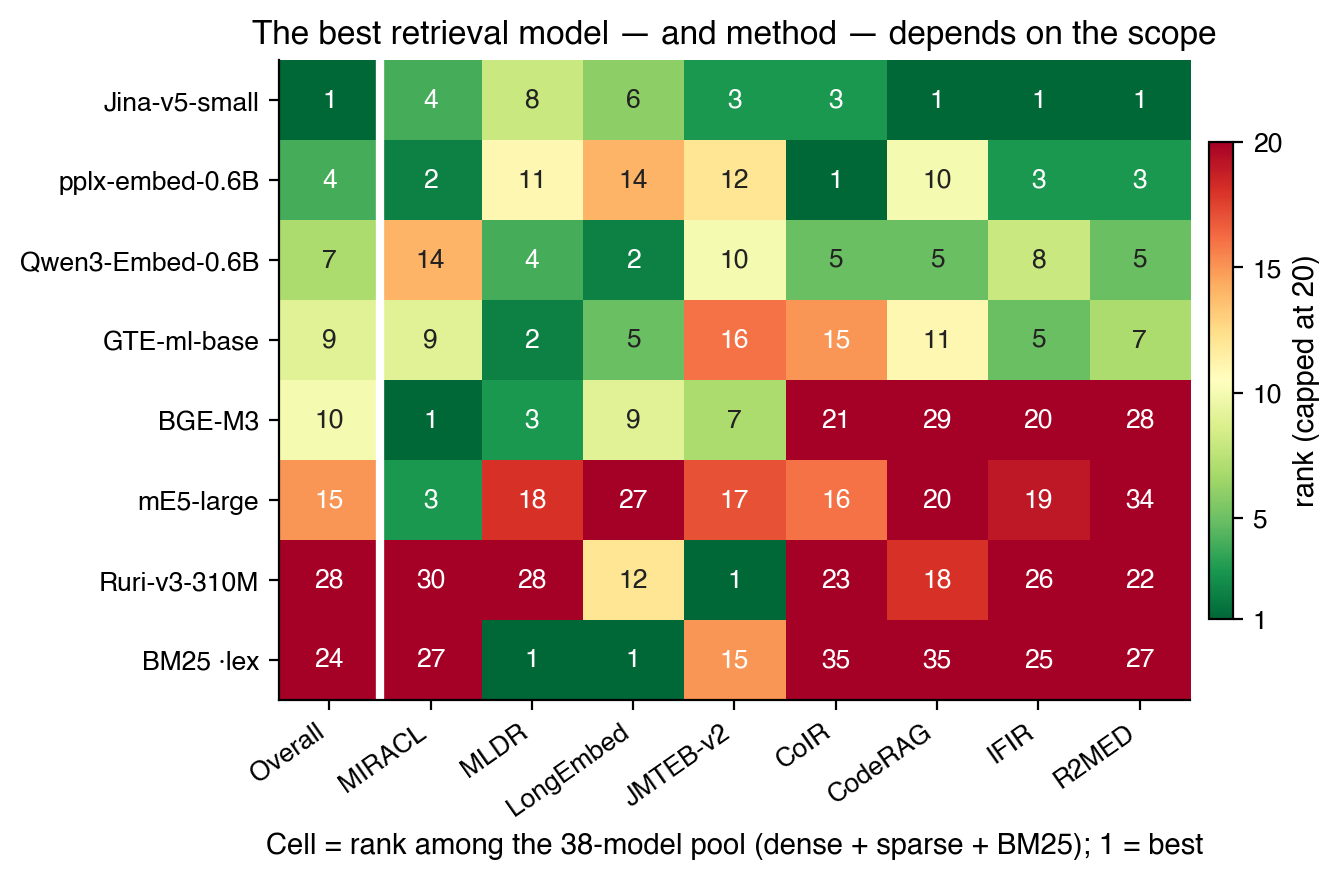}
\caption{Per-scope ranks of $38$ first-stage retrieval systems ($1$ = best; saturated display at rank $20$).}
\label{fig:retr-scope}
\end{figure}

We ranked $38$ first-stage retrieval systems by overall macro and by representative benchmarks (Figure~\ref{fig:retr-scope}). English-specialized models (\model{nomic-ai/nomic-embed-text-v1.5}, \model{naver/splade-v3}) degrade greatly on multilingual tasks and are excluded from this figure, treated in F.2.

No single dominator exists in any column, and the winners cross architectures. (i) On multilingual semantic search (NanoMIRACL: $18$ languages, short queries and short passages fitting a $512$-token window), \model{BAAI/bge-m3} (overall rank $10$) is rank $1$ and \model{intfloat/multilingual-e5-large} (overall rank $15$) rank $3$; this is exactly the setting these models are tuned for, and within this range they are at least on par with the latest top general models. (ii) On the two long-document series, \textbf{BM25 is rank $1$ on both} (overall rank $24$): NanoMLDR has $\approx 5$K--$28$K-char and NanoLongEmbed $\approx 28$K--$326$K-char documents, and many dense models truncate documents at the max sequence length so relevant passages fall outside, whereas BM25 matches the whole document lexically, independent of length. Among dense, the long-context-trained \model{Qwen/Qwen3-Embedding-0.6B} is best (LongEmbed rank $2$, MLDR rank $4$). (iii) On Japanese (NanoJMTEB-v2), the Japanese-specialized \model{cl-nagoya/ruri-v3-310m} (overall rank $28$) is rank $1$. Thus the best retrieval system is scope-dependent across architectures, and a single overall score is neither necessary nor sufficient for a practitioner's target scope.

\subsection{English NanoBEIR: late interaction and learned sparse become first-class choices}
\label{app:F2}
\begin{figure}[t]
\centering
\includegraphics[width=\linewidth]{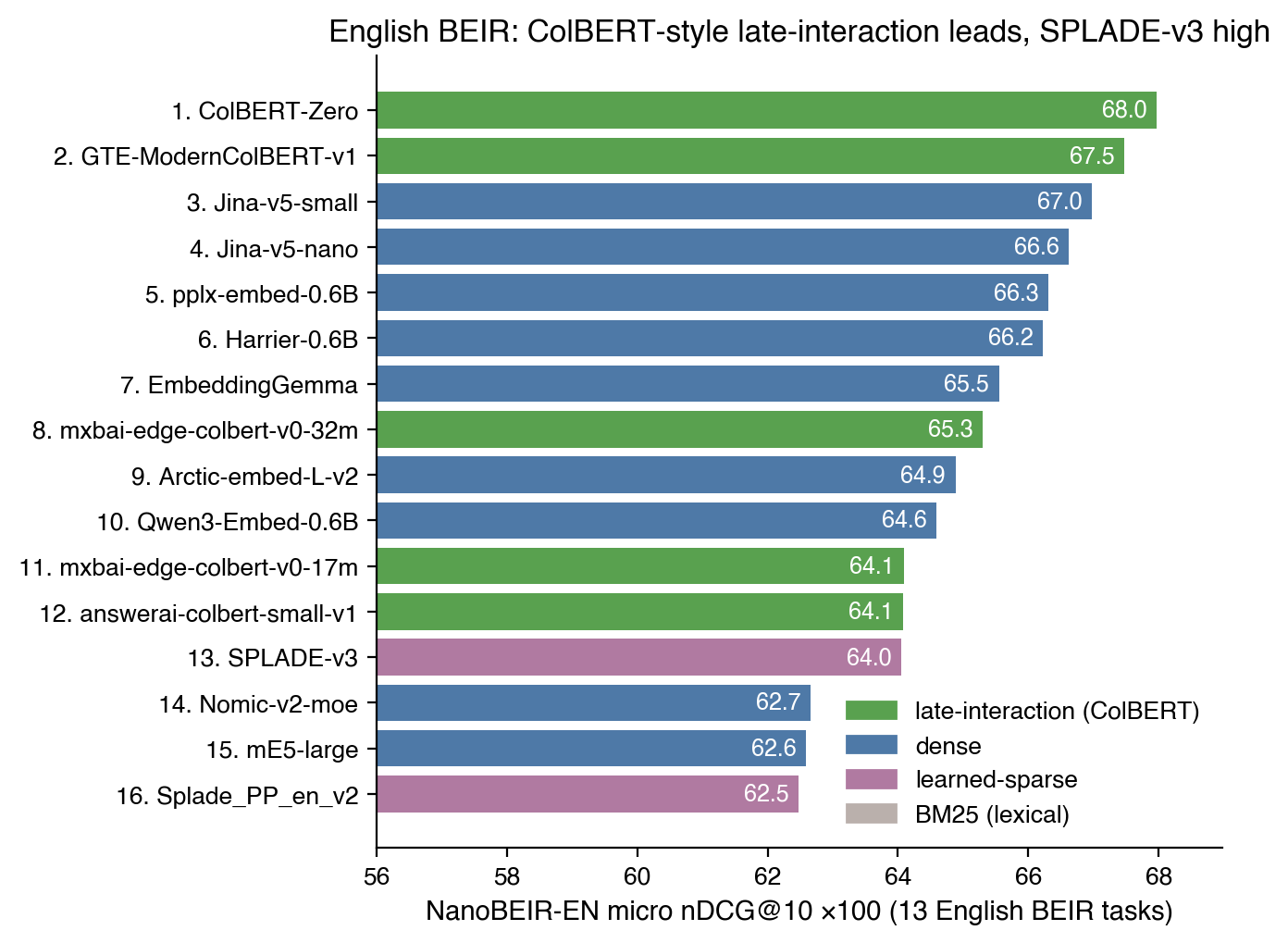}
\caption{Top of the English NanoBEIR ($13$ tasks, NanoBEIR-en) micro leaderboard.}
\label{fig:beiren}
\end{figure}

Architectures of English IR origin (ColBERT-family late interaction, SPLADE-family learned sparse) are English-centric and sink low on multilingual macro. Restricting to the $13$ English tasks within MNanoBEIR (NanoBEIR-en), we ranked $44$ systems including late interaction (dense $33$, learned sparse $4$, late interaction $6$, BM25 $1$) by micro average (Figure~\ref{fig:beiren}). The top is occupied by two late-interaction models, \model{lightonai/ColBERT-Zero} ($67.97$) and \model{lightonai/GTE-ModernColBERT-v1} ($67.47$), above the best dense (\model{jinaai/jina-embeddings-v5-text-small}, $66.97$). Five of the top $12$ are late interaction, and the learned sparse \model{naver/splade-v3}, trained only on MS MARCO, is rank $13$ of $44$ ($64.05$), within the top quartile. By contrast, BM25, the long-document winner of F.1, stays at rank $36$ ($57.15$) on these short-passage-centric tasks. Token-level MaxSim matching fits the exact-word-match-plus-local-context that many BEIR tasks require, and SPLADE's learned vocabulary expansion fits vocabulary-centric relevance. These are English-specialized architectures, weak on multilingual scopes, but that is precisely the demonstration that ``the right architecture depends on the language/task family.'' Evaluating dense, sparse, late interaction, and lexical on the same basis makes this trade-off visible rather than hidden in the aggregate.

\subsection{Reranking: the reranker advantage concentrates in the semantic-search scope}
\label{app:F3}
\begin{figure}[t]
\centering
\includegraphics[width=\linewidth]{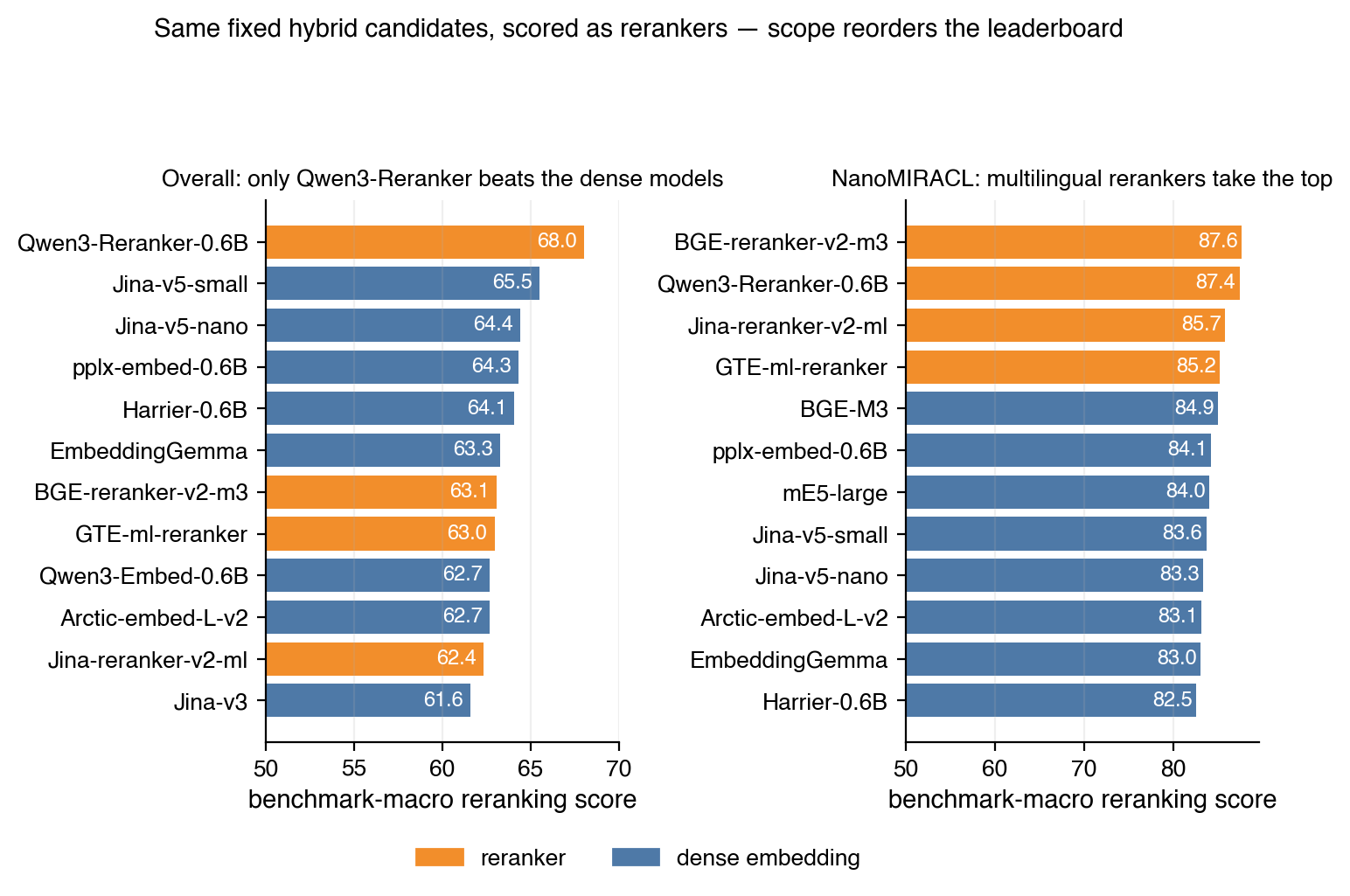}
\caption{Top composition of overall reranking (left) and NanoMIRACL (right).}
\label{fig:rerankscope}
\end{figure}

As in \secref{sec:reranker-method}, the benchmark scores all models as rerankers over the same fixed hybrid candidate set, so embedding models and rerankers can be compared directly. On the overall ($54$ models excluding BM25; macro, with safeguard), only the modern general reranker \model{Qwen/Qwen3-Reranker-0.6B} ($68.03$) exceeds the dense top, with ranks $2$--$6$ occupied by dense embedding models (\model{jinaai/jina-embeddings-v5-text-small} $65.51$, etc.). Classical multilingual cross-encoders (\model{BAAI/bge-reranker-v2-m3} rank $7$, \model{Alibaba-NLP/gte-multilingual-reranker-base} rank $8$, \model{jinaai/jina-reranker-v2-base-multilingual} rank $11$) stay in the middle. This is because rerankers trained on MS MARCO-style semantic-search data do not generalize as broadly as the top dense to the full diversity (code, reasoning, instruction-following, long documents, $40+$ languages). Restricting to NanoMIRACL, however, the top $4$ are all multilingual cross-encoders (\model{BAAI/bge-reranker-v2-m3} rank $1$ at $87.57$, overall $7 \to 1$), overtaking dense outright on their design-target multilingual semantic-search scope (Figure~\ref{fig:rerankscope}). The top-$6$ architecture composition of overall vs NanoMIRACL is opposite (dense $5$ + reranker $1$ vs reranker $4$ + dense $2$), showing the reranker advantage concentrates in scope. Note these multilingual rerankers may have used MIRACL training data, in which case---even without directly knowing the test/dev positives---indirect adaptation to the MIRACL-domain query/document distribution may inflate the NanoMIRACL score (same premise as the \secref{sec:modelscope} contamination discussion). Historically, dedicated reranking benchmarks were small, whereas this benchmark can measure reranker generalization itself over $500+$ tasks, which can be a signal for developing more general-purpose multilingual rerankers.

\subsection{Dimensionality reduction and quantization: mild, uniform, and model-specific costs}
\label{app:F4}
\begin{figure}[t]
\centering
\includegraphics[width=\linewidth]{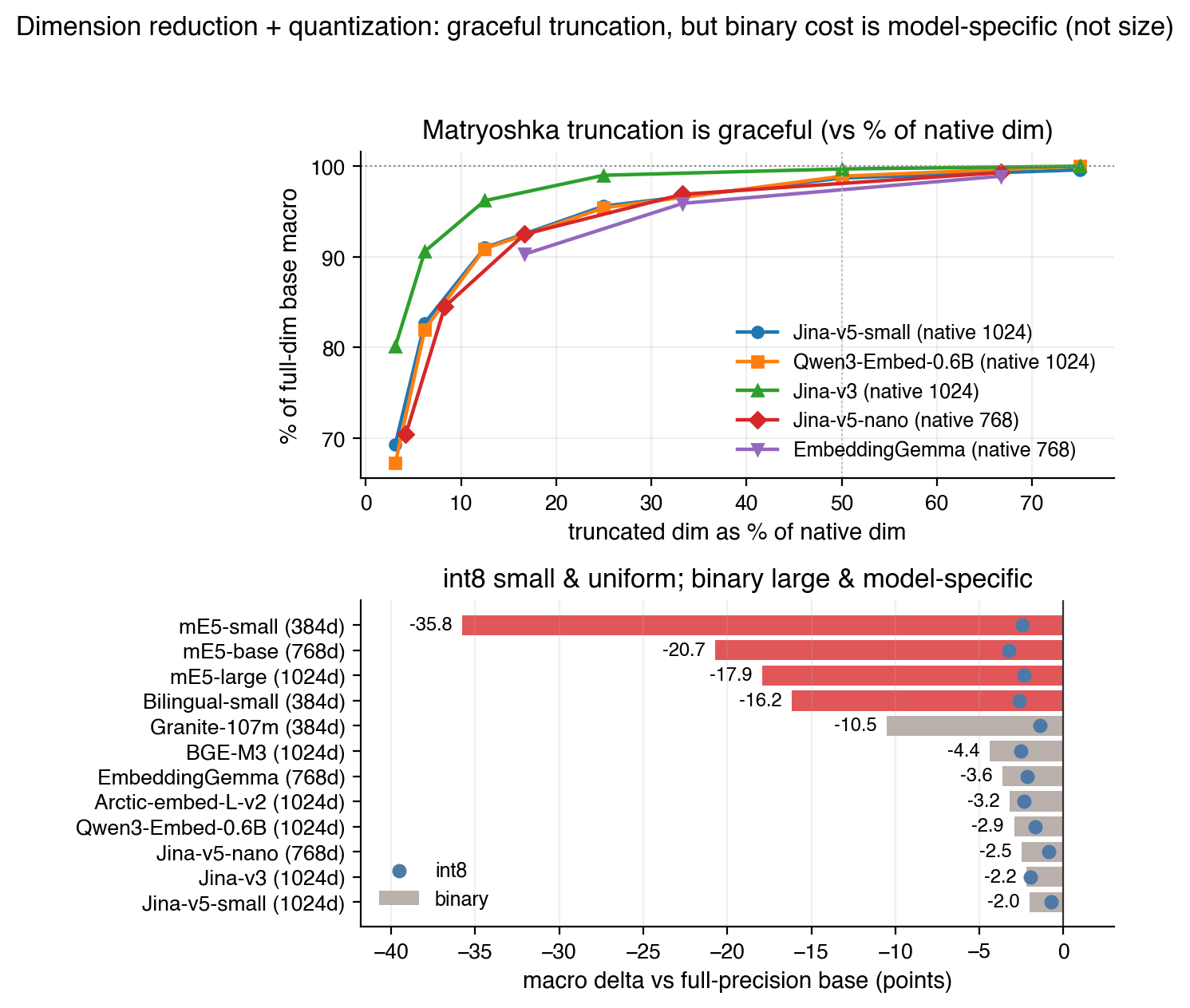}
\caption{Matryoshka dimensionality-reduction retention (left, native-dimension ratio) and int8/binary quantization degradation (right).}
\label{fig:dimquant}
\end{figure}

Applying the same efficiency settings (Matryoshka dimensionality reduction, int8/binary quantization) to $33$ dense models, we measured the macro delta vs.\ base (Figure~\ref{fig:dimquant}; rescore is treated separately in F.5). Note the \secref{sec:dimquant} deltas (binary $-6.50$, int8 $-1.95$, etc.) are equal-weight micro averages over all tasks, differing slightly from this appendix's per-benchmark macro averages. We describe dimensionality reduction, int8, and binary separately because their cost natures differ.

\textbf{Matryoshka dimensionality reduction is mild but must be read in native-dimension ratio.} ``$512$ dimensions'' is $50\%$ for a native-$1024$ model but $67\%$ for a native-$768$ model, so absolute-dimension comparison conflates truncation ratios. Aligned by native ratio, each model's retention curve nearly overlaps, keeping about $99\%$ of base macro at $50\%$ native (e.g., $1024 \to 512$) and about $95\%$ at $25\%$ ($1024 \to 256$). The flattest, \model{jinaai/jina-embeddings-v3} (native $1024$), keeps $96\%$ even at $12.5\%$ ($128$ dimensions).

\textbf{int8 is a small, uniform cost.} $33$-model average $-1.90$, worst $-3.25$; the quality drop is tiny regardless of model. int8 drops each dimension from a $4$-byte float to $1$ byte, reducing storage to about $1/4$ (\secref{sec:dimquant-method}). Because the quality drop ($\approx -1.9$ points) is nearly negligible while storage is reliably cut, on the quality--efficiency trade-off it can be used routinely as an ``almost free'' setting (meaning a setting that greatly saves storage/compute while keeping the retrieval-quality drop within error).

\textbf{binary is model-specific.} Average $-6.87$, but the range is wide, $-2.01$ to $-35.79$. The degradation is especially conspicuous in the \textbf{multilingual-E5 family} (mE5-small $-35.8$, mE5-base $-20.7$, mE5-large $-17.9$, and the E5-derived \model{Lajavaness/bilingual-embedding-small} $-16.2$): the five E5/E5-derived models average about $-19$, far below the other group (roughly $-2$ to $-10$). That English E5 (\model{intfloat/e5-base-v2}, \model{intfloat/e5-small-v2}; a different series from this benchmark's multilingual-E5) also degrades greatly under binary quantization is reported independently (\citealp{thoresen2026vespa}: ``E5-base-v2 drops to $92\%$, E5-small-v2 to $87\%$''). Conversely, models trained for quantization robustness (\model{jinaai/jina-embeddings-v5} family, \model{google/embeddinggemma-300m}, \model{Snowflake/snowflake-arctic-embed-l-v2.0}, \model{Qwen/Qwen3-Embedding-0.6B}) stay within $2$--$4$ points. This degradation is not explained by model size or embedding dimension: the correlation between binary degradation and dimension is weak ($+0.32$), and at the same $384$ dimensions mE5-small is $-35.8$ vs \model{sentence-transformers/all-MiniLM-L6-v2} $-4.4$, and at the same $1024$ dimensions \model{intfloat/multilingual-e5-large} is $-17.9$ vs \model{jinaai/jina-embeddings-v5-text-small} $-2.0$. That is, binary robustness is determined by \textbf{training characteristics}, not size or dimension. Only by applying the same setting to all supporting models can universal costs (dimensionality reduction, int8) and model-specific costs (binary) be separated this way.

\subsection{float rescore: an operation that preserves cross-model comparison, and its exception}
\label{app:F5}
rescore is the simplest two-stage retrieval, retrieving the top $100$ with the quantized embeddings and rescoring just those $100$ with the pre-quantization float embeddings (\secref{sec:dimquant-method}). Here we read it not as per-model recovery but as the \textbf{effect on cross-model ranking (which model to choose)}. The point is simple: \textbf{binary alone greatly reshuffles cross-model superiority, but adding rescore returns it almost to the float ranking.} On the $\approx 10$K-document corpus (this Nano-set), under binary + rescore, ``the best model in float = the best model in binary operation'' holds.

Table~\ref{tab:f1} shows representative dense models' full-dimension macro scores under float (base), int8, int8+rescore, binary, and binary+rescore. The Spearman rank correlation with the float ranking over all $33$ dense models differs by quantization method:
\begin{itemize}[leftmargin=*]
\item \textbf{int8 only $0.995$ / int8 + rescore $1.000$.} int8 nearly preserves the float ranking even without rescore, so for model selection the float leaderboard can be used as-is.
\item \textbf{binary only $0.937$ / binary + rescore $0.988$.} binary alone greatly disturbs the ranking but rescore returns it almost to float.
\end{itemize}

\begin{table}[t]
\centering
\caption{Macro scores by quantization method for representative dense models ($\nDCG{}\times 100$; \best{teal} marks the severe binary-only collapse of the multilingual-E5 family). Model IDs are abbreviated.}
\label{tab:f1}
\footnotesize
\setlength{\tabcolsep}{6pt}
\renewcommand{\arraystretch}{1.12}
\rowcolors{2}{rowgray}{white}
\begin{tabular}{@{}>{\ttfamily}lrrrrr@{}}
\toprule
\rowcolor{headcol} \hdr{\normalfont Model} & \hdr{float} & \hdr{int8} & \hdr{int8+rs} & \hdr{binary} & \hdr{binary+rs} \\
\midrule
jina-v5-small & 64.93 & 64.18 & 64.86 & 62.92 & 64.85 \\
Qwen3-Embed-0.6B & 61.77 & 60.07 & 61.68 & 58.84 & 61.68 \\
gte-ml-base & 59.89 & 58.87 & 59.78 & 55.20 & 59.69 \\
bge-m3 & 59.77 & 57.23 & 59.61 & 55.37 & 59.62 \\
mE5-large & 58.18 & 55.81 & 58.13 & \best{40.27} & 54.67 \\
mE5-base & 55.90 & 52.66 & 55.90 & \best{35.19} & 51.52 \\
mE5-small & 53.60 & 51.17 & 53.55 & \best{17.82} & 38.75 \\
\bottomrule
\end{tabular}
\end{table}

Table~\ref{tab:f1} can be read directly. \textbf{int8 is nearly equal to float regardless of method, and int8+rescore is practically identical to float.} On the other hand, \textbf{binary alone behaves very differently by model}: robust models (jina-v5-small $64.93 \to 62.92$) sink only slightly, but \textbf{the multilingual-E5 family collapses search recall itself under binary quantization} (mE5-large $58.18 \to 40.27$, mE5-small $53.60 \to 17.82$). \textbf{Adding binary+rescore recovers most models to near float} (jina-v5-small $64.85$, bge-m3 $59.62$), but the E5 family does not fully return because positives are sometimes missing from the binary-search top $100$ (mE5-small recovers only to $38.75$).

That is, \textbf{the float leaderboard can be used as-is for int8 model selection, but is unreliable for binary operation unless rescore is used.} Assuming binary + rescore, choosing the float-best model remains valid in binary operation. Furthermore, comparing $11$ MRL-capable models at the fixed operating point $256$ dimensions + binary + rescore ($32$ bytes/vector, $1/128$ the size of float $1024$ dimensions), they keep $88$--$98\%$ of base, and the truncation/quantization-most-robust \model{jinaai/jina-embeddings-v3} rises from float rank $6$ to operating-point rank $4$. Note rescore recovery depends on positives remaining in the quantized-search top $100$; the Nano-set scale ($\le \approx 10$K documents/task) makes this likely, so the recovery may shrink on larger corpora.

\subsection{learned sparse pruning: the document side is a cheap knob, the query side an expensive knob}
\label{app:F6}
We read \model{naver/splade-v3}'s independent query-side/document-side \texttt{max active dims} sweep (\secref{sec:sparse-method}, \secref{sec:sparse}) as a percentage of the un-pruned base row (macro $35.6$) (Table~\ref{tab:f2}; the $34.16$--$29.31$ of \secref{sec:sparse} are micro averages of the pruned variants, a different basis). Recomputing on English tasks only gives nearly the same percentages, so the pruning behavior is robust to task composition.

\begin{table}[t]
\centering
\caption{Pruning grid for \model{naver/splade-v3} (\% of the un-pruned base macro; \best{teal} marks the $\ge 99\%$ operating envelope, $q\ge 24$ and $d\ge 128$).}
\label{tab:f2}
\footnotesize
\setlength{\tabcolsep}{9pt}
\renewcommand{\arraystretch}{1.25}
\begin{tabular}{@{}lrrrr@{}}
\toprule
\rowcolor{headcol} \hdr{query $\backslash$ document} & \hdr{$d{=}64$} & \hdr{$d{=}128$} & \hdr{$d{=}256$} & \hdr{$d{=}512$} \\
\midrule
$q{=}8$ & 86.6\% & 88.8\% & 89.7\% & 89.8\% \\
$q{=}16$ & 92.6\% & 96.1\% & 97.1\% & 97.2\% \\
$q{=}24$ & 93.9\% & 98.3\% & \best{99.4\%} & \best{99.5\%} \\
$q{=}32$ & 94.5\% & \best{99.1\%} & \best{100.6\%} & \best{100.6\%} \\
\bottomrule
\end{tabular}
\end{table}

The grid is clearly asymmetric. \textbf{The document-side knob ($d$) is directly tied to the search engine's storage/index size, and the query-side knob ($q$) to search-time compute (posting-list processing, latency).} The document side can be cut aggressively: $512 \to 256$ is lossless ($100.6\%$), $512 \to 128$ keeps $99.1\%$, and only $d{=}64$ causes about a $5.5\%$ drop. The query side is sensitive: $q{=}32 \to 24$ keeps $99.5\%$, but $q{=}16$ drops about $3\%$ and $q{=}8$ about $10\%$. The operating envelope keeping $\ge 99\%$ of base is \textbf{$q \ge 24$ and $d \ge 128$}.

As an example max-cap setting, choosing \textbf{document side $d{=}256$, query side $q{=}24$} keeps \textbf{$99.4\%$ of base}. At this point, limiting the document side to $256$ to suppress index size while limiting the query side to $24$ to suppress search-time compute keeps the quality drop at effectively $0.6\%$. Thus this result is material for deciding, for a given sparse model, how far the index-time document-side dimension and the search-time query-side cap can be reduced with little quality loss. From a single uniform sweep, one obtains a deployable operating guideline: the document footprint can be cut cheaply, but the query-side cost is hard to reduce.

\section{Availability and licensing}
\label{app:G}
The project source code (the evaluation and visualization implementation) is released under the MIT license on GitHub. The evaluation data (Nano-sets) is released on Hugging Face Datasets. The license of each Nano-set follows the license of the original dataset it is built from; users must comply with the license terms of each original source. Among the Nano-sets, the NanoBEIR-\{lang\} family reuses already-published Nano-sets (to keep evaluation consistent with the original), while the other Nano-sets were constructed by the author of this paper.

\end{document}